\documentclass[twocolumn]{aastex631}

\received{March 1, 2023}
\revised{June 18, 2023}
\accepted{June 20, 2023}

\shorttitle{The Milky Way halo in APOGEE\,DR17}
\shortauthors{Feltzing et al.}
\graphicspath{{./}{figures/}}

\begin{document}

\title{The metal-weak Milky Way stellar disk hidden in the
  \textit{Gaia}-Sausage-Enceladus debris: the APOGEE\,DR17 view.}

\correspondingauthor{Sofia Feltzing}
\email{sofia.feltzing@geol.lu.se}
\email{sofia.feltzing@astro.lu.se}

\author[0000-0002-7539-1638]{Sofia Feltzing}
\affiliation{Lund Observatory, Department of Astronomy and Theoretical  Physics, Box 43,
SE-221\,00 Lund, Sweden}
\affiliation{Lund Observatory, Department of Geology, S\"olvegatan 12,
  SE-223\,62 Lund, Sweden}

\author[0000-0002-3101-5921]{Diane Feuillet}
\affiliation{Lund Observatory, Department of Astronomy and Theoretical   Physics, Box 43,
SE-221\,00 Lund, Sweden}
\affiliation{Lund Observatory, Department of Geology,
S\"olvegatan 12, SE-223\,62 Lund,  Sweden}




\begin{abstract}

  
  We have for the first time identified the early stellar disk in the
  Milky Way by using a combination of elemental abundances and
  kinematics. Using data from APOGEE DR17 and \textit{Gaia} we select
  stars in the Mg-Mn-Al-Fe plane with elemental abundances indicative
  of accreted origin and find stars with both halo-like and disk-like
  kinematics. The stars with halo-like kinematics lie along a lower
  sequence in [Mg/Fe], while the stars with disk-like kinematics lie
  along a higher sequence. Through with asteroseismic observations, we
  determine the stars with halo-like kinematics are old, 9--11 Gyr and
  that the more evolved stellar disk is about 1--2 Gyr younger.

  We show that the in situ fraction of stars on deeply bound orbits is
  not small, in fact the inner Galaxy likely harbours a genuine
  in-situ population together with an accreted one. 

In addition, we show that the selection of
\textit{Gaia}-Sausage-Enceladus in the $E_{\rm n}-L_{\rm z}$-plane is
not very robust. In fact, radically different selection criteria give
almost identical elemental abundance signatures for the accreted stars.
  
\end{abstract}


\section{Introduction}
\label{sec:intro}

The formation and evolution of the Milky Way can be studied via the
ages, elemental abundances and kinematics of its stars
\citep{2002ARA&A..40..487F}. The stars present in the stellar
components of the Galaxy are thought to have formed via two main
processes: in situ formation and accretion. In situ stars would have
formed in the body of the main progenitor of the galaxy and the
accreted stars fin the galaxies that later merged with the main
progenitor. The accreted stars would have chemical signatures
different from those formed in situ since they have formed in
shallower potential wells. However, some of the oldest stars formed in
a Milky Way-like progenitor may occupy the same elemental abundance
space as the accreted stars \citep{2022MNRAS.tmp.3011H}.

The Milky Way halo is thought to comprise stars formed in the early
Milky Way as well as stars formed in nearby low-mass dwarf galaxies
that have been accreted over time
\citep{2019A&A...632A...4D,2020ARA&A..58..205H,2020MNRAS.493..847F,2022arXiv221207441H}. Historically,
these accreted stellar populations have been identified as spatial
overdensities or kinematically associated stars
\citep{1999Natur.402...53H,2006ApJ...642L.137B,2013NewAR..57..100B,2020ARA&A..58..205H}
and have mainly been limited to mergers that have not yet dissolved
into the Milky Way field populations.  Following \textit{Gaia} Data
Release 1 \citep[DR1,][]{2016A&A...595A...2G} and Data Release 2
\citep[DR2,][]{2018A&A...616A...1G}, substantial accreted material was
identified in the Milky Way.

The most prominent, newly discovered stellar population is the
\textit{Gaia}-Sausage-Enceladus \citep{2018Natur.563...85H,
  2018MNRAS.478..611B, 2019MNRAS.488.1235M, 2020ApJ...901...48N}.
Apart from the \textit{Gaia}-Sausage-Enceladus, several other debris
have been identified using \textit{Gaia} data
\citep[e.g.][]{2019MNRAS.488.1235M, 2019A&A...631L...9K,
  2020ApJ...901...48N}. Most of these newly identified populations
generally have metallicities lower than those of the Milky Way disk
and bulge \citep[see e.g.][Fig. 7 for some
examples]{2022ApJ...926..107M}.

On a larger scale, it was found that the high velocity Milky Way halo
has a dual sequence in the colour-magnitude diagram
\citep{2018A&A...616A..10G}. \citet{2018Natur.563...85H,
  2018ApJ...863..113H, 2019ApJ...881L..10S, 2019NatAs...3..932G} show
that for stars with large transversal velocities (defined as
$V_T > 200$ km s$^{-1}$), these two populations show different
metallicity distribution functions, tentatively different elemental
abundance trends, and different ages pointing to different origins,
potentially we are seeing an in situ population (the red sequence) and
an accreted population (the blue sequence).  In addition,
\citet{2020MNRAS.494.3880B} identified a structure they dubbed the
Splash, which they take to be the result of the merger of the
\textit{Gaia}-Sausage-Enceladus. This merger perturbed stars in the
existing stellar disk creating a heated
component. \citet{2020MNRAS.494.3880B} identify the Splash as stars
with $|V_{\theta}| < 100$ km\,s$^{-1}$ and [Fe/H]~$> -0.8$~dex.

\citet{2019A&A...632A...4D} argue that there is no in situ halo just
an accreted one. In this interpretation, the two sequences in the halo
colour-magnitude diagram are the cumulative accreted populations and
the heated early disk. However, \citet{2020ApJ...891L..30A} provide
simulations that show that it is feasible to create something akin to
the Splash without any merger taking place. They argue for a formation
of stars in a heated medium. It appears that observational data has not yet been
sufficient to distinguish between different formation scenarios.

Although many accreted stellar populations have been identified in the
Milky Way halo through their kinematics, the main criteria
distinguishing stars as having formed outside the Milky Way is a
difference in the elemental abundances. The elemental abundance ratios
measured in the atmospheres of low-mass stars have long been used as a
tool to probe the evolutionary history of galaxies. Not only does the
abundance of individual elements increase over time as evolved stars
contribute enriched material to the interstellar medium, but the star
forming conditions of a given galaxy will affect the elemental
abundance trends measured in its stellar populations.

Theoretically, it is well understood that the enrichment of
$\alpha$-elements is dependent on the star formation rate
\citep{2012ceg..book.....M}. This is confirmed in observations of
Local Group dwarf galaxies where star formation proceeds more slowly
\citep{2009ARA&A..47..371T}. However, recently other elements have
empirically been found to show significant differences in stars formed
in high-mass or low-mass stellar systems. The APOGEE survey
\citep{2017AJ....154...94M} has provided aluminum abundance
measurements for $\sim 570,000$ stars in the Milky Way and nearby
dwarf galaxies. From this dataset, it has been shown that the accreted
\textit{Gaia}-Sausage-Enceladus stellar population and nearby dwarf
galaxies have [Al/Fe] abundance ratios $\sim 0.5-1.0$ dex lower than
the Milky Way disk across a large [Fe/H] range
\citep[e.g.][]{2021arXiv210905130H, 2022MNRAS.tmp.3011H}.

\citet{2015MNRAS.453..758H} and \citet{2020MNRAS.493.5195D} suggest
that the Mg-Mn-Al-Fe-plane could have high diagnostic power for
identifying accreted populations in the Milky Way. Observational
studies have found that many of the kinematically identified accreted
stellar populations in the Milky Way lie within the proposed accreted
region of this Mg-Mn-Al-Fe-plane \citep[e.g.][]{2021MNRAS.508.1489F,
  2022MNRAS.tmp.3011H}. Initial attempts to understand this plane
using chemical evolution models have found a difference in the
expected distribution of low-mass \textit{Gaia}-Sausage-Enceladus-like
galaxies, present day dwarf galaxies and Milky Way-like galaxies.
\citet{2021MNRAS.500.1385H,2023MNRAS.519.3611F} provide chemical
evolution models in the Mg-Mn-Al-Fe-plane for these different types of
galaxies.

In this paper we use the latest APOGEE dataset combined with
astrometry from \textit{Gaia} and the Mg-Mn-Al-Fe-plane to explore a
subset of stars that have elemental abundances indicative of accreted
or early Milky Way origin.

This paper is organised as follows: Section\,\ref{sect:data} describes
our selection of data from APOGEE\,DR17 and \textit{Gaia}.
Section\,\ref{sect:chem} discusses the chemical signatures of accreted
populations, in particular the Mg-Mn-Al-Fe-plane and potential
problems using this plane related to our shortcomings in deriving
elemental abundances reliably.  Section\,\ref{sect:pop_def} defines
our samples using the Mg-Mn-Al-Fe-plane while
Sect.\,\ref{sect:kinprop} explores the kinematic and chemical
properties of specific samples.  Section\,\ref{sect:dating} attempts
to date the different samples using ages available in the literature
while Sect.\,\ref{sect:discussion} provides a discussion.  The paper
concludes with a summary in Sect.\,\ref{sect:conclusion}.

  \section{Data}
  \label{sect:data}

\begin{figure*}
    \centering
    \includegraphics[width=16cm]{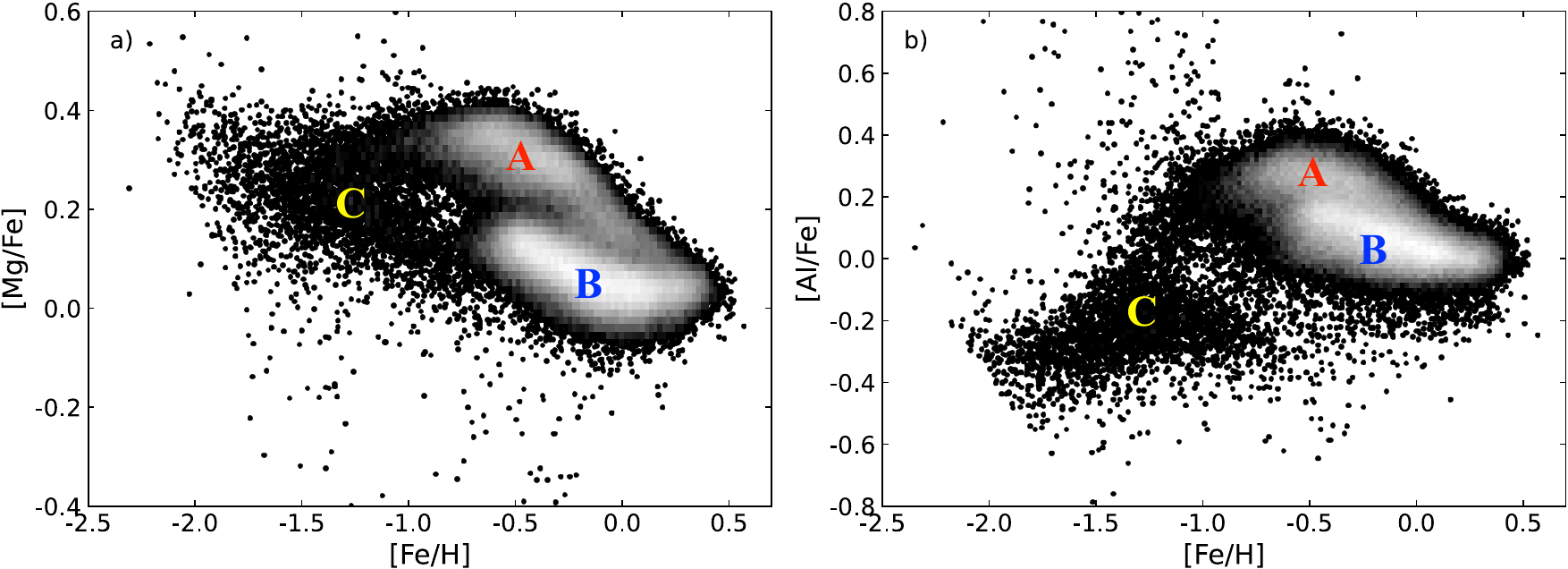}
    \caption{Elemental abundance distribution for our full sample
      selected from APOGEE\,DR17. {\bf a)} [Mg/Fe] vs [Fe/H] and {\bf
        b)} [Al/Fe] vs [Fe/H]. Details of the selection of data can be
      found in Sect.\,\ref{sect:data} and Table\,\ref{tab:data}
      provides an overview of the flags used in the selections from
      the main catalogues of  APOGEE\,DR17 and \textit{Gaia}.
    The
      labels A, B, C are inserted to help the reader to identify the three
    main components seen in this plot in relation to the description
    given in Sect.\,\ref{sect:data}: A -- classical thick disk, B --
    classical thin disk, C -- additional low-$\alpha$ component.}
    \label{fig:fullabund}
\end{figure*}

  We use data from the 17th Data Release (DR17) of the high-resolution
  spectroscopic SDSS APOGEE survey \citep{2017AJ....154...94M,
   2021arXiv211202026A}\footnote{The data was retrieved from the
   SDSS-IV archive at \url{https://www.sdss4.org/dr17/irspec/spectro_data/}.} combined with the {\it Gaia} Early Data Release 3
  \citep[Gaia EDR3,][]{2016A&A...595A...1G,2021A&A...649A...1G}
  astrometric data\footnote{For the purpose of the analysis present in
  this paper there is no difference between using EDR3 and DR3 for
  \textit{Gaia} as the astrometric solution is the same. \textit{Gaia}
  parameters additional to those provided by APOGEE were queried via
  the ESA archive \url{https://gea.esac.esa.int/archive/}}. To ensure
  robust stellar parameters, elemental abundances, and kinematic
  measurements, we impose the  selection criteria listed in
  Table\,\ref{tab:data}. 

\begin{table}
\caption{Selection criteria used to select data from APOGEE\,DR17 and
  \textit{Gaia}. Further details can be found in Sect.\,\ref{sect:data}.}  
\begin{center}
\begin{tabular}{lllll}
 \multicolumn{3}{l}{\textit{APOGEE\,DR17}}\\
  Flag  & \multicolumn{2}{l}{Criterium} \\
  \hline
    \tt{SNREV} & $>$ & 80\\
 \tt{TEFF} &$<$& 6000\\
 \tt{TEFF} &$>$& 4000 \\
  \tt{LOGG} &$<$& 2.8  \\
 \tt{FE\_H\_FLAG} &=& 0\\
 \tt{MG\_FE\_FLAG} &=& 0\\
 \tt{MN\_FE\_FLAG} &=& 0\\
 \tt{AL\_FE\_FLAG} &=& 0\\
 \tt{STARFLAG} & $\neq$ & \tt{VERY\_BRIGHT\_NEIGHBOR} or \tt{PERSIST\_HIGH}\\
 \tt{ASPCAPFLAG} &$\neq$& {\tt STAR\_BAD}, {\tt CHI2\_BAD}, {\tt M\_H\_BAD}, or {\tt CHI2\_WARN} \\
 \tt{EXTRATARG} & $\neq$ & \tt{DUPLICATE} \\
\hline
  \\
  \multicolumn{3}{l}{\textit{Gaia}}\\
  Property & \multicolumn{2}{l}{Criterium}\\
  \hline
  {$\sigma_{\pi}/\pi$} &  $< $& 0.2\\
 \hline
  \\
  \multicolumn{3}{l}{\textit{Additional selection for sample with no RC stars}}\\
  Property & \multicolumn{2}{l}{Criterium}\\
  \hline
  \tt{LOGG} &  $<$ & 2\\
 \tt{LOGG} &  $>$ & 1\\
 \hline
\end{tabular}
\end{center}
  \label{tab:data}
 \end{table}

 The sample is limited to giants to avoid an apparently anomalous
 feature in the [Al/Fe] vs [Fe/H] distribution of metal-rich dwarf
 stars. Although we focus mainly on lower metallicity stars in this
 work, aluminum is one of the four diagnostic elements used in our
 analysis and high data quality is important. The limits on effective
 temperature ensure only giants with reliable elemental abundance
 measurements are included. For more details on the APOGEE\,DR17 data
 quality see the SDSS DR17 documentation available on the
 web\footnote{\url{https://www.sdss.org/dr17/}} and a detailed discussion of
 DR16 in \citet{2020AJ....160..120J}.  In addition, the sample is
 cleaned of stars in fields targeting known star clusters and dwarf
 galaxies. A full list of the field and program names removed can be
 found in App.\,\ref{app:fields}.  This constitutes our full
 sample. In what follows we, following \citet{2019ApJ...874..102W},
 often limit the data set further to exclude red clump stars, i.e.
 $1 < \log g < 2$.

 The [Mg/Fe] and [Al/Fe] abundance distributions as a function of
 [Fe/H] of the full sample are shown in Fig.\,\ref{fig:fullabund}.  In
 Fig.\,\ref{fig:fullabund} we have indicated the classical thick disk
 sequence with a {\tt A} in red and the classical thin with a {\tt B}
 in blue.  An additional low-[Mg/Fe] component can be seen at the
 lower metallicities, which has been identified by previous studies as
 a signature of an accreted population
 \citep{2010A&A...511L..10N,2018ApJ...852...49H,2018ApJ...863..113H,2018Natur.563...85H}. In
 Fig.\,\ref{fig:fullabund} we indicate the position of this sequence
 with a {\tt C} in yellow. These stars also have [Al/Fe] lower than
 the Milky Way disk, another potential signature of an accreted origin
 \citep{2015MNRAS.453..758H,2021MNRAS.500.1385H,2021MNRAS.508.1489F}.

Full kinematics and 3D positions are calculated for the sample using
Astropy \citep{2013A&A...558A..33A, 2018AJ....156..123A} and {\it
  galpy} \citep{2015ApJS..216...29B}. Input data used are
\textit{Gaia} astrometric measurements and magnitudes, photogeometric
distances derived by \citet{2021AJ....161..147B}, and APOGEE
spectroscopic radial velocity measurements. We use the
\textit{actionAngleStaeckel} approximation \citep{2013ApJ...779..115B,
  2012MNRAS.426.1324B} with the {\tt MWPotential14} Milky Way
potential \citep{2013ApJ...779..115B}, delta of 0.4, and default
values for all other parameters.

We estimated the uncertainties in the kinematics through a Monte Carlo
analysis using 10 000 iterations. For each iteration, full kinematics
were calculated using parameters drawn from an uncertainty
distribution. A Gaussian distribution was used for the distance and
radial velocity. For the RA proper motion and Dec proper motion, a
multivariate Gaussian distribution was used accounting for the
\textit{Gaia} covariance in RA proper motion, Dec proper motion, and
parallax as well as the uncertainty in the individual parameter. The
uncertainty in the kinematic parameters for each star were taken to be
the standard deviation in each parameter over the 10 000 iterations.

  As the main focus of this paper in on a few subsamples, we calculate
  kinematic uncertainties for all of Sample I (see below) and a
  representative selection of Sample III. We find the largest
  kinematic uncertainties in stars on high-energy, low-$L_z$ orbits as
  these stars tend to be at larger distances. However, the typical
  uncertainties are less than 10\% in all kinematic parameters and do
  not affect our conclusions. Effects of select kinematics are
  discussed where relevant in the text and the distributions of
  kinematic uncertainties are shown in Appendix \ref{app:mcmillan}.

\section{Defining and naming stellar samples selected in the Mg-Mn-Al-Fe-plane}
\label{sect:pop_def}

In this section we make an empirical definition of the samples we
wish to study. These definitions are based on the appearance of the
data in the Mg-Mn-Al-Fe-plane. In Sect.\,\ref{sect:chem} we will
ascertained that the division of Milky Way stars into two samples
using the Mg-Mn-Al-Fe-plane is robust against shortcomings in the
elemental abundance analysis . We thus proceed with a discussion of
how to best define and name our samples.

\begin{figure*}
\begin{centering}
\includegraphics[width=18cm,trim={0 0cm 0 1cm},clip]{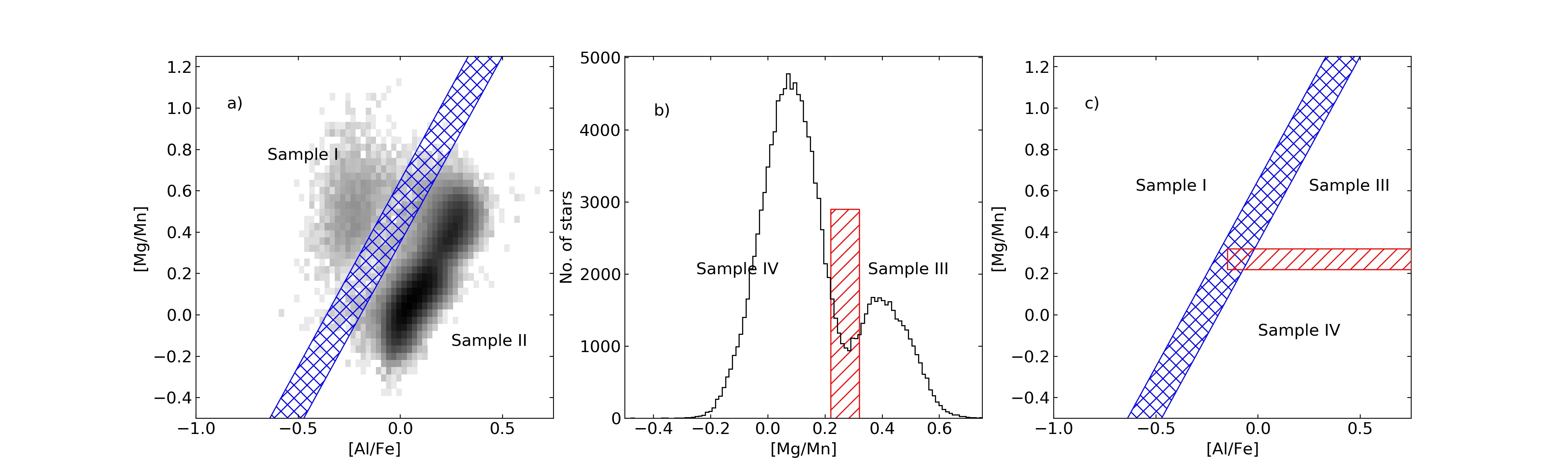}
\caption{{\bf a)} Full sample, showing definition of the diagonal cut,
  which results in Sample\,I and II. {\bf b)} Histogram of the [Mg/Mn]
  values for stars belonging to Sample\,II. The histogram identifies a
  suitable cut to define Sample\,III and IV.  {\bf c)} Shows the two
  cuts used and the position of Sample\,I, III, and IV in the
  Mg-Mn-Al-Fe-plane. The exact definitions of the cuts are summarised
  in Table\,\ref{tab:samples}.}\label{fig:cuts}
\end{centering}
\end{figure*}

Figure\,\ref{fig:cuts}\,a) empirically defines our cut in the
Mg-Mn-Al-Fe-plane, splitting the sample into two (Sample\,I and
Sample\,II).  The validity of the diagonal cut is extensively
discussed in Sect.\,\ref{sect:chem}. Figure\,\ref{fig:cuts}\,b) shows a
histogram of [Mg/Mn] for the stars in Sample\,II. We see a clear dip
in the number counts (also visible in the 2D histogram in panel a). We
define a cut such that stars with [Mg/Mn] $>$ 0.32 is one sample
(Sample\,III) and stars with [Mg/Mn] $<$ 0.22 another
(Sample\,IV). Table\,\ref{tab:samples} summarises our empirical
definitions of the samples. The current study is mainly concerned with
Sample\,I and to some extent Sample\,III. We plan to return to the
full Sample\,II and also Sample\,IV in other studies, but we show the
full sample selection here for completeness. We are thus left with
four different samples, where Sample\,III and IV are subsets of
Sample\,II. Sample\,Ia and Ib are defined in
Sect.\,\ref{sect:kinprop2}.

Naming these samples is somewhat problematic. One option would be to
call them halo, thick and thin disk or accreted and in-situ
populations. However, it is likely that the thick disk (and halo)
overlap to some larger or smaller extent with the Splash
\citep{2020MNRAS.494.3880B}. Another option would be to follow
\citet{2021MNRAS.500.1385H} and refer to the stars to the left of the
diagonal cut as accreted stars. We are quite certain we will find some
accreted stars there, but are all stars accreted?

Our aim is to investigate how these samples interact and how the Milky
Way was assembled. By giving the samples names we may inadvertently
guide our thinking. To avoid this we have opted to simply refer to the
samples defined in the Mg-Mn-Al-Fe-plane as Sample\,I, II, III and IV
(compare Fig.\,\ref{fig:cuts}). If sub-samples are defined to these
major samples, for example by imposing a kinematic constraint, they
will be given a letter in addition to identify them
(e.g. Sample\,Ia). The median uncertainty in $L_z/J_{tot}$ for
  Sample I is 0.06. We note that we have left a substantial gap in the
  selection of Sample Ia and Ib.

\begin{table}
  \caption{Stellar sub-samples used in this work. The selection
    criteria for stars to be included in each of the four main samples
    are given on the first line for each sample (compare
    Fig.\,\ref{fig:cuts}). Criteria listed for each sub-sample (e.g.,
    Sample\,{I}a) lists the criterion added to the main criterion
    (e.g. adding a kinematic criterion to the elemental abundance
    criterion).}\label{tab:samples}
    \begin{tabular}{lll}
\hline
 Name & Selection criteria\\
\hline
   {\bf Sample\,I}: & \\
      \smallskip
         S\,{I} & [Mg/Mn] $ > 0.65+1.6\ast$[Al/Fe]  \\
                   S\,Ia &{\tt plus}  $|L_z/J_{\rm tot}| < 0.25$\\
                   S\,Ib &{\tt plus} $L_z/J_{\rm tot} > 0.6$\\
\hline
    {\bf Sample\,II}: &\\
          S\,{II} & [Mg/Mn] $ < 0.35+1.6\ast$[Al/Fe] \\
\hline
    {\bf Sample\,III}: & \\
          S\,{III}& [Mg/Mn] $ < 0.35+1.6\ast$[Al/Fe] \\
               & {\tt and} [Mg/Mn] $>$0.32 \\
\hline
    {\bf Sample\,IV}: & \\
          S\,{IV} & [Mg/Mn] $ < 0.35+1.6\ast$[Al/Fe] \\
               & {\tt and} [Mg/Mn] $<$0.22\\
\hline
\end{tabular}
\label{tab:samples}
\end{table}

\section{Chemical signatures of accreted stellar populations}
\label{sect:chem}

It has long been understood that the elemental abundance ratios as
observed in the stellar atmospheres of long-lived stars hold
information about the conditions in the gas from which the stars
formed \citep{1989AIPC..183..168L,1997ARA&A..35..503M}. By analysing
the spectra of such stars we can trace the chemical evolution of a
stellar population over time or we can use the abundance ratios as
indications of the relative stellar ages. Different elements are
released to the inter-stellar medium by stars with different masses
and therefore on different timescales \citep[see
e.g.][]{1996snih.book.....A,1997NuPhA.621..467N,2020ApJ...900..179K}. Comparing
the elemental abundances therefore gives a timing of the events.

Stellar systems of different mass and gas densities will evolve on
different timescales. This means that smaller galaxies, like the
Magellanic Clouds and the dwarf spheroidal (dSph) galaxies, are
expected to show different elemental abundance trend. This is born out
by observations \citep[see
e.g.][]{2009ARA&A..47..371T,2021arXiv210905130H}. This in turn means
that stellar populations accreted to the Milky Way should show
elemental abundance trends that differ from those seen for stars
formed in the main body of the Milky Way. Hereafter, we will refer to
those stars as ``in situ'' stars while stars now present in the Milky
Way that formed in (smaller) systems that have been accreted at some
point onto the main body of the Milky Way will be referred to as
``accreted''.

It is interesting to consider which elements will give us the best
possibilities to identify accreted stellar populations. The best
elements to use will depend on several factors, beyond those of purely
chemical evolution concern, such as availability of atomic lines of
sufficient strength (but not too strong), our ability to model the
strengths of lines in the stellar spectra as a function of the
elemental abundance, e.g., due to deviation from local thermal
equilibrium.

Much focus has been attached to the $\alpha$-elements as it was early
understood that they should show different trends depending on the
star formation rate in the system \citep[][and references
therein]{1997ARA&A..35..503M} and since such elements have atomic
lines of good strength across the optical spectrum. Another focus has
been the neutron-capture elements as they show distinct patterns for
very metal-poor stars and for metal-poor stars in present day dSph
galaxies
\citep{2009A&A...501..519B,2013A&A...554A...5K,2016AJ....151...82R}. The
neutron-capture elements are, however, difficult to study as in many
stars the available lines are weak. Several of the spectral lines
arising from these elements also have complex structures including
hyper-fine-splitting and NLTE effects \citep[see e.g.][and references
therein]{2016A&A...586A..49B} making them difficult to analyse.

Here we follow upon the work of \citet{2015MNRAS.453..758H} and
explore the possibilities to use the elemental abundances of
magnesium, manganese, aluminum and iron to identify accreted stars.

\subsection{Exploring the Mg-Mn-Al-Fe-plane}
\label{sect:explore}

\begin{figure*}[]
   \begin{centering}
   \includegraphics[width=18cm,trim={0 0.5cm 0 1.5cm},clip]{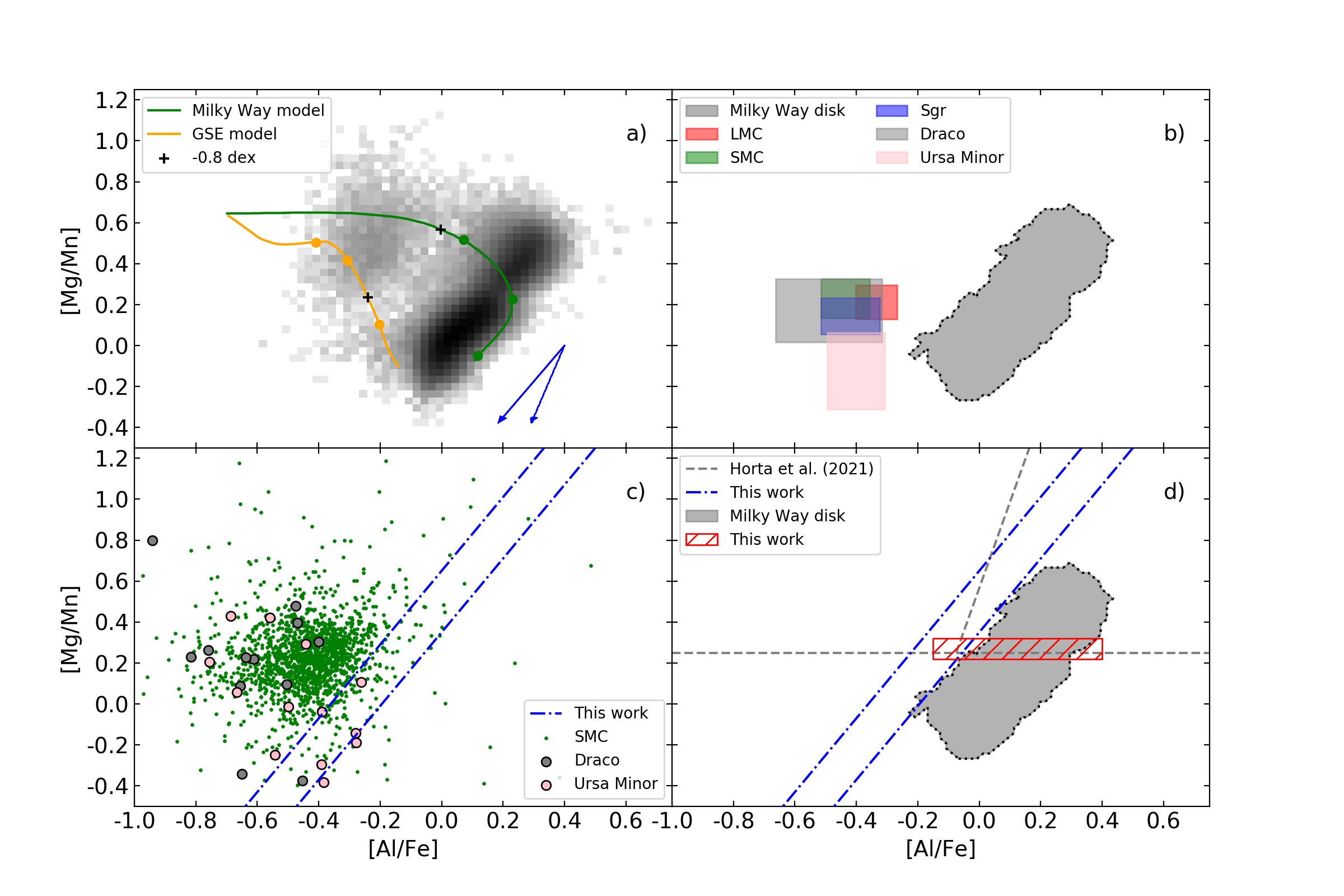}
   \caption{The Mg-Mn-Al-Fe-plane. {\bf a)} This figure shows the
     placement of the APOGEE\,DR17 data in the Mg-Mn-Al-Fe-plane. Two
     chemical evolution models presented in
     \citet{2021MNRAS.500.1385H} are also shown. The black $+$-signs
     indicate when the model has reached a metallicity of $-0.8$\,dex.
     The blue arrows show the sizes and directions of 3D+NLTE effects
     discussed in Sect.\,\ref{sect:nlte3d}. {\bf b)} This figure
     illustrates the placement of six different galaxies in the
     Mg-Mn-Al-Fe-plane using APOGEE\,DR17 data. Each of the five dwarf
     galaxies are shown as a box centered on the mean values of the
     data for all red giant stars in that galaxy and the size is
     simply half of the full width at half maximum.  The Milky Way
     disk is shown as a grey area. {\bf c)} This figure shows the
     individual stellar data for three of the dwarf galaxies in panel
     b). In addition, the division between traditional halo and disk
     used in this paper are shown as two lines (compare
     Fig.\,\ref{fig:cuts}).  {\bf d)} This figure reproduces the
     division of the Mg-Mn-Al-Fe-plane as used in
     \citet{2021MNRAS.500.1385H} and the cuts used in this work to
     define Sample\,I, II, III and IV (see also
     Fig.\,\ref{fig:cuts}). The Milky Way disk sample from figure b)
     is reproduced as a faint grey area. }
     \label{fig:mgmnalfe_ill}
   \end{centering}
\end{figure*}

\citet{2015MNRAS.453..758H} and later \citet{2020MNRAS.493.5195D}
proposed to use the Mg-Mn-Al-Fe-plane to identify stellar populations
formed in smaller systems and later accreted onto the main body of the
Milky Way. Later works show that this works well for example in
identifying the \textit{Gaia}-Sausage-Enceladus members
\citep{2021MNRAS.508.1489F}. To the best of our understanding, the
works by \citet{2015MNRAS.453..758H} and \citet{2020MNRAS.493.5195D}
were essentially empirical, inspired by the large and high quality
dataset from APOGEE\,DR14. \citet{2021MNRAS.500.1385H} made a first
theoretical illustration of how this plane can work to identify
different stellar population. For completeness and because of the
importance of these findings for our work we here repeat some of the
arguments from \citet{2021MNRAS.500.1385H} for and against using the
Mg-Mn-Al-Fe-plane to identify accreted stellar populations in the
Milky Way.

We start by exploring the
Mg-Mn-Al-Fe-plane. Figure\,\ref{fig:mgmnalfe_ill}~a) shows the density
map in this plane of the data selected in Sect.\,\ref{sect:data}. We
note that the stars are not evenly distributed. We see a lighter,
round concentration centered at ([Al/Fe],[Mg/Mn]) = (--0.3,+0.5)\,dex
and a more elongated structure to the right. In between these
structures there is a clear depression in the number densities. We
will, \citet{2015MNRAS.453..758H} and \citet{2020MNRAS.493.5195D} did,
use this lack of stars as a natural place to separate traditional halo
and disk stars. The stars in the structure to the right may also show
two rather than a single structure with one round concentration at
(0.3,0.45) and a more elongated structure below. We will return to
these features in more detail in future studies.

The figure also shows evolutionary tracks for two different systems;
the Milky Way and the \textit{Gaia}-Sausage-Enceladus.  The trends
shown are replications from \citet{2021MNRAS.500.1385H}, their
Fig.\,1. \citet{2021MNRAS.500.1385H} uses the Milky Way solar
neighborhood chemical evolution model as their fiducial model (shown
here in red) that is representative of in situ chemical evolution. The
model shown in blue is tailored to represent the
\textit{Gaia}-Sausage-Enceladus progenitor. The latter then shows what
an accreted stellar population may look like. For each model the
following time stamps are indicated by filled circles: 300\,Myr,
1\,Gyr, and 5\,Gyr. In addition the point when each model has reached
a chemical enrichment of --0.8\,dex is indicated. Further details on
the model parameters can be found in \citet{2017ApJ...835..224A} and
\citet{2021MNRAS.500.1385H}. As can be understood from these model
predictions, the region to the left will inevitably include both
accreted stars as well as in situ stars \textit{if} the in situ
population has a reasonably large amount of metal-poor stars. On the
other hand \textit{if} the in situ population mainly forms from gas
that is already enriched then the region to the left will essentially
be void of in situ stars.

Figure\,\ref{fig:mgmnalfe_ill} b) shows where stars in different known
systems fall. The Milky Way disk is shown by using the APOGEE\,DR17
data selected as disk stars for this study (shown as contours, see
Fig.\,\ref{fig:kinfeh}). In addition, data from APOGEE\,DR17 for the
Large Magellanic Cloud (LMC), the Small Magellanic Cloud (SMC), the
core and tidal stream of the Sagittarius (Sgr) dwarf spheroidal
galaxy, and the Draco and Ursa Minor dSph galaxies are shown as
coloured boxes. The selection of the stars in these galaxies are
described in Appendix \ref{app:dwarfs}. The centres of the boxes are
the mean values of [Mg/Mn] and [Al/Fe] for the data for each galaxy
and the size of the boxes is simply half of the standard deviation for
each data set around their mean values.  As can be seen, the Milky Way
stellar disk (the heaviest system) has the highest [Al/Fe] values
whilst lighter systems have lower [Al/Fe] values. Specifically, the
SMC is lower in [Al/Fe] than the LMC. The Milky Way disk spans a large
range in [Mg/Mn] while the other systems have a somewhat smaller
range.

The smaller systems all have [Al/Fe] systematically lower than the
Milky Way disk, indicating that it is not unreasonable to assume that
stars from a systems that has formed in relative isolation and then
accreted onto the main body of the Milky Way will be possible to
identify thanks to their low [Al/Fe]. This is further supported by the
models from \citet{2021MNRAS.500.1385H} and
\citet{2023MNRAS.519.3611F} that show that a lighter system will reach
a maximum [Al/Fe] value that is smaller than the main body of the
Milky Way disk.  See also \citet{2023MNRAS.519.3611F} for a full
characterization of low-mass systems in APOGEE.

Figure\,\ref{fig:mgmnalfe_ill} c) shows the full data-set used to
calculate the boxes in Fig.\,\ref{fig:mgmnalfe_ill} b) for SMC, and
the Draco and Ursa Minor dSph galaxies. As can be seen scatter is
relatively high for the two dSph galaxies whilst the data is more
compactly distributed for SMC. The two lines illustrate the cut that
we use to separate traditional halo from disk
(Fig.\,\ref{fig:cuts}). Dwarf galaxies surviving to this day are
clearly on the left-hand side of these cuts and illustrate that we may
reasonably assume to find accreted stars in those regions.

Figure\,\ref{fig:mgmnalfe_ill} d) replicates the cuts used by
\citet{2021MNRAS.500.1385H} to divide the plane into \textit{Accreted}
and \textit{In situ} high- and low-$\alpha$ stars (their
nomenclature). We also show the cuts used in this study defined in
Sect.\,\ref{sect:pop_def}.  The main difference is that we divide the
sample not considering any selection based on other elemental
abundances (i.e. $\alpha$). The near vertical cut that separate the
samples differ somewhat between the two studies but the general idea
is the same. In our selection, we also leave a gap in order to provide
clean samples, while \citet{2021MNRAS.500.1385H} prefers a single cut
with no gap.  \citet{2021MNRAS.500.1385H} discuss how, based on their
models, a Milky Way-sized galaxy will inevitably create some stars
with high [Mg/Mn] and low [Al/Fe] that will contaminate any accreted
population identified in the Mg-Mn-Al-Fe-plane (compare
Fig.\,\ref{fig:mgmnalfe_ill} a).

\subsection{Comments on NLTE and 3D effects on the elemental abundances }
\label{sect:nlte3d}

In the vast majority of studies of stellar elemental abundances the
abundances are derived under assumption of Local Thermal Equilibrium
(LTE). Our understanding of how to treat departures from LTE (NLTE)
has been growing for a long time and is now commonly implemented in
small studies as well as in large spectroscopic surveys. Departures
from LTE can have a significant effects, in particular when studying
stars which span a wide range of [Fe/H]-values \citep[see e.g.][]{2012MNRAS.427...27B,2012MNRAS.427...50L}. The stars in the
samples we are studying are giant stars. For the most evolved stars,
the assumption of plane parallel model atmospheres is also under
question \citep{2006A&A...452.1039H} as well as the possibility that
the stellar atmosphere is not homogenous but instead a full 3D
modeling is necessary.

In large spectroscopic surveys it is difficult to take full account of
NLTE and 3D effects, mainly because the calculation of these effects
are time consuming. However, recently first instances of including
e.g. departure from the LTE assumption have been performed in the 2nd
and 3rd data release from the GALAH survey \citep{2018MNRAS.478.4513B,
  2021MNRAS.506..150B}. In the future, more surveys will undoubtedly
do so.

In APOGEE\,DR17, NLTE corrections have been applied to the magnesium
abundance measurements following \citet{2020A&A...637A..80O}, but not
for aluminum, iron, or manganese. This means that it is important for
us to check whether the abundance measurements used in our study may
be effected by NLTE and/or 3D effects to such a degree that our
conclusions fail.  Below we discuss possible effects on the two
elemental abundance ratios that we consider in this work.

\subsubsection{NLTE and 3D effects on [Al/Fe]}
\label{sect:al}

\citet{2017A&A...607A..75N} studies NLTE and 3D effects on the
derivation of [Al/Fe] in late type stars. They show that corrections
can be substantial for some stars whilst for others the corrections
are minor. In addition, the corrections vary significantly for a given
star depending on which atomic lines are being used.

APOGEE uses three near-infrared Al\,I lines to derive the aluminum
abundances. The lines in the APOGEE linelist by
\citet{2015ApJS..221...24S,2021AJ....161..254S} are listed as 16723.5,
16755, and 16768.\,{\AA}. These are given as measured in vacuum,
whilst the lines listed as APOGEE lines in \citet{2017A&A...607A..75N}
are given as 16718.9, 16750.5, and 16763.3\,{\AA}, which are measured
in air. The line at 16750.5\,{\AA} (air) also has hyperfine
splitting. This leads \citet{2017A&A...607A..75N} to recommend that
this line should not be used in derivation of aluminum abundances for
metal-poor cool giant stars such as Arcturus.

For Arcturus, a red giant, with {($T_{\rm eff}/\log g$/[Fe/H]) =
  (4247/1.59/--0.52)}, \citet{2017A&A...607A..75N} find that average
3D and NLTE together gives a correction of about (--0.2,--0.3, -0.1
)\,dex for the three lines used in APOGEE (as read from their
Fig.\,10). This should be compared to an average using all available
lines across the optical and near-infrared stellar spectrum which give
a difference of --0.13\,dex (their Table\,1). They also analyse
HD\,122563, another metal-poor giant, {($T_{\rm eff}/\log g$/[Fe/H]) =
  (4608/1.61/--2.64). In this star the Al\,I lines are weak because
  the star is so metal-poor and they can therefore only analyse the
  lines at 16718.9 and 16750.5\,{\AA} (air). They find excellent
  agreement between the [Al/Fe] abundance in HD\,122563 derived from
  these two lines and the blue UV line at 3961\,{\AA}, as well as an
  average difference of +0.09\,dex in the sense 1D LTE - 3D NLTE
  (i.e. the value should be added to the 1D LTE value to achieve the
  corrected value).

  Arcturus and HD\,122563 span the full range of [Fe/H] for our sample
  and have $\log g$ and $T_{\rm eff}$-values that are representative
  of our sample. The size of a 3D plus NLTE correction to [Al/H]
  abundance in Arcturus derived from near-infrared lines should be
  negative and the size about 0.2\,dex. For HD\,122563, which is much
  more metal-poor, the correction is also in the negative sense but
  less substantial, around 0.1\,dex.

  Although we do not have the possibility to apply line-by-line NLTE
  and 3D corrections to the data from APOGEE\,DR17 the discussion in
  the preceding text leads us to conclude that for the full sample the
  maximum corrections to [Al/H] from NLTE and 3D is no more than
  0.2\,dex (perhaps slightly more) and not less than 0.1\,dex.

  For optical spectra it is found that NLTE and 3D
  effects on derived iron abundances are small or negligible for stars around --1\,dex and with surface gravities close to 1--2
  \citep{2012MNRAS.427...27B,2012MNRAS.427...50L,2016MNRAS.463.1518A}. However,
  more recent studies indicate that the corrections could be larger
  than previously perceived \citep{2022A&A...668A..68A}. For iron
  lines in the wavelength region covered by APOGEE there exists no
  full 3D-NLTE calculation that can be trusted
  \citep{2021A&A...647A..24M}. NLTE caclulation are readily available
  and show to be essentially non-existent for the stellar parameter
  range we are interested in. Taking these aspects into account we
  find that there is no need to consider NLTE-3D effects for iron for our
  sample. Hence, the effects for [Al/Fe] are the same as for [Al/H].
  
\subsubsection{NLTE and 3D effects on [Mg/Mn]}
\label{sect:mgmn}

For magnesium, we make use of the calculations by
\citet{2017ApJ...847...15B} and \citet{2017ApJ...835...90Z} to
estimate the size of the effects of departures from LTE, as well as 3D
effects. \citet{2018ApJ...866..153A} also study departures from LTE
for magnesium but they do not include the near-infrared lines used in
APOGEE\,DR17.

\citet{2017ApJ...835...90Z} study NLTE effects on derived magnesium
abundances based on Mg\,I lines in the H-band. Their Table\,3 shows
that the maximum correction for 1D NLTE for red giant stars is
0.35\,dex for [Mg/Fe].  Table\,5 in \citet{2017ApJ...847...15B} shows
that the difference between 1D LTE and $<$3D$>$ NLTE in [Mg/H] from
NIR Mg lines is of the order zero.

Departures from 1D LTE for Mn was studied by
\citet{2019A&A...631A..80B}, who found that 3D NLTE Mn abundances
derived from the Mn\,{\sc i} lines used in APOGEE are, on average,
significantly higher than those calculated in 1D LTE: for stars with
($T_{\rm eff}, \log g,$[Fe/H]) = (4500/2.0/--1) and (4500/2.0/--2.0)
they find a correction of +0.3 and +0.35\,dex, respectively \citep[top
panel in Fig.\,17 in][]{2019A&A...631A..80B}.

Combining this information, we conclude that the [Mg/Mn] elemental
abundances derived from the NIR Mn\,{\sc i} lines used in APOGEE has a
negative correction of up to 0.35\,dex.

\subsubsection{Summary: How well can stellar populations be separated in the Mg-Mn-Al-Fe-plane?}

Based on the literature review provided in Sects.\,\ref{sect:al} and
\ref{sect:mgmn} we find that for a typical giant star in our sample
the NLTE combined with 3D corrections amount to about --0.1 to
--0.2\,dex in [Al/Fe] and --0.35\,dex in [Mg/Mn]. This implies a
movement down and to the left in the Mg-Mn-Al-Fe-plane. The slope of
the correction is roughly aligned with the empirical, diagonal cut we
adopt to split our sample into two (see Fig.\,\ref{fig:mgmnalfe_ill}
and \ref{fig:cuts}). This implies that although stars would
collectively move within the diagram if the NLTE and 3D corrections
were applied, in fact the gap between the two populations would
remain.

We conclude that, in the case of the data selected in this study from
APOGEE\,DR17, we can indeed make use of the Mg-Mn-Al-Fe-plane to
separate accreted stars from those formed in the main body of the
Galaxy.  For individual stars there might be differences but for the
populations as such it is sufficient to use the uncorrect elemental
abundances from APOGEE\,DR17.

It is important to remember that this statement is valid for the stars
selected for this study, i.e. stars on the Red Giant Branch (compare
Table\,\ref{tab:data}). Stars in other evolutionary stages may not be
as readily separable and a fresh assessment should be made for each
study taking exact atomic lines used in the derivation of the
elemental abundances into account.

\section{Kinematic properties and elemental abundances for Sample\,I and III}
\label{sect:kinprop}

\begin{figure}
\centering
\includegraphics[width=8cm,trim={0 0cm 0 1cm},clip]{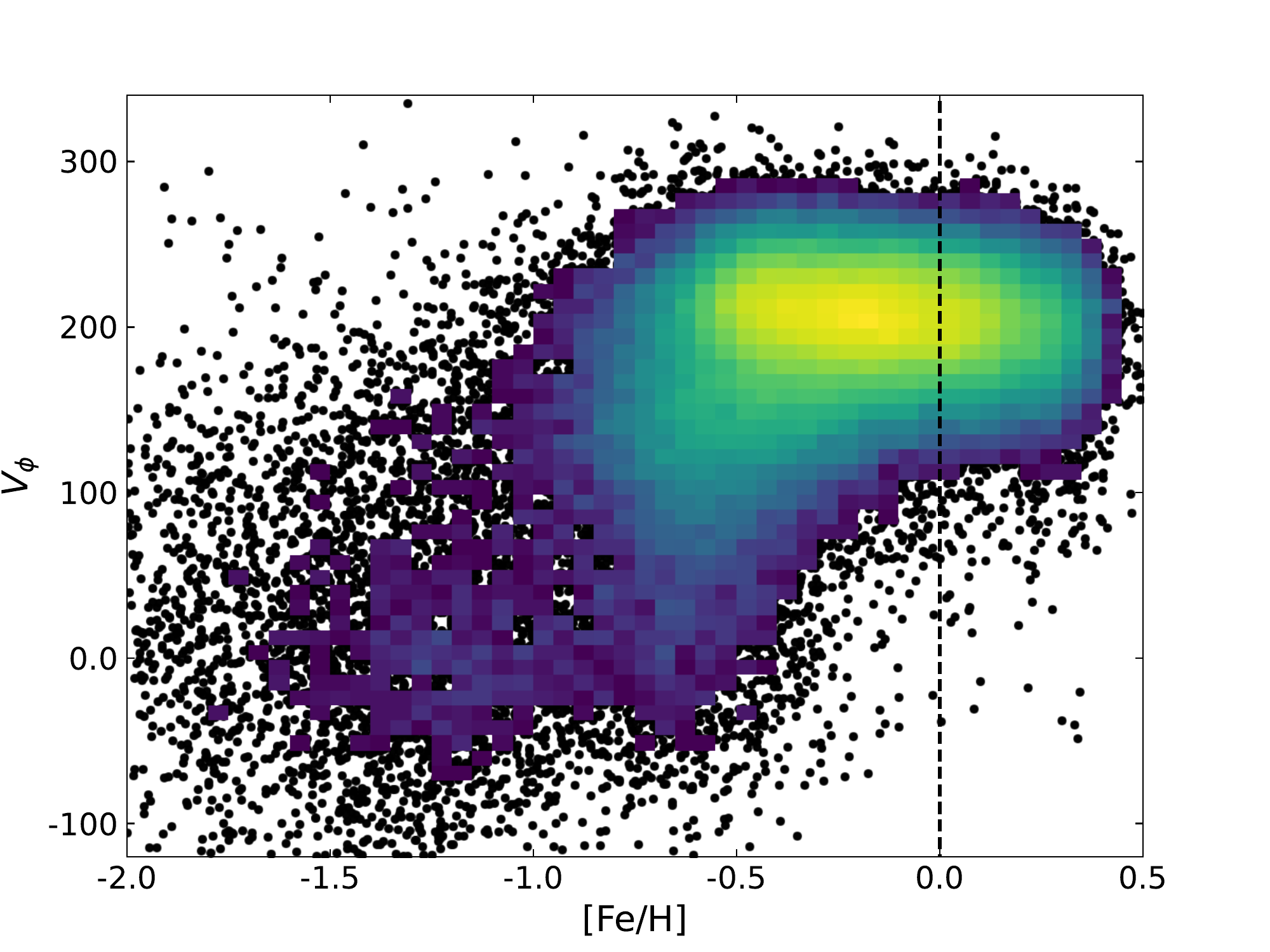}
\caption{$V_{\phi} -$[Fe/H] plane for the full sample from  APOGEE\,DR17 used in this study.}
\label{fig:kinfeh}
\end{figure}

Figure\,\ref{fig:kinfeh} shows the density plot of $V_{\phi} $ as a
function of [Fe/H] for our full sample. We see that the majority of
the stars have a velocity compatible with that of the stellar disk
\citep{2016ARA&A..54..529B}. There is also a prominent downward trend
at [Fe/H] of about $-0.7$\,dex as well as substructure at lower
metallicities with kinematics typical of the stellar halo
\citep{2016ARA&A..54..529B}.

In this section we first analyse the kinematical properties of the
full sample as well as the chemically defined sub-samples. Secondly,
we combine selection in elemental abundance space with the kinematic
properties and identify the  stellar disk.

  \subsection{Kinematics of  Sample\,I and III}
  \label{sect:kinprop2}
  
\begin{figure*}
\begin{centering}
\includegraphics[width=16cm,trim={0 1cm 0 2cm},clip]{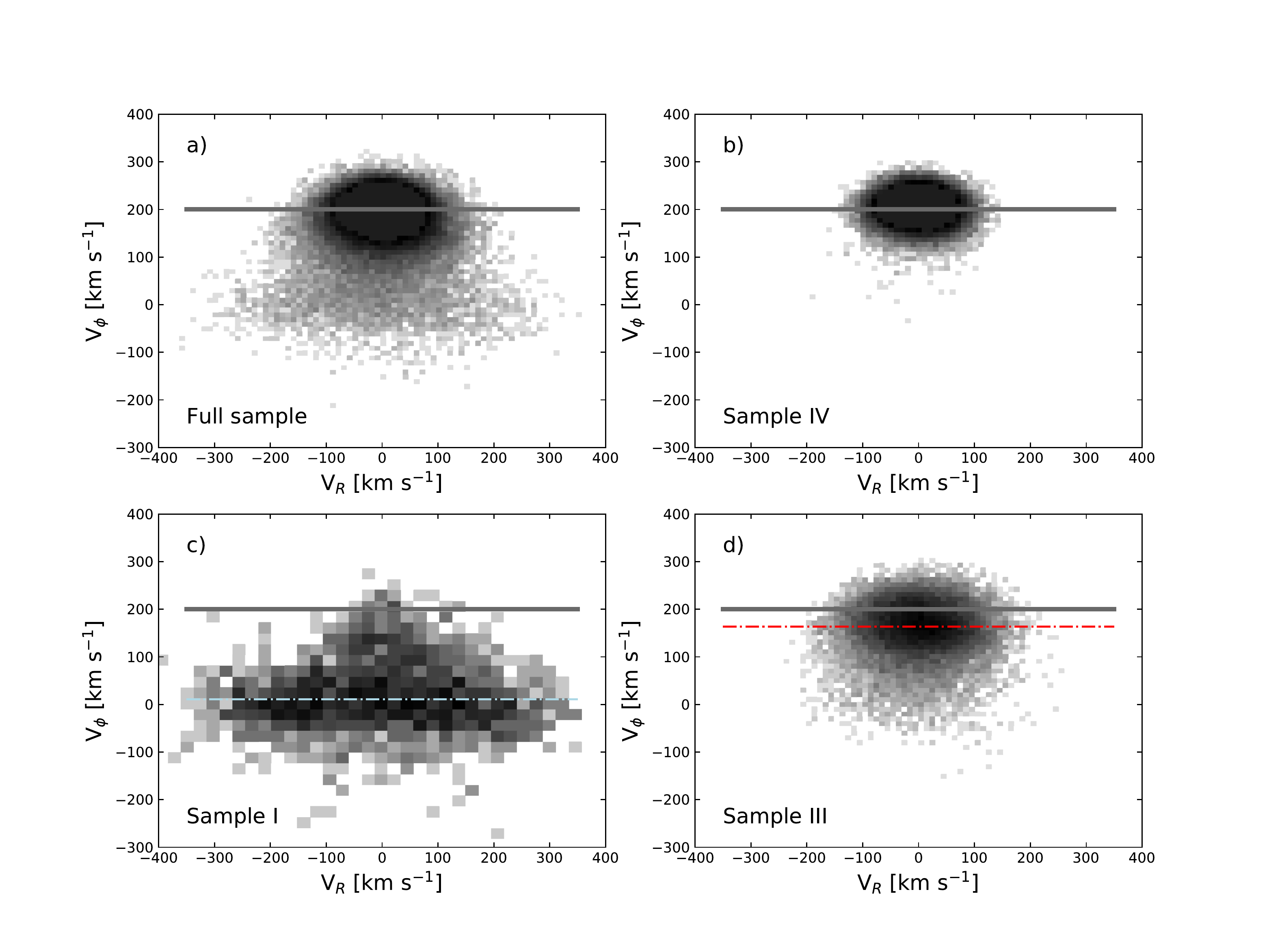}
\caption{ $V_{\phi}$ as a function of $V_{R}$ for the full sample and
  the samples defined via cuts in the Mg-Mn-Al-Fe-plane. See
  Fig.\,\ref{fig:cuts} and Table\,\ref{tab:samples} for the definition
  of the samples. Median value for $V_{\phi}$ for the full sample
  shown as a thick grey line, repeated in all four panels.  {\bf a)}
  Full sample.  {\bf b)} Sample\,IV.  {\bf c)} Sample\,I. Median value
  of $V_{\phi}$ for this sample shown as a light-blue dash-dotted
  line.  {\bf d)} Sample\,III. Median value of $V_{\phi}$ for this
  sample shown as a red dash-dotted line. }
\label{fig:vphi_vr} 
\end{centering}
\end{figure*}

\begin{figure*}
\begin{centering}
\includegraphics[width=14cm,trim={0 2cm 0 2.5cm},clip]{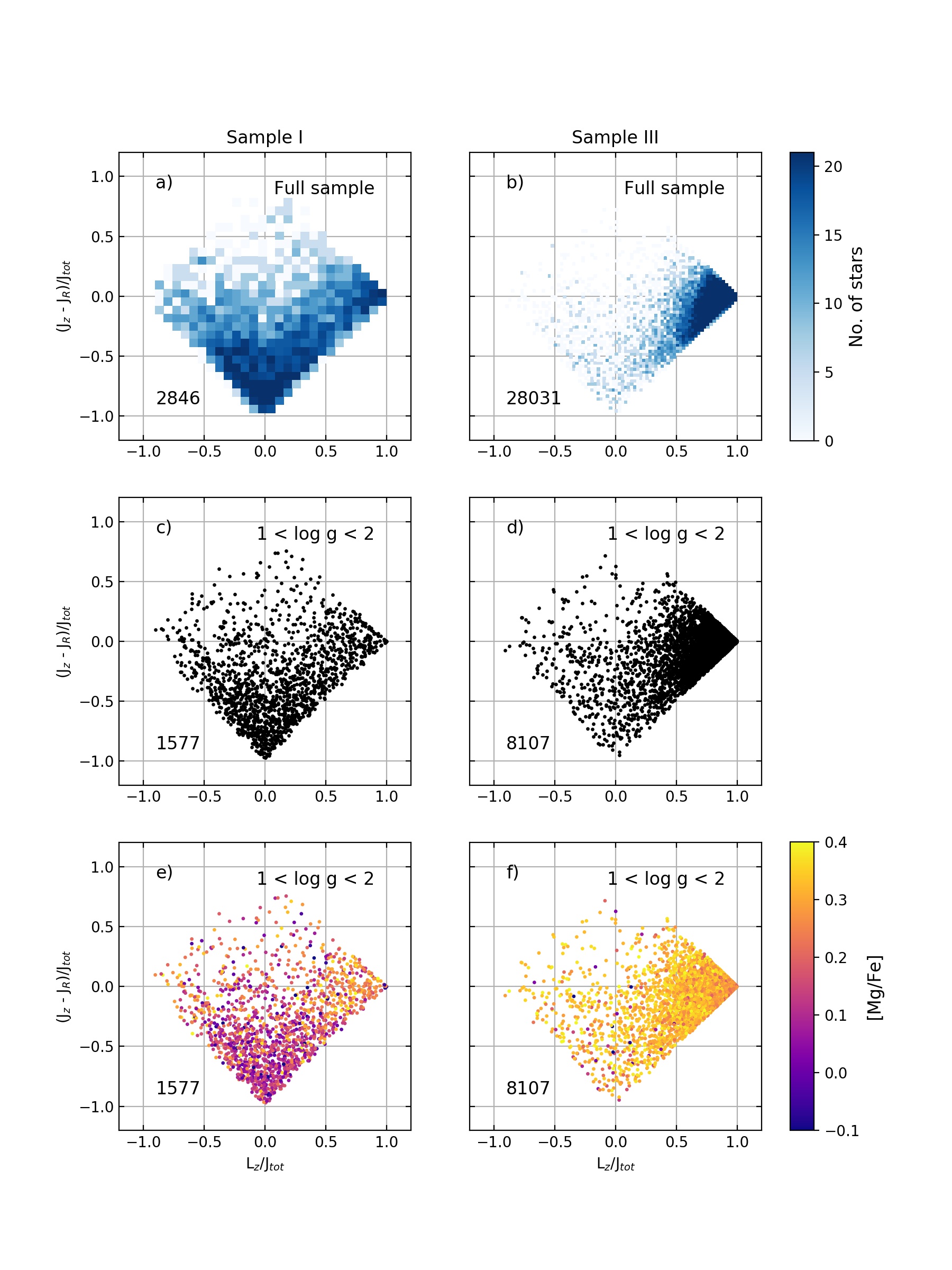}
\caption{Action diamonds for Sample\,I and III. 
{\bf a)} Action diamond for Sample\,I for stars with $1 < \log g < 2$. 
{\bf b)} Action diamond for Sample\,III for stars with $1 < \log g < 2$. 
{\bf c)} Action diamond for Sample\,I for stars with $1 < \log g < 2$  colour coded by [Mg/Fe]. 
{\bf d)} Action diamond for Sample\,III for stars with $1 < \log g < 2$ colour coded by [Mg/Fe]. 
{\bf e)} 2D histogram action diamond for the full Sample\,I. 
{\bf f)} 2D histogram action diamond for the full Sample\,III. }
\label{fig:romb1} 
\end{centering}
\end{figure*}

There are several kinematic spaces used in the literature to analyse
the stellar content of the Milky Way halo. The Toomre diagram has been
extensively used to make a separation of stellar disk and halo stars
\citep[examples
include][]{2018Natur.563...85H,2014A&A...562A..71B,2010A&A...511L..10N}. A
simple plot of $V_{\phi}$ as a function of $V_{R}$ can help to study
the connection between disk and halo
\citep{2020MNRAS.494.3880B}. Other quantities such as angular momenta
or actions are (near) preserved and can be used to identify, e.g.,
streams \citep{1999Natur.402...53H}.

We start by studying our stellar samples defined in the
Mg-Mn-Al-Fe-plane in various kinematic spaces.
Figure\,\ref{fig:vphi_vr} shows $V_{\phi}$ as a function of $V_{R}$ for
the full sample and for Sample\,I, III, and IV as defined in
Fig.\,\ref{fig:cuts}. In each panel the median value for $V_{\phi}$ of
the full sample is shown as a grey line, whilst the median of the
selected sample is shown as a coloured, dashed line. For Sample\,IV
the grey and the coloured lines have the same value.  Sample\,I shows
an elongated structure centred at $V_{\phi}=0$ and an extension of
stars to higher (pro-grade) $V_{\phi}$-values which is centred at
$V_{R}=0$. Sample\,IV is, as already noted, very similar to the bulk
of the full sample, whilst Sample\,III has a distinctly lower mean
$V_{\phi}$-value as well as an extension towards $V_{\phi}=0$ and even
negative values (retro-grade).

To summarize, the main sample is well concentrated around ($V_{R}$,
$V_{\phi}$) = (0, $\sim 200$)\,km\,s$^{-1}$. When divided into
sub-samples using the Mg-Mn-Al-Fe-plane we find that one sample takes
up the bulk of the stars centred at the same values as the main sample
whilst the other two samples contain the stars making up the down-ward
flow of stars towards $V_{\phi}=0$ as well as flaring out to larger
(pos/neg) values of $V_{R}$. The three samples are distinct in the
($V_{R}$, $V_{\phi}$)-plane.

Recently, another depiction of the stellar kinematics has been used --
the action diamond
\citep{2019MNRAS.484.2832V,2019MNRAS.488.1235M,2022MNRAS.510.5119L}.
The diamond is constructed from the actions and angular momenta of the
stellar orbits. On the $x$-axis is $L_z/J_{\rm tot}$ and on the
$y$-axis $(J_z - J_R)/J_{\rm tot}$. As explained in
\citet{2022MNRAS.510.5119L} this space is particularly intuitive to
interpret; the left- and right-hand corners occur when the angular
momentum in the $z$-direction and the total action are equal, i.e. a
pro- or retro-grade orbit in the plane. The bottom corner contains the
stars on purely radial orbits while the top corner gathers the stars
on polar orbits.
  
Figure\,\ref{fig:romb1}\,a) and c) show the action diamonds of
Sample\,I for the full sample and for only stars with $1 < \log g <2$
, respectively. In both cases, the sample is mainly concentrated to
the bottom corner. The bottom corner of the action diamond is
associated with radial orbits \citep[see Sect. 4.1.1
in][]{2022MNRAS.510.5119L}.  However, the figures also show that there
is a smaller concentration of stars in the right hand corner. This
corner is associated with pro-grade disk-like orbits.
Figure\,\ref{fig:romb1}\,b) and d), show the action diamonds of
Sample\,III for the full sample and for only stars with
$1 < \log g <2$, respectively.  Sample\,III populates the right-hand
corner of the action diamond associated with disk-like orbits
\citep{2022MNRAS.510.5119L}.  Fig.\,\ref{fig:romb1}\,e) and f) show
the action diamonds of Sample\,I and III for stars with
$1 < \log g <2$, colour-coded according to the [Mg/Fe]-ratios of the
stars. Sample\,I shows mainly low(er) [Mg/Fe]-ratio than
Sample\,III. However, the stars in the left- and right-hand corners of
the action diamond for Sample\,I have high-[Mg/Fe] ratios suggesting
there may be two stellar populations in Sample\,I.

\begin{figure*}
\begin{centering}
\includegraphics[width=17cm,trim={0 3cm 0 3.5cm},clip]{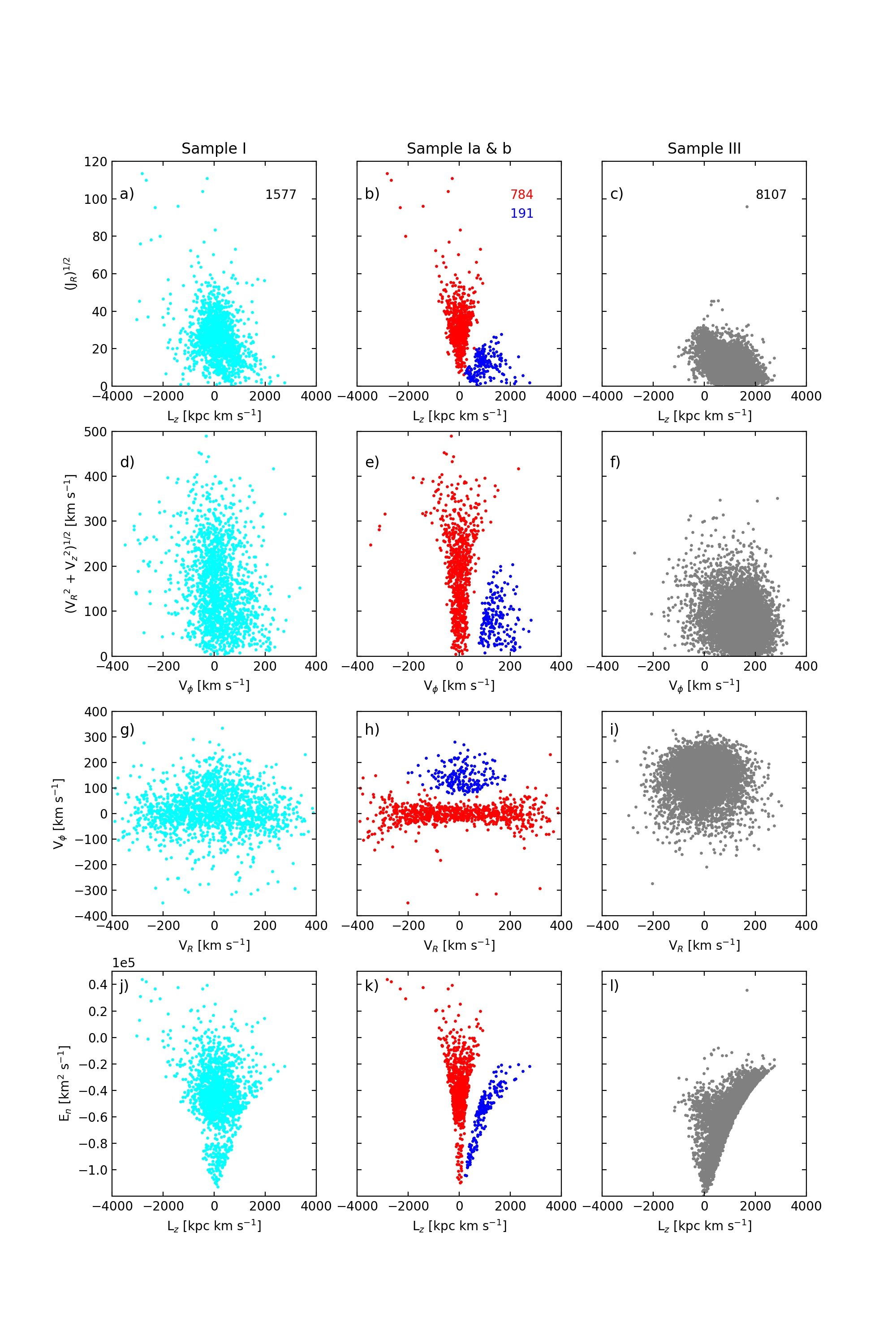}
\caption{Four kinematic spaces showing the properties of the stars
  selected for Sample\,I (first column) and Sample\,III (third column)
  using the Mg-Mn-Al-Fe-plane (Fig\,\ref{fig:cuts}). Only stars with
  $1 < \log g < 2 $ are included (plots look very similar with all
  stars included). The middle column shows the two sub-samples of
  Sample\,I defined using the action diamond: Sample\,Ia (red) are
  stars with $-0.25 < L_z/J_{tot} < 0.25$ and Sample\,Ib (blue) stars
  with $L_z/J_{tot} > 0.6$. The number of stars in each sample are
  indicated in the panels in the top-row.}
  \label{fig:kin_sample1} 
\end{centering}
\end{figure*}

To ensure that our interpretation of the properties of Sample\,I as
seen in the action diamond actually implicates two stellar populations
with different kinematical status we show four commonly used kinematic
planes.  We define two sub-samples in Sample\,I based on the stars'
$L_z/J_{tot}$. Sample\,Ia is defined as stars with
$-0.25 < L_z/J_{tot} < 0.25$ and Sample\,Ib stars with
$L_z/J_{tot} > 0.6$. Figure\,\ref{fig:kin_sample1} shows the four
kinematic spaces for the two new Sample\,Ia (red) and Ib (blue), as
well as Sample\,I and III.  For all samples, only stars with
$1 < \log g < 2 $ are shown\footnote{We note that for the orbital
  calculations with \textit{galpy} we have used the {\tt
    MWPotential2014} in
  \citet{2013ApJ...779..115B,2015ApJS..216...29B}. To be compatible
  with other studies we also present the same plots using the
  \citet{2017MNRAS.465...76M} potential in
  Appendix\,\ref{app:mcmillan}. The conclusions remain, the main
  difference being the values of $E_{\rm n}$, which are shifted.}.

We observe that the kinematical properties of Sample\,I and III are
distinct. Remember that these selections are \underline{only} based on
elemental abundances (Fig\,\ref{fig:cuts}). Sample\,I has mainly
radial orbits of the type associated with the halo and Sample\,III can
best be described as a somewhat heated disk, with most stars on
prograde orbits.  Sample\,Ia and Ib indeed show the expected dichotomy
-- Sample\,Ia has a radial orbit (as per design) while Sample\,Ib
clearly is picking up essentially all the disk-like stars in
Sample\,I. Note that there will be a gap between the samples as per
the selection.

 To summarise, Sample\,I contains two stellar samples identified in
 the action diamond, one on radial orbits and one on disk-like
 pro-grade orbits. Sample\,III contains an essentially disk, pro-grade
 stellar sample.

 \subsection{Chemical properties of Sample\,I and Sample\,III}
\label{sect:chem_SI}

\begin{figure*}
\begin{centering}
\includegraphics[width=15cm,trim={0 0.5cm 0 1cm},clip]{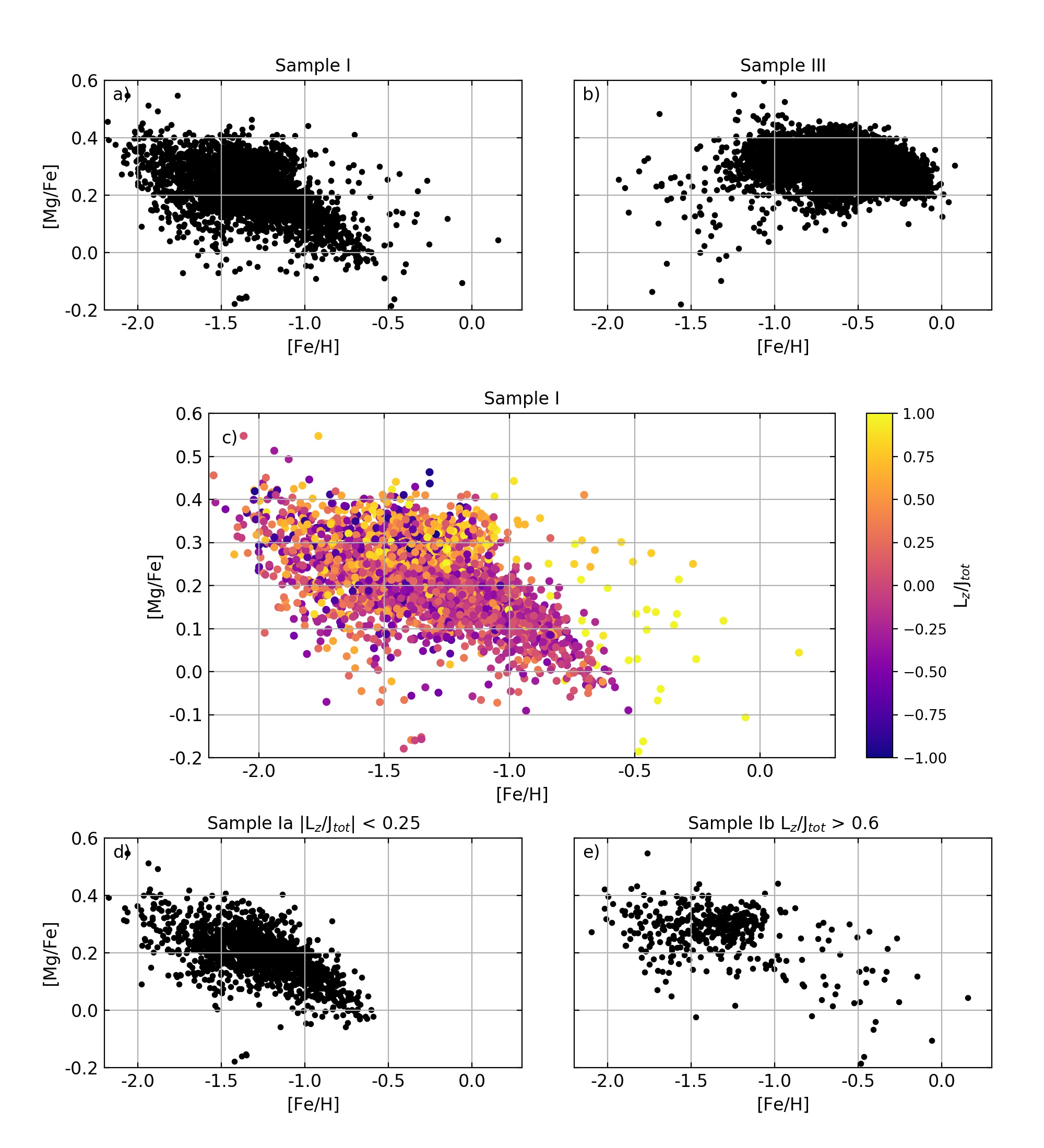}
\caption{[Mg/Fe] as a function of [Fe/H] for Sample\,I, Ia, Ib, and
  III. Samples are restricted to stars with $1 < \log g < 2$.  {\bf
    a)}~[Mg/Fe] as a function of [Fe/H] for Sample\,I.  {\bf b)}~[Mg/Fe]
  as a function of [Fe/H] for Sample\,III.  {\bf c)}~[Mg/Fe] as a function
  of [Fe/H] for Sample\,I colour-coded by $L_z/J_{\rm tot}$. Values as
  indicated by the colour bar.  {\bf d)}~[Mg/Fe] as a function of
  [Fe/H] for Sample\,Ia.  {\bf e)}~[Mg/Fe] as a function of [Fe/H] for
  Sample\,Ib.  The same plots but for [Ni/Fe] can be found in
  App.\,\ref{app:abun_extra}. }
\label{fig:mgfe_sample1}
\end{centering}
\end{figure*}

We now turn to the chemical properties of Sample\,I, Ia, Ib, and III.
We start by noting that Fig.\,\ref{fig:romb1}\,e) and f) show that
Sample\,I and Sample\,III have distinct chemical properties where
Sample\,III contains almost exclusively stars with [Mg/Fe] $>$0.25. In
Sample\,I, stars with [Mg/Fe] $<$0.2 are mainly concentrated towards
the bottom corner of the action diamond. The right- and left-hand
corners are dominated by stars with [Mg/Fe] $>$0.25 There are also
some stars spread into the rest of the action diamond with 
high [Mg/Fe]-ratios.

To summarise, it appears that the two kinematical sub-samples in
Sample\,I identified in Sect.\,\ref{sect:kinprop2} have distinct
chemical signatures with the stars on disk-like orbits being elevated
in [Mg/Fe] in comparison to the stars with radial, halo-like orbits.

Fig.ure\,\ref{fig:mgfe_sample1}\,a) and b) show [Mg/Fe] as a function
of [Fe/H] for Sample\,I and III. These are clearly distinct with
Sample\,III being more metal-rich (median [Fe/H] = --0.55) and showing
high [Mg/Fe] for all stars. Sample\,I has lower [Fe/H] (median [Fe/H] =
--1.33) and a downward trend for [Mg/Fe], starting from the highest
values and continuing down to about 0.0\,dex.

Figure\,\ref{fig:mgfe_sample1}c) shows the stars in Sample\,I
colour-coded according to $L_z/J_{\rm tot}$. Here we find that almost
all stars with positive $L_z/J_{\rm tot}$ have high [Mg/Fe]-ratios,
while stars with $L_z/J_{\rm tot}$ around zero follow the down-ward
trend. To more easily see this, Fig.\,\ref{fig:mgfe_sample1}\,d) and
e) show Sample\,Ia and Ib, i.e. $-0.25 < L_z/J_{\rm tot} < 0.25$ and
$L_z/J_{\rm tot} > 0.6$, respectively. Sample\,Ib has a high
[Mg/Fe]-ratio with only a sprinkle of stars with lower ratios. The
sample also stops quite abruptly at [Fe/H] about --1.1\,dex. We note
that Sample\,III starts roughly at the same iron abundance as
Sample\,Ib stops, suggesting Sample\,III could be a later stage
evolution of Sample\,Ib. Figure\,\ref{fig:nife_sample1} shows that
[Ni/Fe] behaves in the same way as [Mg/Fe]. In fact, the picture is
even a bit clearer when using [Ni/Fe].

\subsection{Potential selection effects}
\label{sect:seleff}

\begin{figure*}
\begin{centering}
\includegraphics[width=18cm,trim={0 2cm 0 3cm},clip]{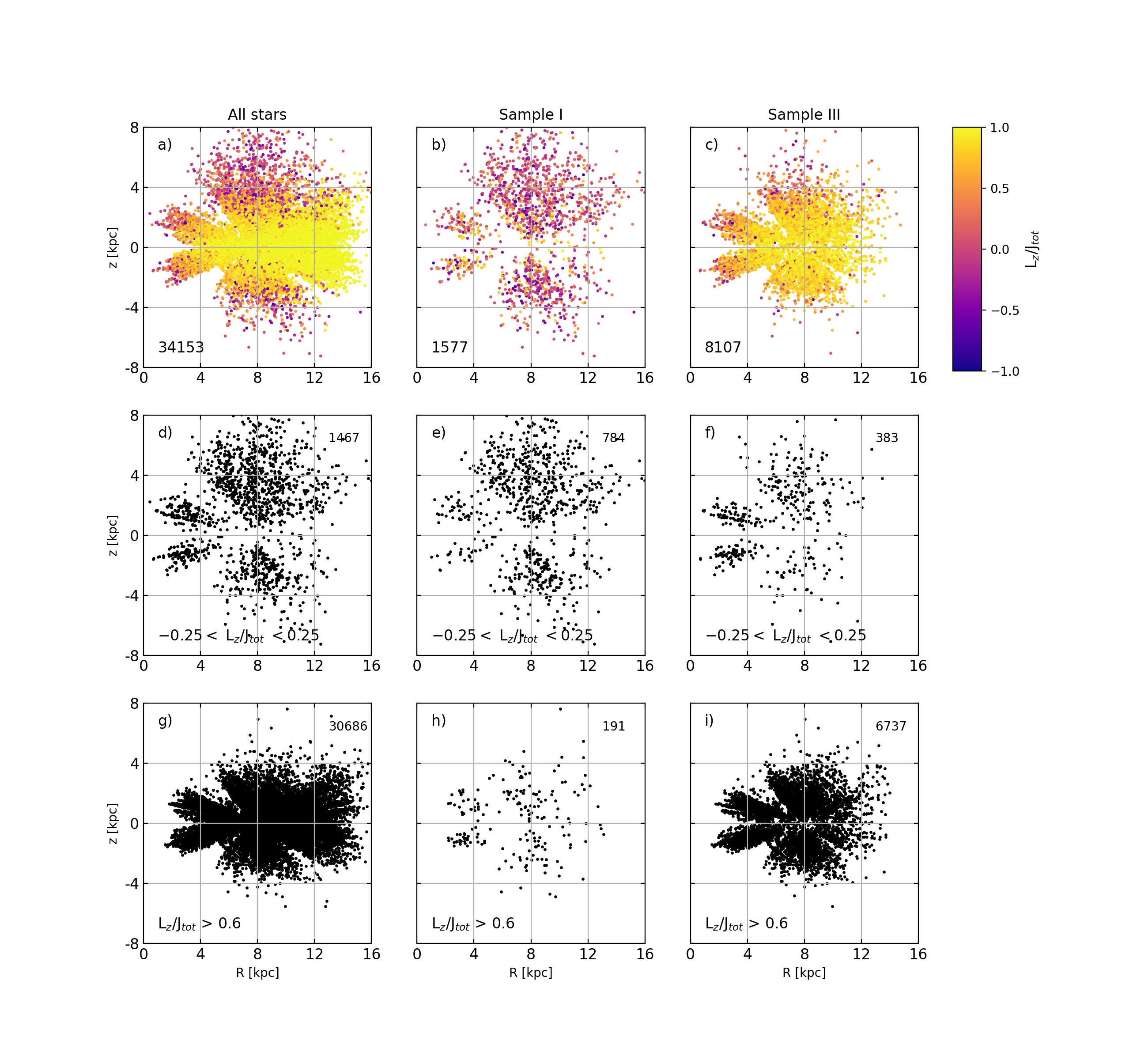}
\caption{Galactocentric radius ($R$) as a function of height above and below the Galactic plane ($z$). Only stars with $1 < \log g < 2$ are included. The number of stars in each panel are indicated, for the top row in the left-hand bottom corner and for middle and bottom rows in the upper right-hand corner. The top-row is colour-coded by $L_z/J_{\rm tot}$ according to the colour-bar to the right of the plots. 
{\bf a)} Full sample. 
{\bf b)} Sample\,I. 
{\bf c)} Sample\,III. 
{\bf d)} Full sample, only stars with $-0.25 < L_z/J_{\rm tot} < 0.25$.  
{\bf e)} Sample\,I, only stars with $-0.25 < L_z/J_{\rm tot} < 0.25$, i.e. Sample\,Ia. 
{\bf f)} Sample\,III, only stars with $-0.25 < L_z/J_{\rm tot} < 0.25$. 
{\bf g)} Full sample, only stars with $0.6 < L_z/J_{\rm tot}$. 
{\bf h)} Sample\,I, only stars with $0.6 < L_z/J_{\rm tot} $, i.e. Sample\,Ib. 
{\bf i)} Sample\,III, only stars with $0.6 < L_z/J_{\rm tot}$.}
\label{fig:check_bias} 
\end{centering}
\end{figure*}

With large spectroscopic surveys, inevitably the selection function of
the survey may influence the perceived properties of a stellar
population
\citep[][]{2019A&A...621A..17M,2016MNRAS.460.1131S}. Preferably it
should be possible to correct for the introduced biases or model their
effect on the data in order to capture the underlying truth. This,
however, can be more or less difficult to do and the examples in the
literature are few. In the present work we are foremost concerned with
identifying and characterising stellar components with the help of
elemental abundance trends. The interpretation of such data does not
need a complete sample but it is important to understand if certain
parts of the Galaxy or parameter space have been excluded thanks to
the selection function of the original survey or via a too vigorous
down selection of objects in the study itself.

To ensure that our conclusions are robust we have looked at the
spatial properties of our samples and also at what effects the
original APOGEE selection function may have on the phase space data we
are using.

\paragraph{Spatial properties of our samples}
The top-row of Fig.\,\ref{fig:check_bias} shows the $R$-$z$-plane for
our full sample, Sample\,I and III, colour-coded by $L_z/J_{\rm
  tot}$. Visual inspection shows that Sample\,I covers the full extent
in $z$ of the full sample, that it also covers a fair portion of the
$R$, but that there is a lack of stars with these chemical signatures
in the disk beyond the solar position
(Fig.\,\ref{fig:check_bias}\,b). Sample\,III is more confined to the
plane ($z < 4$\,kpc, Fig.\,\ref{fig:check_bias}c) and covers the disk
better than Sample\,I.

Rows two and three further divide the data into
$-0.25 < L_z/J_{\rm tot} < 0.25$ (i.e. halo-like orbits) and
$0.6 < L_z/J_{\rm tot}$ (i.e. disk-like orbits). In all three cases
this division shows that stars with halo-like orbits occupy the full
space spanned by all stars (potentially with some lack of stars
outside the solar orbit in the plane) and stars with disk-like orbits
are more confined to the plane.

As we observe the same behavior for the full sample and for
Sample\,I and III, we conclude that there is no direct indication that
our results should not be valid for the whole Galaxy, i.e. the
properties of the stellar populations we observe are not just a local, but a
global phenomenon.

\paragraph{Selection effects showing up in the $E_{\rm n} - L_{\rm z}$
  plane}

\begin{figure*}
  \centering
  \includegraphics[width=18cm,trim={0 1.5cm 0 2cm},clip]{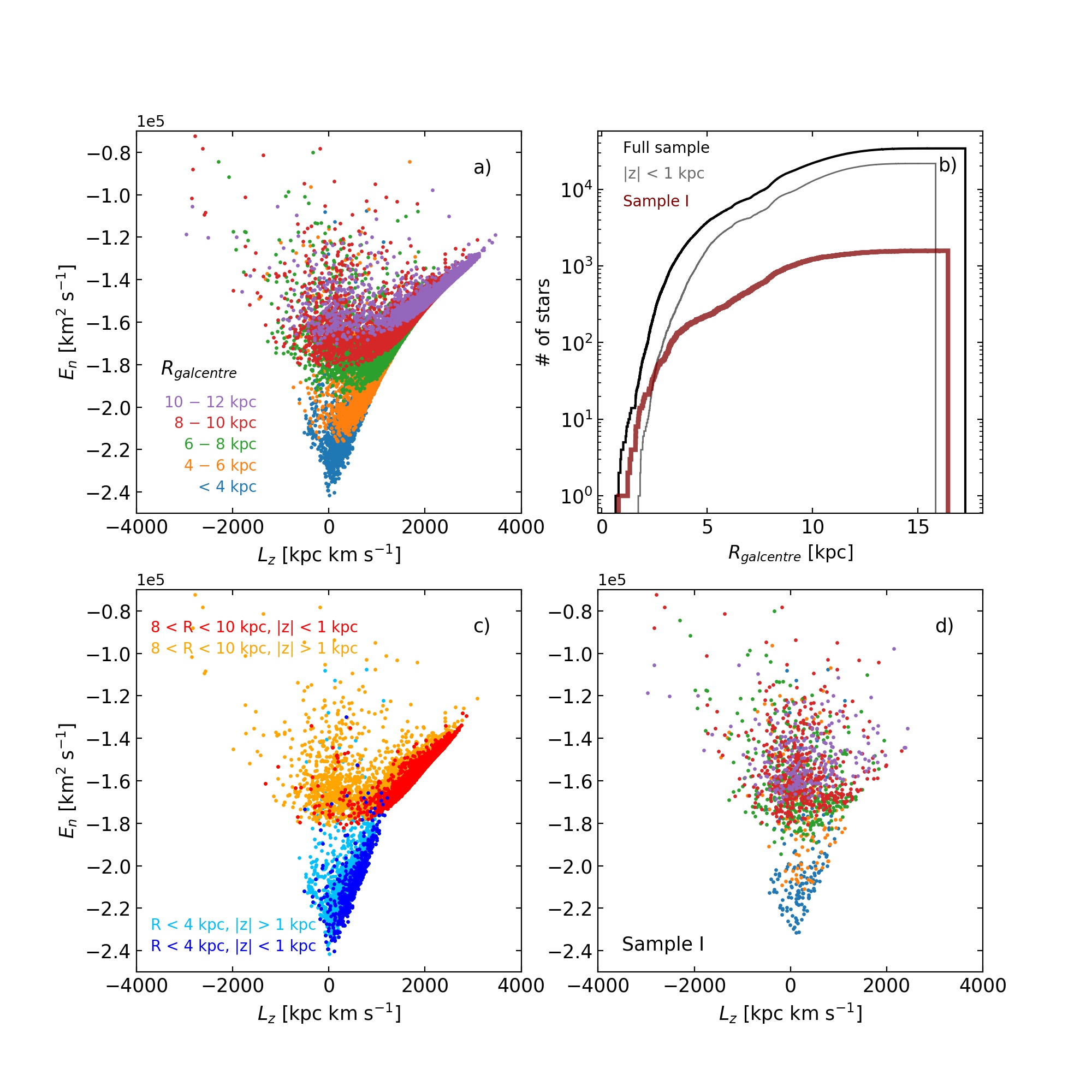}
  \caption{\textbf{a)} $E_{\rm n}$ as a function of $ L_{\rm z}$ for
    stars in our full sample selected at different galactocentric
    radii, as indicated in the legend.  \textbf{b)} Cumulative
    distribution functions for our full sample, stars with z
    $<$1\,kpc, and Sample\,I.  \textbf{c)} $E_{\rm n}$ as a function
    of $ L_{\rm z}$ for stars in two galactocentric bins ($<4$, blue,
    and $8-10$\,kpc, red). Stars with $|z| < 1$\,kpc are shown in
    bright colours while stars with $|z| >1$\,kpc are shown with more
    transparent colours (see legend).  \textbf{d)} $E_{\rm n}$ as a
    function of $ L_{\rm z}$ for stars in our Sample\,I selected at
    different galactocentric radii as indicated in the legend in panel
    a). }
    \label{fig:check_EnLz}
\end{figure*}

Figures\,\ref{fig:check_bias} and \ref{fig:check_EnLz} highlights an
underlying selection effect from the original APOGEE\,DR17 sample.  In
Fig.\,\ref{fig:check_EnLz} a) our sample stars are shown colour-coded
according to their galactocentric distances. Stars in the plane at a
given distance in a galactic potential follow a certain parabola in
the $E_{\rm n} - L_{\rm z}$ plane \citep[see Fig.\,5
in][]{2022MNRAS.510.5119L}. As expected, stars at a certain radius
follow a parabola. The upward scatter in each sample is due to the
sampling of stars at different heights above the Galactic plane. This
is illustrated in panel c) for two radial bins ($<4$ and $8-10$\,kpc)
where stars with $|z|>1$\,kpc are shown in a fainter colour. At a
given $L_{\rm z}$ these stars are more spread in $E_{\rm n}$ than
stars at lower $z$.

In Fig.\,\ref{fig:check_EnLz} d) we show the data for our
Sample\,I. Remember that this sample is simply selected based on the
elemental abundances of the stars (Table\,\ref{tab:samples}). The
$E_{\rm n} - L_{\rm z}$ plane shows a clear lack of stars for
$E_{\rm n} \simeq -2\cdot 10^5$ to $ -1.8\cdot 10^5$. Is this gap real
or part of a selection effect? In panels a) and d) of
Fig.\,\ref{fig:check_EnLz} we have colour coded the stars according to
their galactocentric distances. We can see that the energies for the
gap corresponds to radial distances of about 5\,kpc.  As discussed in
\citet{2022MNRAS.510.5119L} the observing strategy of APOGEE (for DR16
but also for DR17) includes some extra deep fields towards the
Galactic bulge. These result in an excess of stars observed closer to
the Galactic center, at a lower $E_{\rm n}$ than the nominal disk
survey (their Fig.\,6) If we refer to Fig.\,\ref{fig:check_bias}, we
can see that the galactocentric distance of 5\,kpc is less populated
than other radii thanks to the placement of the two deep pointings at
low latitude while the main survey observed at high latitudes reaching
well above the Galactic plane. This survey strategy has left a clear
gap in stellar distribution seen edge-on. This gap is present in all
our samples and is thus not a feature of the selection in the
Mg-Mn-Al-Fe-plane.

 We can thus assume that our results are not biased, however, we are
 missing objects around 5\,kpc (compare Fig.\,\ref{fig:check_EnLz}
 b). There is no reason to assume these objects do not exist but will be
 found in future surveys.  

\section{Dating the stellar components}
\label{sect:dating}

It would be interesting to date the Sample\,Ia and Ib to further
understand their role in the formation of the Milky Way. We are using
RGB stars for our studies and hence it is not feasible to obtain good
ages for individual stars using isochrone fitting
\citep{2010ARA&A..48..581S,2019MNRAS.482..895S}. However,
asteroseismology offers an alternative possibility to derive ages for
RGB stars \citep[see e.g.][]{2017AN....338..644M,2021A&A...645A..85M}.
We searched the literature for ages derived from asteroseismological
observations for stars with similar characteristics to those we find
in Sample\,I and found two studies: \citet{2021NatAs...5..640M} and
\citet{2022MNRAS.514.2527B}. Although these are relatively small
studies they can still give us some first hints as to the nature of
the ages of our samples and also point to which specific studies would
help to better constrain our observations.

We first took a look at the overlap between the two datasets and how
many of the stars in the two studies would fall into our Sample\,I. We
found that the overlap is, for the objective of our study and focus on
Sample\,I, sufficiently large that nothing is gained by using both
samples and we thus selected \citet{2022MNRAS.514.2527B} as being the
sample with more stars falling in Sample\,I.

\subsection{Data}

\begin{figure*}
\centering
\includegraphics[width=17cm,trim={0 13.5cm 0 3cm},clip]{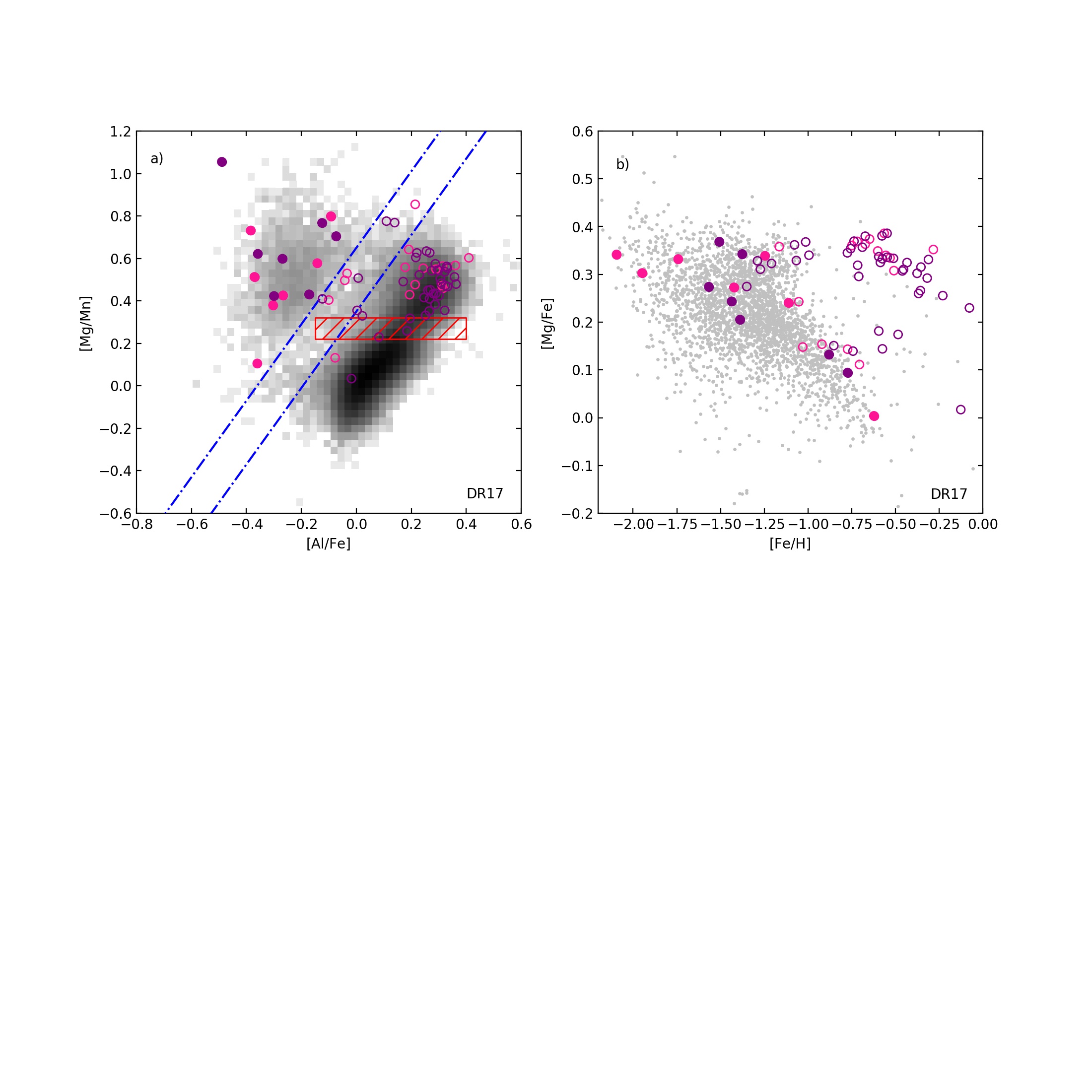}
\caption{Elemental abundances from APOGEE\,DR17 for the stars with
  ages from \citet{2022MNRAS.514.2527B}. Purple symbols refer to stars
  that fulfill our quality criteria for APOGEE\,DR17 and
  \textit{Gaia}\,DR3, while pink symbols do not fulfill those
  criteria. Filled symbols indicate stars that we would associate with
  Sample\,I.  {\bf a)} The Mg-Mn-Al-Fe-plane. The grey 2D histogram
  shows our full sample and the coloured lines our selection criteria
  for the different samples as defined in Fig.\,\ref{fig:cuts}.  {\bf
    b)} [Mg/Fe] as a function of [Fe/H]. Grey dots indicate our
  Sample\,I. }
  \label{fig:seis_elem} 
\end{figure*}

\citet{2022MNRAS.514.2527B} studied a sample compiled from a
cross-match between asteroseismic data from the \textit{Kepler}
mission, astrometric data from the \textit{Gaia} mission,
elemental abundances from APOGEE\,DR16, and the Two Micron All Sky
Survey (2MASS). For references and details of the selection of stellar
data, final target assembly, and calculation of stellar ages and space
motions we refer the reader to Sect.\,2 in
\citet{2022MNRAS.514.2527B}.

In total, \citet{2022MNRAS.514.2527B} provide ages for 70 stars based
on photometric, asterometric, and asteroseismic data (individual
frequencies or $\nu_{\rm max}$ and $\Delta \nu$).  Here we are
interested in the ages provided by this study and less with their
target selection. However, it is worth noting that they start from a
kinematic selection (defined in $L_{\rm z}$). This results in a large
range of [Fe/H] values.

\citet{2022MNRAS.514.2527B} used elemental abundances from APOGEE\,16
\citep{2020ApJS..249....3A}. In our work we are using APOGEE\,DR17
\citep{2021arXiv211202026A}. The exact values of [Fe/H] as well as
[$\alpha$/Fe] will to some extent influence the derived stellar ages.
Figure\,10 in \citet{2022MNRAS.514.2527B} provide a comparison of
[Fe/H] values used in their study and in \citet{2021NatAs...5..640M},
who use APOGEE\,DR14, showing offsets ranging from 0 to about
0.2\,dex. In the same figure there is also a comparison of the ages
derived for the stars which shows the ages to have small differences
(from 0 to about 2\,Gyr). In all cases, the ages derived in the two
studies agree well within the error bars.  We conclude that the small
offsets in [Fe/H] between APOGEE\,DR16 used to derived the ages and
APOGEE\,DR17 used in our elemental abundance selection are negligible,
and that the choice of \citet{2022MNRAS.514.2527B} ages over
\citet{2021NatAs...5..640M} will not influence our conclusions.

The elemental abundances for this sample are shown in
Fig.\ref{fig:seis_elem}.

\subsection{Age of the stars in Sample\,I}
\label{sect:age_s1}

\begin{table*}
\centering
\caption{Stars from \citet{2022MNRAS.514.2527B} that fulfill our
  quality criteria, Table\,\ref{tab:data}, and fall in our Sample\,I
  region in the Mg-Mn-Al-Fe-plane based on elemental abundances from
  APOGEE\,DR17. Ages are from \citet{2022MNRAS.514.2527B}, while
  [Fe/H] and [Mg/Fe] are taken from APOGEE\,DR17. The last column
  indicate if the stars falls in Sample\,Ia. No star falls in
  Sample\,Ib.}
    \label{tab:ages}
  \begin{tabular}{lllllll}
     \hline
     \textit{Gaia}\,ID & [Fe/H] & [Mg/Fe] & Age & Age error & $L_{\rm z}/J_{\rm tot}$ & Ia \\
    \hline
       53635139275591040 & --1.38& 0.20 &11.69 & +2.57/--2.82 & +0.038 & \checkmark \\
  2099659187162016512  & --1.37&0.34 & 11.75 & +1.91/--1.20 &--0.113 &\checkmark \\
  2126445115779806976   & --1.43&0.24 & 5.54 & +1.88/--3.18 &--0.252 & -- \\ 
   2127447522484965504  & --1.50& 0.36 & 10.51 & +2.18/--1.82 & --0.480 & --\\ 
  2133314619611880448  & --1.56&0.27 & 10.27 & +1.96/--2.50 & +0.092 & \checkmark \\ 
  2538202737087917184  & --0.87&0.13 & 9.85 & +3.40/--3.45 & +0.170 & \checkmark \\
  2626567188077168896  & --0.77 & 0.09 & 4.59 & +4.23/--2.19   & --0.026 & \checkmark \\ 
        \hline
    \end{tabular}
\end{table*}

In Table\,\ref{tab:ages}, we list those stars that fulfill our quality
criteria and fall in Sample\,I (see Table\,\ref{tab:data} and
\ref{tab:samples}). We note that our quality criteria are more
restrictive than those applied in \citet{2022MNRAS.514.2527B}, this is
likely partly due to different versions of APOGEE being used.  We also
consider those stars that would be selected based on their elemental
abundances if we disregarded the quality flags we applied.

We find an average of $9.2\pm2.7$\,Gyr for the 7 stars that fall in
the Sample\,I region in the Mg-Mn-Al-Fe-plane and fulfill our quality
cuts.  Two of the stars have young ages, typical of the stellar
disk. If those are excluded the age is $10.8\pm0.8$\,Gyr. If we use
our cut in $L_{\rm z}/J_{\rm tot}$ to consider the stars in Sample\,Ia
(see Sect.\,\ref{sect:kinprop2}), we are left with five stars that
have a mean age of $9.6\pm2.6$\,Gyr. If the young star is excluded we
obtain a mean age of $10.9\pm0.8$\,Gyr.

We can conclude that the stars that fall in Sample\,I or Sample\,Ia
and fulfill our quality criteria for the elemental abundances have
an old age. There are two stars that have young ages. Such stars have
been found also in other studies and are sometimes referred to as young
$\alpha$-rich stars \citep[][]{2015A&A...576L..12C}. One explanation
for the presence of such stars is that they are in fact blue
stragglers \citep{2016A&A...595A..60J}.

None of the stars in the seismic sample that fall in the Sample\,I
region in the Mg-Mn-Al-Fe-plane have $L_{\rm z}/J_{\rm tot} > 0.6$,
i.e.  stars in Sample\,Ib. This means that although we can put an age
on Sample\,Ia we are unable from the presently available stellar ages
to derive an age for Sample\,Ib, i.e. the disk.

We undertake the same analysis for Sample\,III as we did for
Sample\,I, see Table\,\ref{tab:agesIII}.  From this we derive an age
of $8.8\pm2.8$\,Gyr for Sample\,III using the ages from
\citet{2022MNRAS.514.2527B}. Thus with this dataset, we find that the
stars in Sample\,I have a mean-age about 1--2\,Gyr older than those in
Sample\,III.

\section{Discussion}
\label{sect:discussion}

\begin{table*}
  \caption{List of selection criteria used to select the
    \textit{Gaia}-Sausage-Enceladus for Fig.\,\ref{fig:ELz_all} and
    \ref{fig:mgfe_all}.  The letter in the first column refers to the
    panel labels in those figures.  The studies are ordered according
    to publication year.}\label{tab:sel}
    \begin{tabular}{ll | rl l rl | llll}
      \hline
 Panel &    Study & \multicolumn{5}{c}{Selection ctriteria} & Units \\
    \hline 
 i) &   \textit{``Sausage''} &$-100 $ & $ <$ & $V_{\phi}$ & $ < $ & $100$ & km\,s$^{-1}$\\
    \hline
  f) &  \citet{2018Natur.563...85H} &$-1500$ & $<$ & $L_{\rm z}$ & $ <$ & $150$ & kpc\,km\,s$^{-1}$   \\
  &        & $-1.8 \cdot 10^{5}$ & $ <$ &  $E_{\rm  n} $ &  & & km$^2$\,s$^{-1}$ \\
    \hline
c) & \citet{2019MNRAS.488.1235M} & & & $|J_{\phi}/J_{\rm tot} |$ &$<$ &$ 0.07$ \\
   &       & & & $\frac{(J_{\rm z} - J_{\rm z})}{J_{\rm tot}}$ & $ <$ & $ -0.3$ \\
    \hline
h) & \citet{2020MNRAS.497..109F} &     30 &$< $ & $ \sqrt(J_{\rm r})$ & $ < $&$50$ & (kpc\,km\,s$^{-1}$)$^{1/2}$\\
   &       & $-500 $ & $<$ & $L_{\rm z}$ & $ < $ & 500 & kpc\,km\,s$^{-1}$\\
    \hline
b) & \citet{2022MNRAS.tmp.3011H} & & & $|L_{\rm z}|$ & $  < $ & $0.15\cdot 10^{3}$ & kpc\,km\,s$^{-1}$   \\
  &        &$-1.6 \cdot 10^{5}$ & $ <$ &  $E_{\rm  n} $ & $ < $ &$-1.1 \cdot 10^{5}$ &km$^2$\,s$^{-1}$ \\
    \hline
e) & \citet{2022ApJ...926L..36N} &  $-1.5 \cdot 10^{5}$ &$ < $ & $E_{\rm n} $ & & & km$^2$\,s$^{-1}$ \\
  &        & 0.7 &  $< $ &{$eccentricity$} &&&\\ 
 &         & 5 &$<$ & \multicolumn{3}{l |}{$R_{\rm   galcentre}$} &   kpc\\
 &         & $0.25 $ & $< $ & [Mg/Mn]  & & &\\
  &        & $0.55 $ & $< $ &   \multicolumn{3}{l|}{[Mg/Mn]  $-4.25\cdot$[Al/Fe] }\\
  \hline
  \end{tabular}
\end{table*}

Following the work by \citet{2015MNRAS.453..758H} and
\citet{2020MNRAS.493.5195D} we have explored the ability of the
Mg-Mn-Al-Fe-plane to distinguish stars from different stellar
populations in the Milky Way. By further combining the elemental
abundance data with kinematical properties of the stars we have found
that the region with low [Al/Fe]-values (our Sample\,I) contains
stellar populations with kinematical properties associated with the
stellar halo and with the stellar disk. Furthermore, we find that
those samples differ in their [Mg/Fe] and [Ni/Fe] abundance trends, indicating
different origins.


\subsection{\textit{Gaia}-Sausage-Enceladus}

When the astrometric data  from the ESA
\textit{Gaia} satellite  are combined with
radial velocities it is possible to study the halo kinematics in great
detail. Using different techniques, research teams have 
found stellar populations, accreted galaxies, and stellar streams.

The \textit{Gaia}-Sausage-Enceladus and other discreet stellar
populations have been found thanks to the astrometric data from the ESA
\textit{Gaia} satellite \citep{2018Natur.563...85H,2018MNRAS.478..611B,2020ARA&A..58..205H,2022ApJ...926L..36N,2022MNRAS.tmp.3011H}. 

In Table\,\ref{tab:sel} we have, from the literature, collected five
different ways of selecting \textit{Gaia}-Sausage-Enceladus stars. We
also include a kind of trivial definition of the Sausage, i.e.
$-100 < V_{\phi} < 100$\,km\,s$^{-1}$. Figure\,\ref{fig:ELz_all}\,a)
shows a 2D histogram of the $E_{\rm n}-L_{\rm z}$ plane for our full
sample. In all remaining panels, our Sample\,I is shown in cyan for
comparison. Our Sample\,Ia and Ib are shown in panels d) and g).  In
the other panels we show the resulting distributions in the
$E_{\rm n}-L_{\rm z}$ plane when we select stars from our main sample
according to the six \textit{Gaia}-Sausage-Enceladus criteria listed
in Table\,\ref{tab:sel}. We show two selections for each sample. Grey
points indicate stars selected from our catalogue only using the
criterium listed in Table\,\ref{tab:sel} and the number of selected
stars indicated in grey in the lower left-hand corner of each panel.
Black points indicate stars that also fall into our Sample\,I
(Table\,\ref{tab:samples}) and the number of such stars is indicated
in black in the lower left-hand corner of each panel. In the upper
right-hand corner of the six \textit{Gaia}-Sausage-Enceladus panels,
the average $L_{\rm z}$ and associated $\sigma$ are shown.

It is interesting to note that the five definitions of
\textit{Gaia}-Sausage-Enceladus taken from the literature show
distinctly different distributions in the $E_{\rm n}-L_{\rm z}$
plane. In all cases, imposing our selection criterium in the
Mg-Mn-Al-Fe plane lowers the number of stars but it does not seem to
change the distribution in the $E_{\rm n}-L_{\rm z}$ plane much apart
from for the trivial definition shown in panel i), where we can see
that the $E_{\rm n}$ gap is less prominent in the grey points than in
the black ones. A similar effect is suggested in panel c) using the
criterium from \citet{2019MNRAS.488.1235M}.

It is immediately obvious that our Sample\,Ib bears little resemblance
in the $E_{\rm n}-L_{\rm z}$ plane to any of the definitions of the
\textit{Gaia}-Sausage-Enceladus. This is not surprising given that we
find it has disk-like kinematics. Our Sample\,Ia on the other hand
occupies many of the same spaces in $E_{\rm n}-L_{\rm z}$ as the six
definitions. The distribution based on selection by 
\citet{2019MNRAS.488.1235M}  is the one most similar to our
Sample\,Ia. In general though the literature definitions avoid the
lower energies included in Sample\,Ia. These are only present in the
trivial definition and in
\citet{2019MNRAS.488.1235M}. \citet{2020MNRAS.497..109F} and
\citet{2022MNRAS.tmp.3011H} have quite similar distributions while
\citet{2022ApJ...926L..36N} has a much wider distribution in
$L_{\rm z}$ than the other two. It also reaches higher
energies. Finally, we note that the selection criteria defined by
\citet{2018Natur.563...85H} creates an essential retrograde population,
which reaches to quite low energies.

Fig.\,\ref{fig:mgfe_all} presents the associated distribution of
[Mg/Fe] as a function of [Fe/H] for Sample\,I, Ia, Ib, and all six
definitions of \textit{Gaia}-Sausage-Enceladus. Colour-coding is the
same as in Fig.\, \ref{fig:ELz_all}, but here we only print the number
of stars fulfilling both the literature criteria and our selection in
Mg-Mn-Al-Fe-plane. In addition, each panel shows the average [Fe/H]
and [Mg/Fe] for the black points.

There are three immediate observations to be made from these plots: 1)
Regardless of selection criteria all selections result in a down-going
trend for [Mg/Fe] with increasing [Fe/H]. The trend coincides with our
Sample\,I and even more clearly with Sample\,Ia. 2) In all cases
adding our selection in the Mg-Mn-Al-Fe-plane removes the high [Mg/Fe]
stars at higher [Fe/H], i.e. the disk. 3) In all but the trivial case
there are stars more enhanced in [Mg/Fe] at a given [Fe/H] in
Sample\,I (cyan). These stars are deselected in all but the trivial
case and overlap with our Sample\,Ib in this plane.

\begin{figure*}
\centering
\includegraphics[width=18cm,trim={0cm 1.5cm 0 2.5cm},clip]{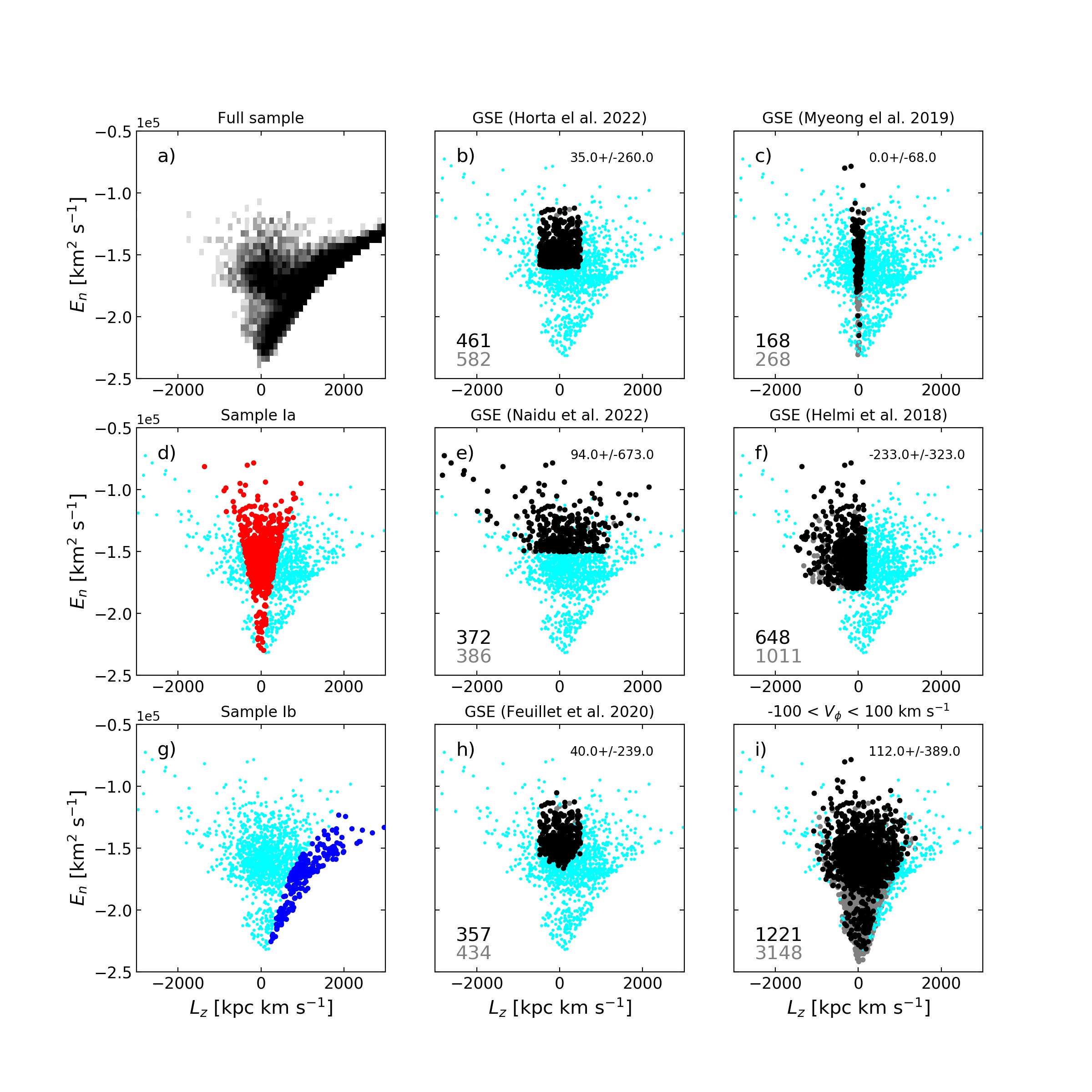}
\caption{$E_{\rm n}$ as a function of $L_{\rm z}$. In all plots the
  data are restricted to stars with $1< \log g <2$.  {\bf a)} 2D
  histogram for our full sample.  {\bf b) -- i)} Samples as indicated
  on the top of each panel. Cyan symbols always refer to our
  Sample\,I. Black symbols refer to stars which fulfill both the
  selection criteria from the study indicated on the top as well as
  our own selection criteria for Sample\,I. Grey symbols refer to
  stars that are selected by the criteria in the indicated study but
  falls outside our Sample\,I. The number of stars in each sample is
  given in the lower left-hand corner of each plot. In the upper right
  hand-corners are the mean $L_{\rm z}$ for each sample indicated in
  black shown. Table\,\ref{tab:sel} shows the selection criteria for
  used for each of the studies. }
\label{fig:ELz_all} 
\end{figure*}

\begin{figure*}
\centering
\includegraphics[width=18cm,trim={0cm 3cm 0 4cm},clip]{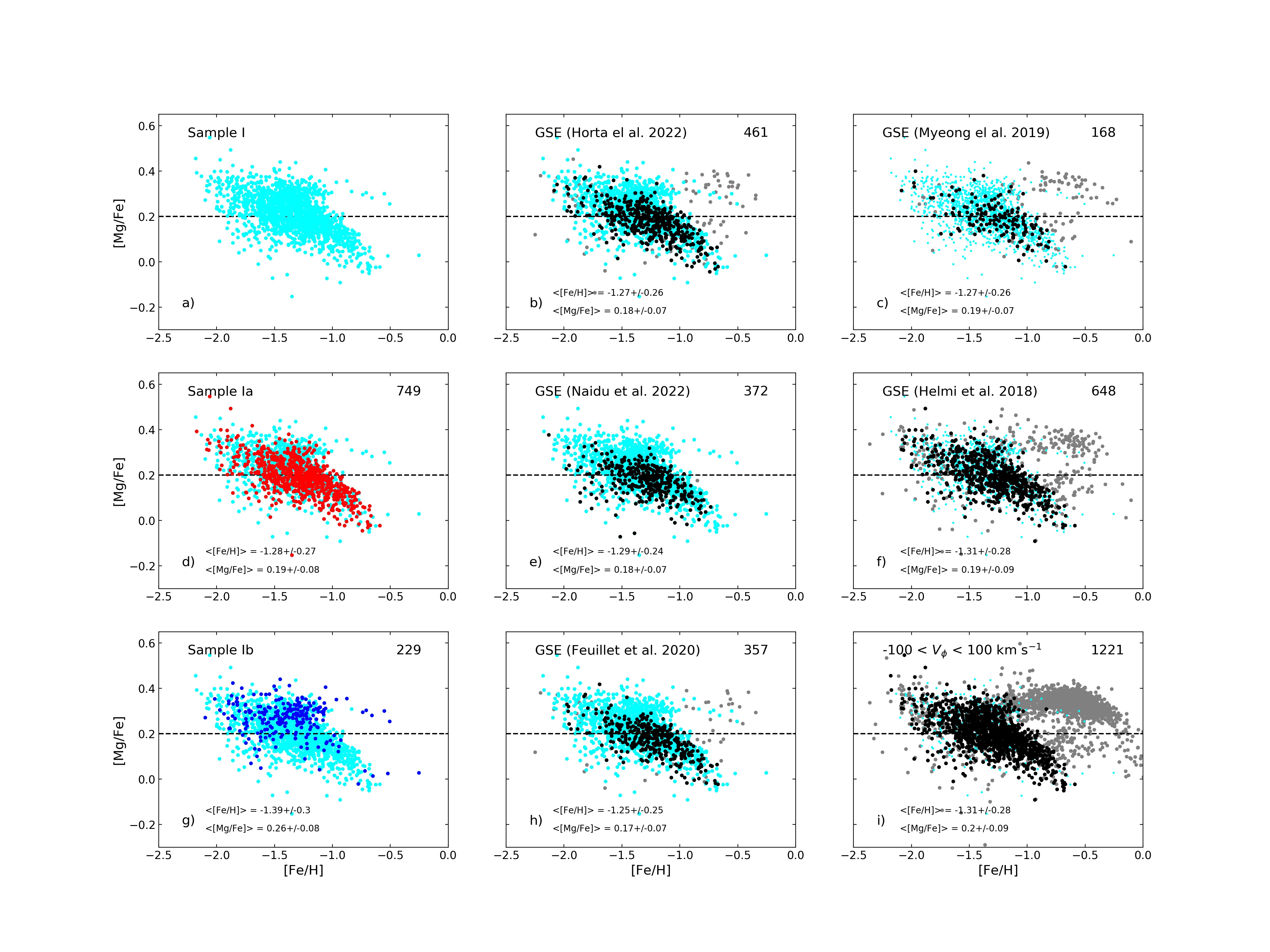}
\caption{[Mg/Fe] as a function of [Fe/H]. In all plots the
  data are restricted to stars with  $1< \log g <2$. 
  {\bf a)} Sample\,I. 
  {\bf b) -- i)} Samples as indicated on the top of each panel. Cyan
  symbols always refer to our Sample\,I. Black symbols refer to stars
  which fulfill both the selection criteria from the study indicated
  on the top as well as our own selection criteria for Sample\,I. Grey 
  symbols refer to stars that are selected by the
  criteria in the indicated study but falls outside our
  Sample\,I. The number of black points in each sample is given in the upper
  right-hand corner of each plot. In the lower left-hand corners are
  the mean [Fe/H] and [Mg/Fe] for the black, blue
  and red points
  shown. Table\,\ref{tab:sel} shows the selection criteria for used for each of the studies. }
 \label{fig:mgfe_all} 
\end{figure*}

Various investigations have used kinematic criteria and/or various
clustering algorithms to identify groups of stars that likely come
from a single accreted progenitor.  We find that the Mg-Mn-Al-Fe-plane
is suitable for identifying stars with halo kinematics and likely
accreted origin, but we find that the same region also includes stars
with typical disk kinematics. When we study the elemental abundance
trends, we find that, regardless of selection method, samples that can
be associated with the \textit{Gaia}-Sausage-Enceladus progenitor show
remarkably similar elemental abundance trends but not always the same
kinematic characteristics. In particular, the accreted stars appear to
be able to have both pro- and retro-grade orbits. It appears
remarkable that regardless of criterium used, we recreate the same
downward [Mg/Fe] trend first observed by
\citet{1997A&A...326..751N,2010A&A...511L..10N}. Thus it appears that
the stars from this merger are well-mixed in kinematics but better
distinguished in elemental abundance space.

The Hestia simulations
\citep{2020MNRAS.498.2968L,2022arXiv220604521K,2022arXiv220604522K,2022arXiv220605491K}
follow three Local Group pairings of galaxies and their respective
merger histories. One important aspect of these simulations is that
the galactic potential evolves in the cosmological context. This means
that after a smaller galaxy merges with the main progenitor the
potential gets deeper with time. The effect is that, regardless of
time of merger, all of the major mergers have roughly the same energy
today. This means that the $E_{\rm n}$ -- $L_{\rm z}$ plane is partly
degenerate when it comes to picking up individual merger
debris. Another important information gleaned from the inspection of
the $E_{\rm n}$ -- $L_{\rm z}$ diagrams for the mergers in the Hestia
simulations is the fact that almost all mergers show stars on both
pro- and retro-grade orbits. This is very similar to what we find when
we are looking at the stars that, in various ways, could be associated
with the \textit{Gaia}-Sausage-Enceladus.

\citet{2019A&A...630L...4M} studied the globular cluster system in the
Milky Way and defined the Main Progenitor as the properties traced by
globular clusters formed in situ in the stellar disk or the Galactic
Bulge. The left panel of Fig.\,2 in \citet{2019A&A...630L...4M} shows
the $E_{\rm n}$ -- $L_{\rm z}$ plane of Milky Way globular clusters.
It is interesting to note that the clusters associated with the Main
Progenitor have a distribution in the $E_{\rm n}$ -- $L_{\rm z}$ plane
that is strongly reminiscent of our Sample\,Ib.
\citet{2019MNRAS.486.3180K}, in a similar manner, used globular
clusters to trace early merger events in the Milky Way and identified
what they claim to be one such event; the deeply bound Kraken
galaxy. We have tentatively looked at the deeply bound portions of our
samples, compare also the trivial definition of the
\textit{Gaia}-Sausage-Enceladus presented earlier. We find some
similarities in elemental abundance space with our Sample\,Ib and a
Kraken sample.

A recent critique of the practice of using the globular clusters to
trace the formation and assembly history of a galaxy can be found in
\citet{2022arXiv221004245P}. They study the accretion of galaxies and
associated globular clusters onto a galaxy where the potential is
allowed to change as the main galaxy grows via mergers over cosmic
time. In agreement with the main findings of the Hestia simulations,
they show that the globular clusters from the merged galaxy do not end
up in the same part of the $E_{\rm n}$ -- $L_{\rm z}$ plane as stars
from their host galaxy.  Only for small galaxies may there remain some
similarities in kinematic properties between the field stars accreted
from the smaller galaxy and its globular clusters.

\subsection{The chemically unevolved stellar disk}
\begin{figure*}
  \centering
\includegraphics[width=18cm,trim={2cm 2cm 4cm 3cm},clip]{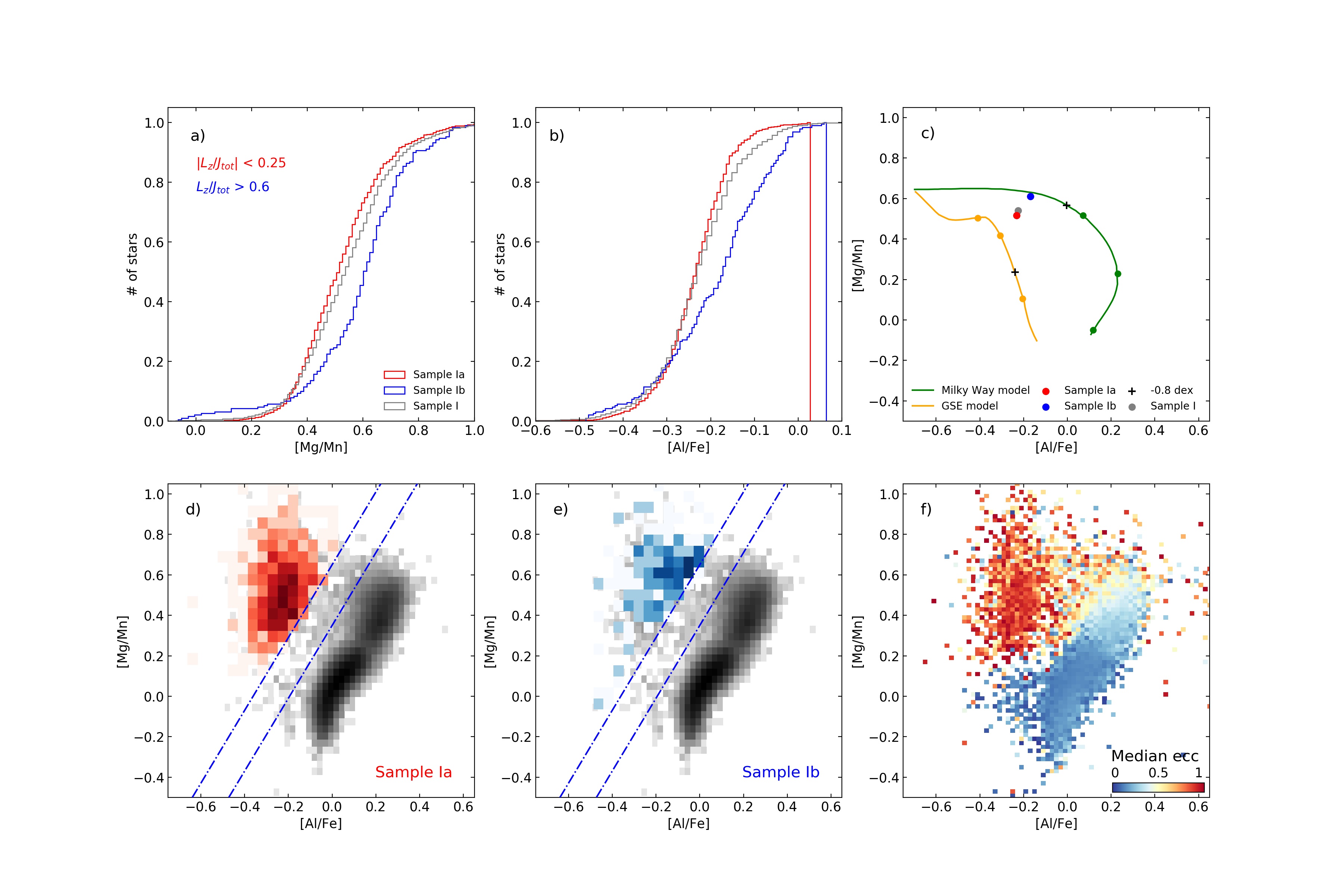}
\caption{{\bf a)} Normalized cumulative distribution of [Mg/Mn] for
  Sample\,I, Ia, and Ib as indicated in the legend. {\bf b)} Normalized cumulative distribution of [Mg/Mn] for
  Sample\,I, Ia, and Ib as indicated in the legend in panel a). {\bf
    c)}  [Mg/Mn] as function of [Al/Fe] showing the positions of the
  median values for Sample\,I, Ia and Ib as indicated in the
  legend. The two models from \citet{2022MNRAS.tmp.3011H} are also
  shown. {\bf d)} 2D histogram [Mg/Mn] as function of [Al/Fe] for the whole sample
  (grey) and Sample\,Ia (red). {\bf e)} 2D histogram [Mg/Mn] as function of [Al/Fe] for the whole sample
  (grey) and Sample\,Ib (blue). {\bf f)} 2D histogram [Mg/Mn] as function of [Al/Fe] for the whole sample
  where the colour indicated the median eccentricity ($ecc.$) for the stars in each
  bin as indicated by the colour-bar.}
\label{fig:disk}
\end{figure*}

Earlier (Sect.\ref{sect:kinprop2}) we found evidence that Sample\,Ib is part of
the disk of the Milky Way and not the halo even though the stars fall
in the region of the Mg-Mn-Al-Fe-plane thought to (mainly) harbour
halo and/or accreted stars. Figure\,\ref{fig:disk} a) and b) shows the
cumulative distributions of [Mn/Mg] and [Al/Fe], respectively, for Sample\,I, Ia and
Ib. It is immediately clear that the kinematically
defined sub-samples differ. Sample\,Ib is more enhanced in both
[Mg/Mn] and [A/Fe]. In Fig.\,\ref{fig:disk} c) we compare the
median values of the elemental abundances for our three samples with
the two chemical evolution models by
\citet{2022MNRAS.tmp.3011H}. Sample\,Ib lies close to the
evolutionary track of the Milky Way model, while the full sample as
well as Sample\,Ia tend more towards the GSE model thus giving
further support to our interpretation that Sample\,Ib really is the
(chemical) beginning of the Milky Way disk.

In Fig.\,\ref{fig:disk} d) and e) we look at the distributions of the
data for Sample\,Ia and Ib in the Mg-Mn-Al-Fe-plane using a 2D
histogram. We find that not only do their median values differ, but also
the distribution of the stars in this plane. Sample\,Ia shows an
elongated, more or less vertical distribution, while Sample\,Ib shows a
horizontal distribution and is a more chemically enriched population. The two
kinematically defined sub-samples in Sample\,I clearly occupy two
chemically distinct populations -- one seemingly following a
trajectory that could describe an accreted dwarf galaxy and the other
a trajectory that would connect well to the chemical evolution of the
main body of the Milky Way when compared with chemical evolution models.

Finally, we look at some of the kinematic information for our whole
sample in Fig.\,\ref{fig:disk} where we show all stars in the
Mg-Mn-Al-Fe-Plane binned into a 2D histogram where the colour
represents the median eccentricity of the stars in each bin. There are
several things to note from this plot. The first one is that the
distribution of eccentricity changes between different regions of the
plane.

In particular, stars that fall virtually in the same place as
Sample\,Ia have a mean eccentricity of 0.90 (with a
$\sigma$=0.10 and median eccentricity uncertainty of 0.03),
i.e. radial orbits. Stars in the region occupied by Sample\,Ib show a
lower mean eccentricity of 0.37 (with a $\sigma$=0.16 and
  median eccentricity uncertainty of 0.01). This carries over into
the region where Sample\,III sits with mean a eccentricity of
0.38 (with a $\sigma$=0.21 and a median eccentricity uncertainty
  $< 0.01$) and then flows down towards lower [Mg/Mn], decreasing in
eccentricity such that Sample\,IV is almost entirely on circular
orbits with a mean eccentricity of 0.15 (with a $\sigma$=0.09).

This figure is a nice illustration that the stars on eccentric
orbits, i.e. accreted stars, are found in a particular region of the
Mg-Mn-Al-Fe-plane, but that they do not occupy the whole area that is
normally assigned to the accreted region. Instead only a specific region is
occupied. This shows that although we originally divided our Sample\,I
into two extremes, as concerns their kinematics, we still picked up the
major accreted component -- presumably the
\textit{Gaia}-Sausage-Enceladus. The stars with disk kinematics in
Sample\,I on the other hand connect kinematically quite well to
Sample\,III which harbours the stars we ordinarily associate with the
thick disk, and possibly with the \textit{Splash} to some extent. 

We summarise that the stellar disk extends into the regions
in the Mg-Mn-Al-Fe-plane 
associated with merger debris. The kinematics of the whole plane shows
that this part of the stellar disk, the chemically unevolved disk, smoothly
connects to the hotter part of the disk that is also significantly
chemically enriched. 

\subsection{Estimating the accreted fraction of stars}

\begin{figure}
\centering
\includegraphics[width=9.5cm,trim={2.5cm 3cm 0 3cm},clip]{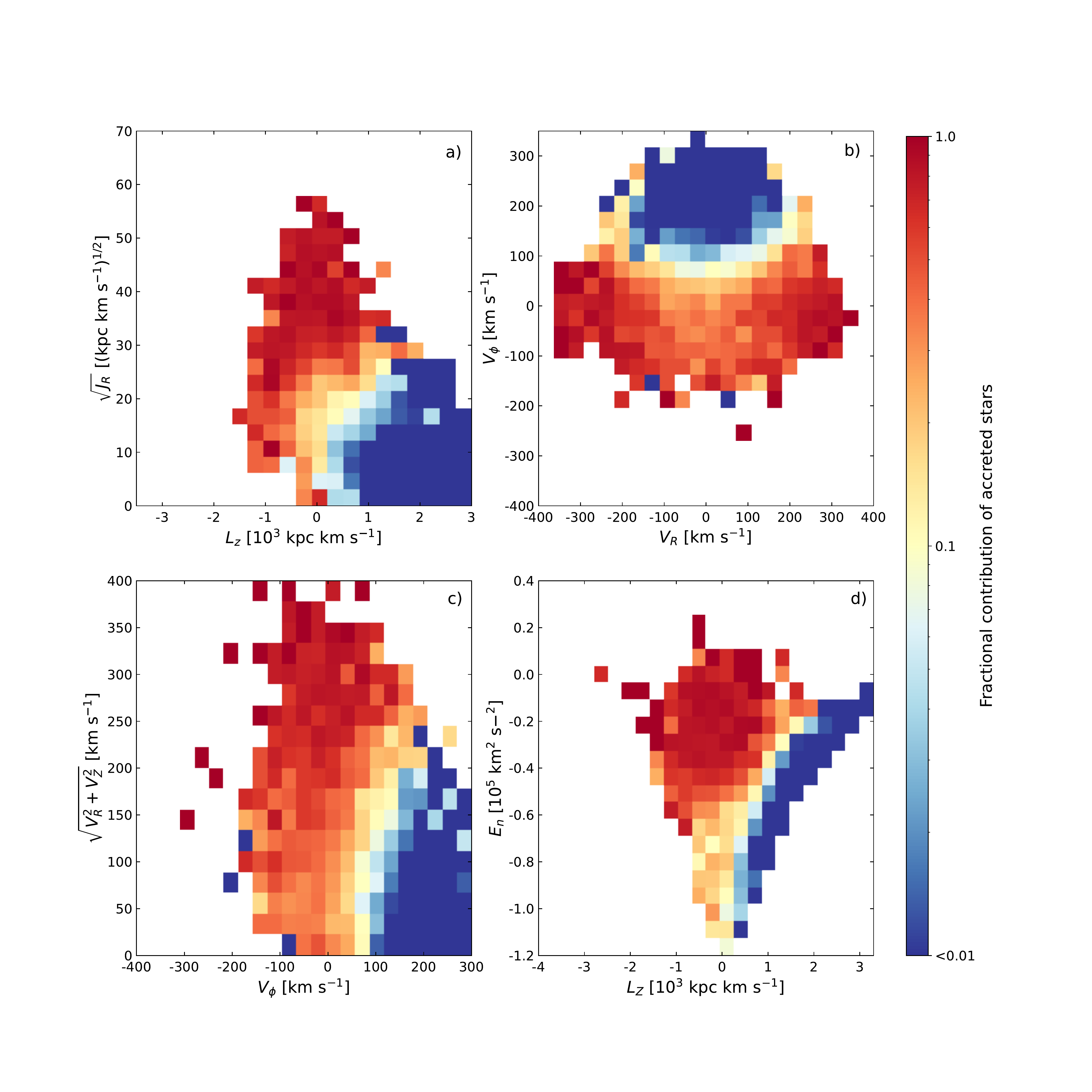}
\caption{Accreted fraction of stars in APOGEE\,DR17 based on selection
  of Sample\,I in the Mg-Mn-Al-Fe-plane, Table\,\ref{tab:samples}. The
  colour-coding indicates how much of the population should be
  considered accreted (log of fractional contribution indicated by the
  colour bar). Pixels with two or fewer stars have been removed.  {\bf
    a)} $\sqrt{J_{\rm R}}$ as a function of $L_{\rm z}$.  {\bf b)}
  $V_{\phi}$ as a function of $V_{\rm R}$.  {\bf c)} Toomre diagram.
  {\bf d)} $E_{\rm n}$ as a function of $L_{\rm z}$.  Units as
  indicated on the axes.}
\label{fig:accfrac} 
\end{figure}

\begin{figure}
\centering
\includegraphics[width=9.5cm]{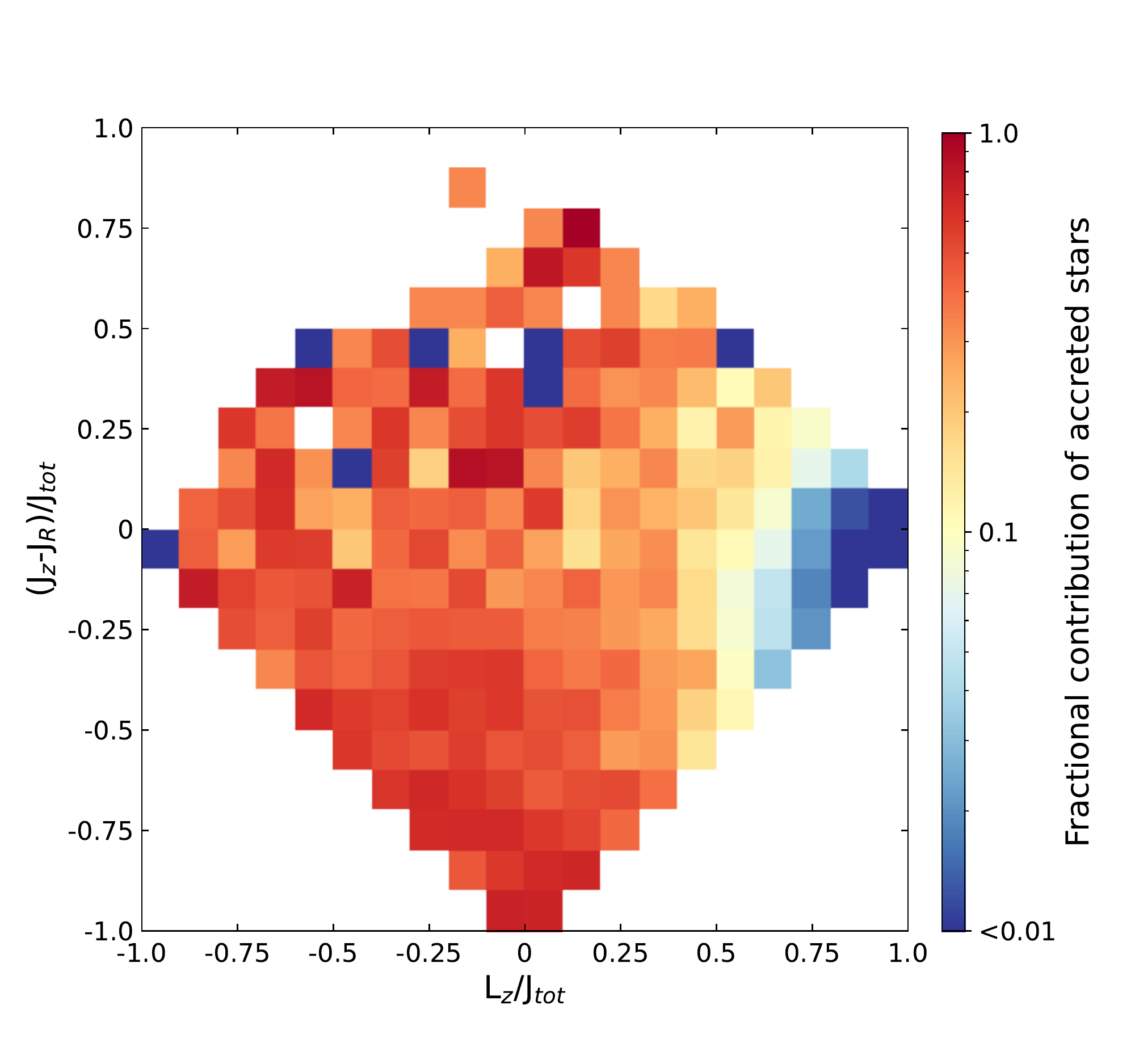}
\caption{Action diamond showing the accreted fraction of stars in
  APOGEE\,DR17 based on selection of Sample\,I in the
  Mg-Mn-Al-Fe-plane, Table\,\ref{tab:samples}. The colour-coding
  indicates how much of the population should be considered accreted
  (log of fractional contribution indicated by the colour bar). Pixels
  with two or fewer stars have been removed.  Units as indicated on
  the axes.}
\label{fig:accfrac_mey} 
\end{figure}

If we want to understand how galaxies form and evolve we would like to
know how much of the stellar mass has formed in the galaxy itself and
how much has been accreted. If stars formed in smaller galaxies have
elemental abundance signatures that make them stand out from those
stars that have formed in the galaxy itself that would be one way of
estimating how many of the stars in a galaxy today have actually
formed in other galaxies and this would then give us constraints on
galaxy formation in general.
 
As we have discussed, the Mg-Mn-Al-Fe-plane is a good place to
identify stars formed in other galaxies. But, we have also shown that
the area associated with the signatures of smaller stellar systems in
the Mg-Mn-Al-Fe-plane also contains stars belonging to the (old)
disk. This is not unexpected given the modeling results by
\citet{2022MNRAS.tmp.3011H}. However, we think it is still of interest
to obtain an upper limit to the number of accreted stars in our sample
as well as obtaining maps that show us where the majority of accreted
and in situ stars are situated.

Here we use Sample\,I in its entirety to represent accreted
stars. This thus gives us an \textit{over} estimate of the fraction of
accreted stars as Sample\,I includes stars on disk orbits.  The
accreted fraction of stars with a kinematic signature is simply the
number of stars in Sample\,I divided by the total number of
stars. Fig.\,\ref{fig:accfrac} shows our results in four kinematic
spaces. In these plots pixels with two or fewer stars have been
removed. Figure\,\ref{fig:accfrac_mey} shows the same type of data but
now presented in the action diamond. 

We see that the kinematic region associated with the
\textit{Gaia}-Sausage-Enceladus shows up as dark blue regions in all
four spaces in Fig.\,\ref{fig:accfrac} \citep[compare e.g.][and
Fig.\,\ref{fig:ELz_all}]{2021MNRAS.508.1489F} but the action diamond
appears to be a lot less sensitive to pick out the accreted stars (see
Fig.\,\ref{fig:accfrac_mey}).  In
Fig.\,\ref{fig:accfrac}\,d) we see that for low energies, i.e. tightly
bound orbits below $\sim -0.7\cdot 10^5$\,km$^2$\,s$^{-1}$, the
fraction of accreted stars is about 10--30\% for stars on orbits with
$L_{\rm z}$ close to zero. This is interesting as it is indicating
that based on the stellar chemical make-up we are seeing stars that
have formed in the main galaxy here, i.e. there is not just accreted
stars but we are seeing evidence for the main progenitor.

 Thus, even though the central parts of the Milky Way may harbor
 accreted components there is a substantial portion of in situ formed
 stars that should belong to the initial galaxy. This would agree with
 our finding that Sample\,Ib is a disk-like population.

 \section{Conclusions}
 \label{sect:conclusion}

 We have for the first time identified the early stellar disk in the
 Milky Way by using a combination of elemental abundances and
 kinematics.

 Stars accreted on to the Milky Way by other (smaller) galaxies
 merging with our Galaxy can be difficult to find. In particular
 kinematic signatures may be erasade more quickly and completely than
 previously though when the evolving Galactic potential is taken into
 account. Instead, we turn our attention initially to the elemental
 abundances in the stars.

 \citet{2015MNRAS.453..758H} and \citet{2020MNRAS.493.5195D}
 empirically found that the elemental abundance plane spanned by
 [Mg/Mn] and [Al/Fe] could be used to identify accreted
 stars. \citet{2021MNRAS.500.1385H} further discussed this possibility
 underpinning their arguments with chemical evolution models. We
 re-address the validity of the Mg-Mn-Al-Fe-plane for identifying
 accreted debris and find that it is useful also when taking issues
 related to the derivation of elemental abundance, such as departures
 from LTE, into account. As illustrated in
 Fig.\,\ref{fig:mgmnalfe_ill}, the selection of clumps of stars in the
 Mg-Mn-Al-Fe plane is robust against departures from NLTE and 3D for
 the red giant branch stars used in this study.

 We proceed to use this abundance plane to identify the accreted/halo
 component solely using elemental abundances, which we refer to as
 Sample\,I. The kinematical properties of the stars in Sample\,I
 contains, as expected, stars with all the kinematic hallmarks of
 being accreted, but we also find stars on clear disk-like orbits.  We
 also find that the spatial distribution of the stars differ. Stars
 with disk-like kinematics are (more) confined to the Galactic plane.

 We further analyse the properties of the two sub-samples and identify
 the accreted stars (mainly) with the \textit{Gaia}-Sausage-Enceladus
 whilst the disk stars are the start of the main body of the Milky Way
 disk (as predicted by chemical evolution models). The stars in
 Sample\,I with disk-like orbits have higher [Mg/Fe] as a given [Fe/H]
 than the stars in Sample\,I with halo-like orbits, see
 Fig.\,\ref{fig:mgfe_sample1}, similar to a thick disk population.

We have thus for the first time identified the early stellar disk by
using a combination of elemental abundances and kinematics.

In addition, we show that the selection of
\textit{Gaia}-Sausage-Enceladus in the $E_{\rm n}-L_{\rm z}$-plane is
not very robust. This is in line with recent numerical simluations
which indicate that merger signatures are erased also in this plane
\citep{2022arXiv220604522K,2022arXiv221004245P}.

Our study shows the need to carefully combine both elemental
abundances and kinematics to make progress understanding the mass
accretion and early history of the Milky Way. The latest Gaia data
release, as well as the new and upcoming massive spectroscopic surveys
(WEAVE, GALAH, APOGEE, SDSS-V, DESI, LAMOST, 4MOST), will provide the
necessary data.

\begin{acknowledgments}
SF and DF were  supported by Swedish Research Council grant
2016-03412 and by a project grant from the Knut and Alice Wallenberg
foundation (KAW 2020.0061 \textit{Galactic Time Machine}).

Funding for the Sloan Digital Sky 
Survey IV has been provided by the 
Alfred P. Sloan Foundation, the U.S. 
Department of Energy Office of 
Science, and the Participating 
Institutions. 

This work has made use of data from the European Space Agency (ESA)
mission {\it Gaia} (\url{https://www.cosmos.esa.int/gaia}), processed
by the {\it Gaia} Data Processing and Analysis Consortium (DPAC,
\url{https://www.cosmos.esa.int/web/gaia/dpac/consortium}). Funding
for the DPAC has been provided by national institutions, in particular
the institutions participating in the {\it Gaia} Multilateral
Agreement.

Funding for the Sloan Digital Sky Survey IV has been provided by the
Alfred P. Sloan Foundation, the U.S. Department of Energy Office of
Science, and the Participating Institutions. SDSS-IV acknowledges
support and resources from the Center for High Performance Computing
at the University of Utah. The SDSS website is {\url{www.sdss4.org}}.

SDSS-IV is managed by the 
Astrophysical Research Consortium 
for the Participating Institutions 
of the SDSS Collaboration including 
the Brazilian Participation Group, 
the Carnegie Institution for Science, 
Carnegie Mellon University, Center for 
Astrophysics | Harvard \& 
Smithsonian, the Chilean Participation 
Group, the French Participation Group, 
Instituto de Astrof\'isica de 
Canarias, The Johns Hopkins 
University, Kavli Institute for the 
Physics and Mathematics of the 
Universe (IPMU) / University of 
Tokyo, the Korean Participation Group, 
Lawrence Berkeley National Laboratory, 
Leibniz Institut f\"ur Astrophysik 
Potsdam (AIP),  Max-Planck-Institut 
f\"ur Astronomie (MPIA Heidelberg), 
Max-Planck-Institut f\"ur 
Astrophysik (MPA Garching), 
Max-Planck-Institut f\"ur 
Extraterrestrische Physik (MPE), 
National Astronomical Observatories of 
China, New Mexico State University, 
New York University, University of 
Notre Dame, Observat\'ario 
Nacional / MCTI, The Ohio State 
University, Pennsylvania State 
University, Shanghai 
Astronomical Observatory, United 
Kingdom Participation Group, 
Universidad Nacional Aut\'onoma 
de M\'exico, University of Arizona, 
University of Colorado Boulder, 
University of Oxford, University of 
Portsmouth, University of Utah, 
University of Virginia, University 
of Washington, University of 
Wisconsin, Vanderbilt University, 
and Yale University.
\end{acknowledgments}

\vspace{5mm}
\facilities{Gaia}

\software{astropy \citep{2013A&A...558A..33A,2018AJ....156..123A},  
    galpy \citep[][\url{http://github.com/jobovy/galpy}]{2015ApJS..216...29B}     }

  \appendix
  \restartappendixnumbering

\section{Additional information on the selection of stars from APOGEE\,DR17}
  
  \subsection{APOGEE fields and programs excluded}
 \label{app:fields} 
 
\begin{table*}
\footnotesize
\caption{APOGEE fields removed from consideration when construcitng
  our sample. The strings listed correspond to the APOGEE~DR17
  parameter {\tt FIELD}.
  Stars labelled with these strings were removed from our final
  catalogue.
  The asterisk ($\ast$) is used as a wild card character, indicating multiple possible names. 
  \label{tab:fields}}
\centering
\begin{tabular}{l l l l l l l l l}
\hline 

  \verb|47TUC| & \verb|ANDR*| & \verb|BOOTES1| & \verb|Berkeley*| & \verb|CARINA| & \verb|COL261| \\
  \verb|CygnusX*| & \verb|DRACO| & \verb|FL_2020| & \verb|FORNAX| & \verb|GD1-*| & \verb|IC342_NGA| \\
  \verb|IC348*| & \verb|INTCL_N*| & \verb|JHelum*| & \verb|LAMBDAORI-*| & \verb|LMC*| & \verb|M10| \\
  \verb|M107| & \verb|M12-N| & \verb|M12-S| & \verb|M13| & \verb|M15| & \verb|M2| \\
  \verb|M22| & \verb|M3| & \verb|M3-RV| & \verb|M33| & \verb|M35N2158| & \verb|M35N2158_btx| \\
  \verb|M4| & \verb|M5| & \verb|M53| & \verb|M54SGRC*| & \verb|M55| & \verb|M5PAL5| \\
  \verb|M67*| & \verb|M68| & \verb|M71*| & \verb|M79| & \verb|M92| & \verb|N1333*| \\
  \verb|N1851| & \verb|N188*| & \verb|N2204| & \verb|N2243*| & \verb|N2264| & \verb|N2298| \\
  \verb|N2420| & \verb|N2808| & \verb|N288| & \verb|N3201*| & \verb|N362| & \verb|N4147| \\
  \verb|N5466| & \verb|N5634SGR2| & \verb|N5634SGR2-RV_btx| & \verb|N6229| & \verb|N6388| & \verb|N6397| \\
  \verb|N6441| & \verb|N6752| & \verb|N6791| & \verb|N6819*| & \verb|N752_btx| & \verb|N7789| \\
  \verb|NGC188_btx| & \verb|NGC2420_btx| & \verb|NGC2632_btx| & \verb|NGC6791*| & \verb|NGC7789*| & \verb|ORION*| \\
  \verb|ORPHAN-*| & \verb|Omegacen*| & \verb|PAL*| & \verb|PLEIADES*| & \verb|SCULPTOR| & \verb|SEXTANS| \\
  \verb|SGR*| & \verb|SMC*| & \verb|Sgr*| & \verb|TAUL*| & \verb|TRIAND-*| & \verb|TRUMP20| \\
  \verb|Tombaugh2| & \verb|URMINOR| & \verb|moving_groups| & \verb|ruprecht147| & \verb|sgr_tidal*|  \\
\hline
\end{tabular}
\end{table*}

\begin{table}
\footnotesize
\caption{APOGEE programs removed from consideration. The strings listed correspond to the APOGEE~DR17 parameter {\tt PROGRAMNAME}. Stars labelled with these strings were removed from our final cataligue.  We list only unique programs but some stars belong to multiple programs.
  \label{tab:programs}}
\centering
\begin{tabular}{l l l l l l l l l}
\hline 
  \verb|Drout_18b| & \verb|Fernandez_20a| & \verb|Geisler_18a| & \verb|beaton_18a| & \verb|cluster_gc| & \verb|cluster_gc1| \\
  \verb|cluster_gc2| & \verb|cluster_oc| & \verb|clusters_gc2| & \verb|clusters_gc3| & \verb|geisler_18a| & \verb|geisler_19a| \\
  \verb|geisler_19b| & \verb|geisler_20a| & \verb|halo2_stream| & \verb|halo_dsph| & \verb|halo_stream| & \verb|kollmeier_19b| \\
  \verb|magclouds| & \verb|monachesi_19b| & \verb|sgr| & \verb|sgr_tidal| & \verb|stream_halo| & \verb|stutz_18a| \\
\verb|stutz_18b| & \verb|stutz_19a| & & & & & & \\
\hline
\end{tabular}
\end{table}

When creating our catalogue from APOGEE\,DR17 we de-selected
observations beloning to specific objects or programs. In particular
we removed open and globular clusters as well as the Magellanic Clouds
and other dwarf galaxies in the Local Group. In addition, we removed
those programs that targetted specific stellar streams.  Tables
\ref{tab:fields} and \ref{tab:programs} list the values of the
APOGEE\,DR17 {\tt FIELD} and {\tt PROGRAMNAME} parameters that were
excluded when selecting the sample of Milky Way field stars used in
our study.

\subsection{Dwarf Galaxy membership selection}
\label{app:dwarfs}

Below we detail the selection of the member stars for the five dwarf
galaxies used in Fig. \ref{fig:mgmnalfe_ill}. We require a minimum SNR
of 10 for all dwarf galaxy members.  We used the 876 Sagittarius core
and stream members reported in \citet{2020ApJ...889...63H} and
cross-matched them with APOGEE\,DR17 to get the updated elemental
abundances.  Other members of dwarf galaxies were selected using the
APOGEE {\tt FIELD} parameter. A further selection was done based on a
combination of radial velocity (RV) and proper motion using limits
empirically found to contain the main distribution of member
stars. The RV was taken from APOGEE\,DR17. The proper motions (PMRA,
PMDEC) are from Gaia EDR3 as provided in APOGEE\,DR17. The limits used
for LMC and SMC as well as for the Draco and Ursa Minor dSph galaxies
are given in Table \ref{tab:dwfgal}. In the last column in the table
we indicate the number of stars selected for each galaxy.

\begin{table}
\footnotesize
\caption{Selection criteria used to find stars that belong to various
  Local Group dwarf galaxies as indicated by their name.
  The asterisk ($\ast$) is used as a wild card character, indicating multiple possible names. 
\label{tab:dwfgal}}
\centering
\begin{tabular}{l l l l l l}
 Galaxy & {\tt FIELD} & RV & PMRA & PMDEC & N Stars \\
\hline 
  LMC & \verb|LMC*| & $> 160$ & [1.0, 2.3] & [-1.0, 2.0] & 5847 \\
  SMC & \verb|SMC*|& $> 90$ & [0.2, 1.6] & [-1.7, -0.9] & 1739 \\
  Draco & \verb|DRACO|& $[-310, -260]$ & -- & -- & 19 \\
  Ursa Minor & \verb|URMINOR|& -- & $[-0.6, 0.2]$ & $[-0.2, 0.4]$ & 29 \\
\hline
\end{tabular}
\end{table}

\section{Addition kinematic spaces and uncertainties} 
\label{app:mcmillan}

\begin{figure*}
\centering
\includegraphics[width=15cm,trim={0 3cm 0 4cm},clip]{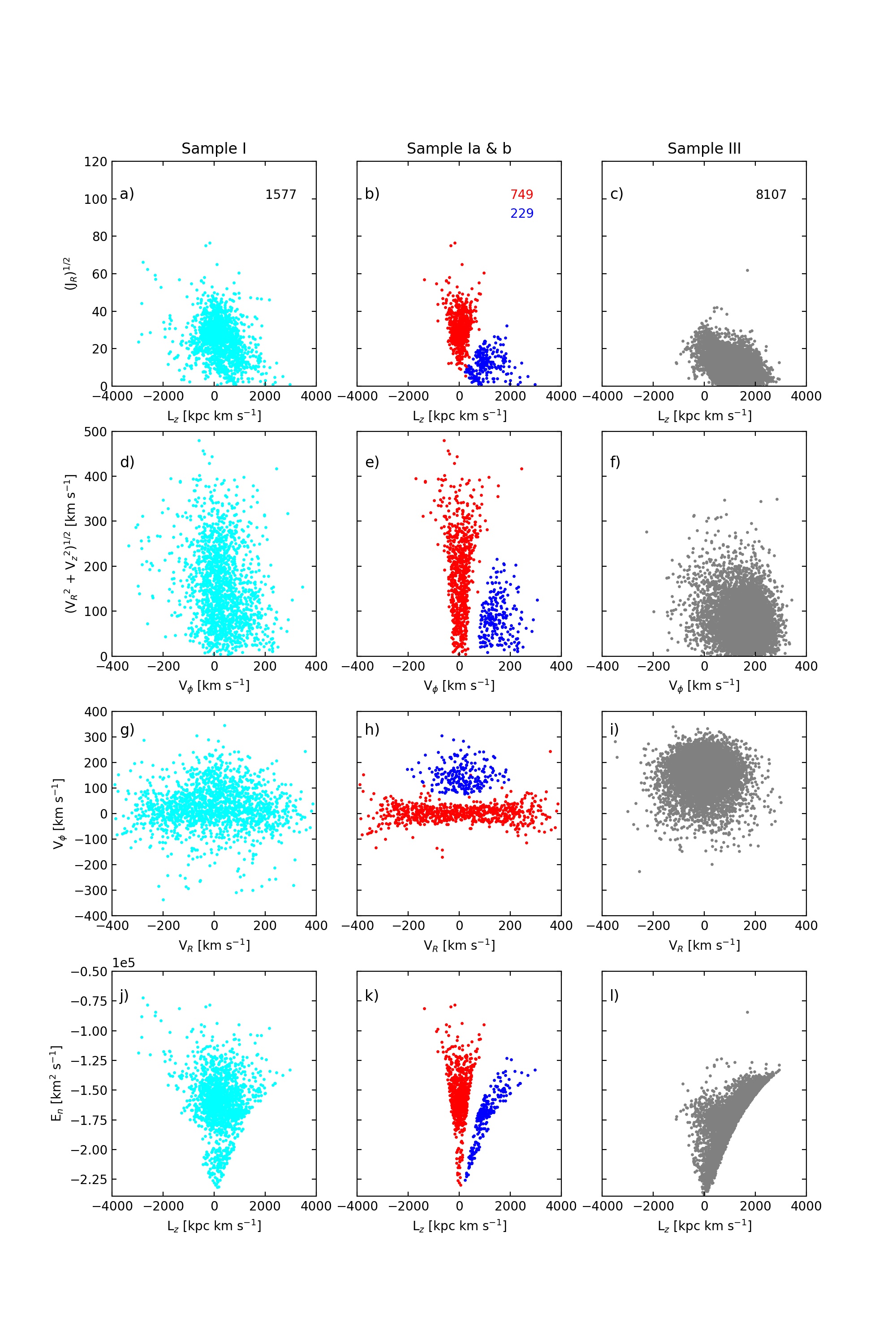}
\caption{Four kinematic spaces showing the properties of the stars
  selected for Sample\,I (first column) and Sample\,III (third column)
  using the Mg-Mn-Al-Fe-plane (Fig\,\ref{fig:cuts}). These plots use
  the \citet{2017MNRAS.465...76M} potential. Only stars with $1 < \log g < 2 $ are included (plots look very similar with all stars included). The middle column shows the two sub-samples of Sample\,I defined using the action diamond: Sample\,Ia (red) are stars with $-0.25 < L_z/J_{tot} < 0.25$ and Sample\,Ib (blue) stars with $L_z/J_{tot} > 0.6$. The number of stars in each sample are indicated in the panels in the top-row. }
  \label{fig:kin_sample1_McM} 
\end{figure*}

In the first part of our analysis we use the {\tt MWPotential2014} in
\citet{2013ApJ...779..115B,2015ApJS..216...29B} to calculate orbital
parameters for the stars. Later we also use the
\citet{2017MNRAS.465...76M} potential in order to be able to apply the
same cuts in for example $E_n$ as used in some other studies. For
completeness we here show the orbital parameter spaces calculated
using the \citet{2017MNRAS.465...76M} potential. Comparing
Fig.\,\ref{fig:kin_sample1_McM} with those calculated with the {\tt
  MWPotential2014} in \citet{2013ApJ...779..115B,2015ApJS..216...29B}
(Fig.\,\ref{fig:kin_sample1}) there are hardly any differences apart
from the expected shift in $E_{\rm n}$. We conclude that our results
are robust against whichever commonly used potential is being
implemented.

\begin{figure*}
\centering
\includegraphics[width=17.5cm,trim={0cm 0cm 0 0cm},clip]{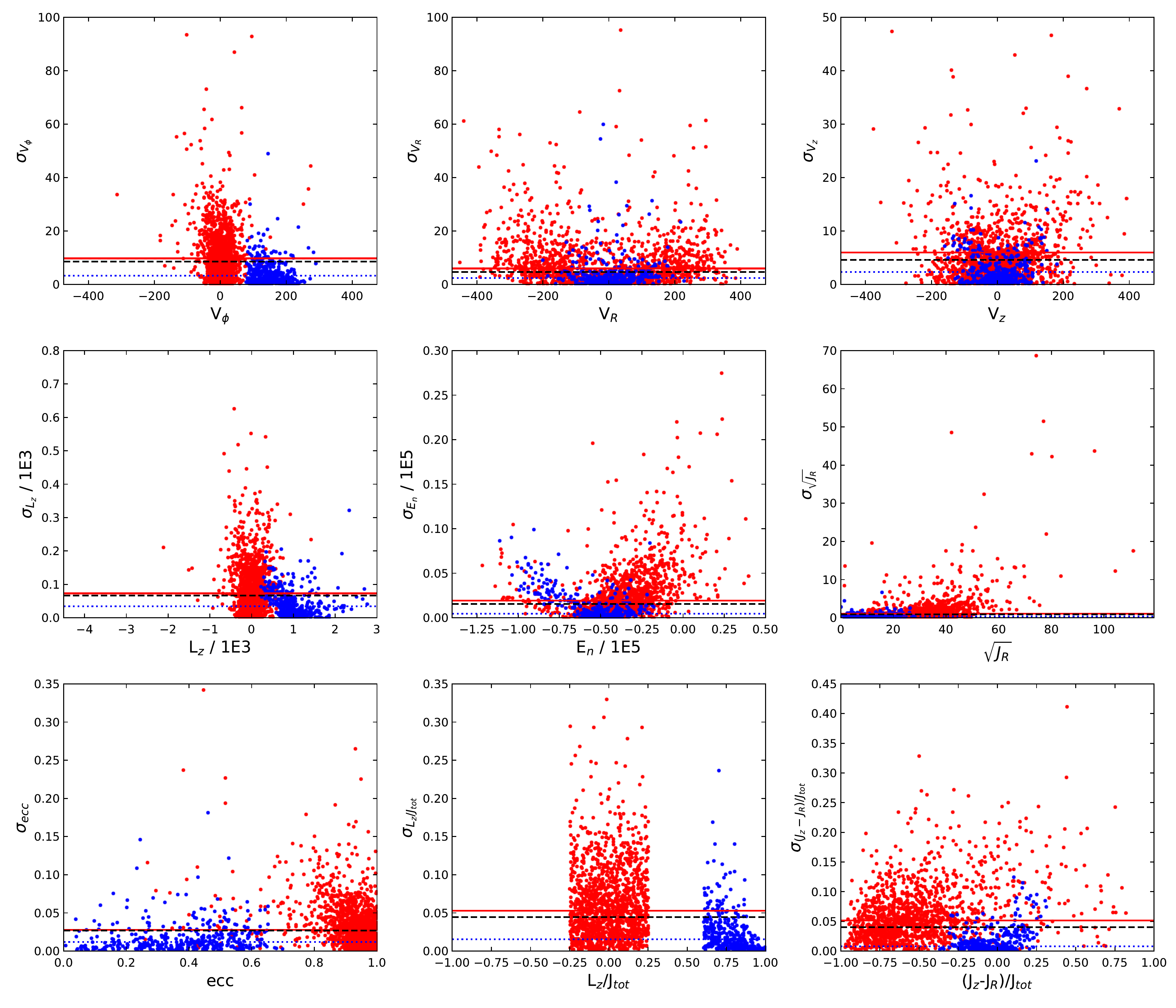}
\caption{Distributions of kinematic uncertainties of stars in Sample
  Ia (red points) and Sample Ib (blue points). Uncertainties in each
  kinematic parameter are shown as a function of the given
  parameter. The calculation of uncertainties is described in Section
  \ref{sect:data}. The horizontal lines indicate the median
  uncertainty in each sample.}
  \label{fig:kin_err} 
\end{figure*}

We also calculated uncertainties in the kinematic parameters as
  described in Section \ref{sect:data}. Here we show the distribution
  of these uncertainties as a function of the given kinematic
  parameter for Sample Ia (red) and Ib (blue),
  Fig. \ref{fig:kin_err}. We note that Sample Ia represents the
  largest kinematic uncertainties as this sample is, on average, more
  distant. The median uncertainty in each parameter for each sample is
  indicted by the horizontal lines. The median uncertainties in all of
  Sample I are also shown as the black line. We note that although
  some individual stars have large uncertainties, the median
  uncertainties are low, typically $< 10$\%.

\section{Additional elemental abundance plots}
\label{app:abun_extra}

\begin{figure*}
\centering
\includegraphics[width=17cm,trim={0 0.5cm 0 1cm},clip]{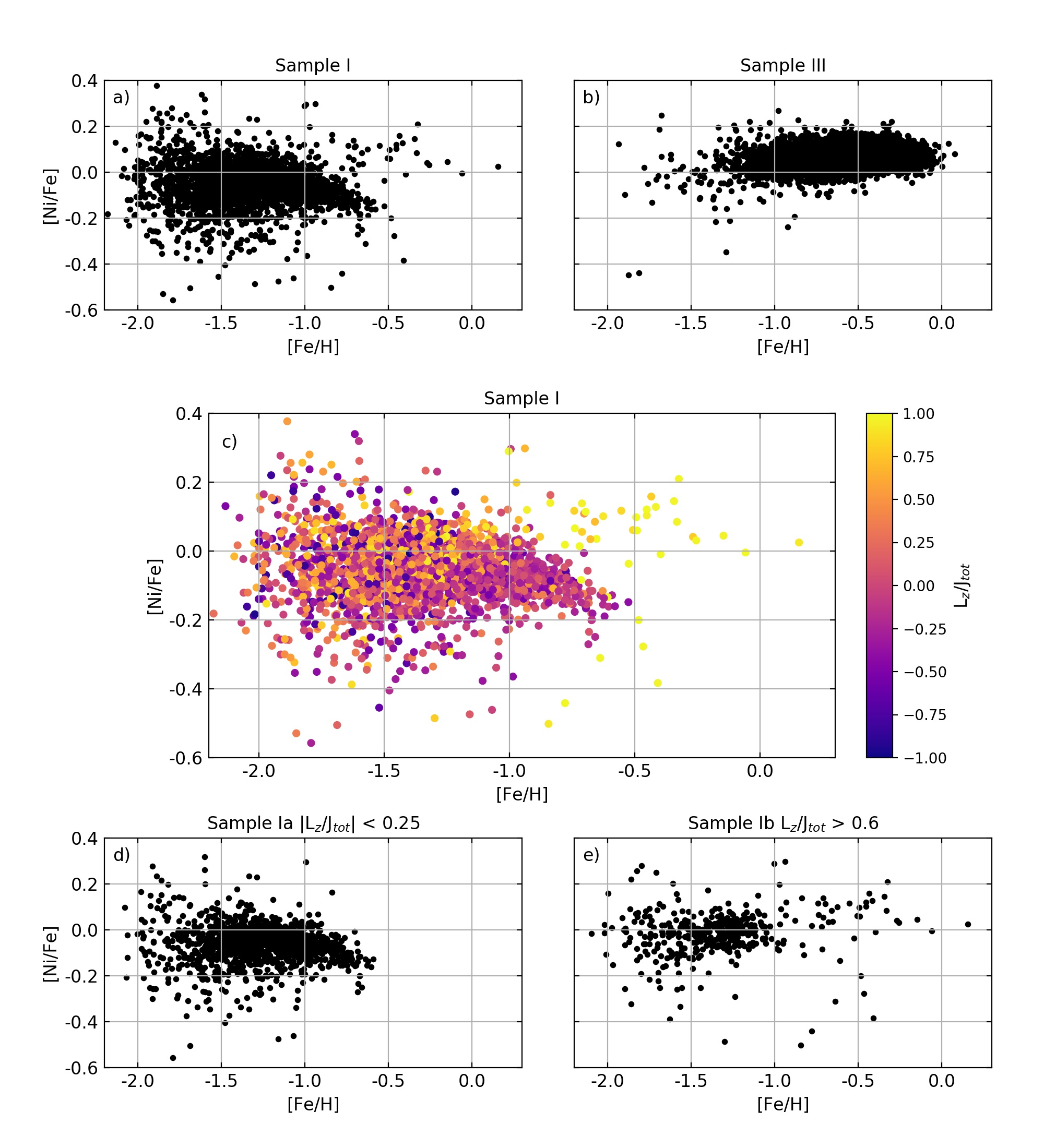}
\caption{[Ni/Fe] as a function of [Fe/H] for Sample\,I, Ia, Ib, and III. Samples are restricted to stars with $1 < \log g < 2$. 
{\bf a)}~[Ni/Fe] as a function of [Fe/H]  Sample\,I. 
{\bf b)}~[Ni/Fe] as a function of [Fe/H]  Sample\,III. 
{\bf c)}~[Ni/Fe] as a function of [Fe/H]  Sample\,I colour-coded by $L_z/J_{\rm tot}$. Values as indicated by the colour bar. 
{\bf d)}~[Ni/Fe] as a function of [Fe/H] for Sample\,Ia. 
{\bf e)}~[Ni/Fe] as a function of [Fe/H] for Sample\,Ib. } 
\label{fig:nife_sample1} 
\end{figure*}

Apart from the $\alpha$-elements, nickel also shows the downward trend
characteristic of \textit{Gaia}-Sausage-Enceladus, as already
demonstrated in \citet{1997A&A...326..751N,2010A&A...511L..10N}. Fig.\,\ref{fig:nife_sample1} shows the same
elemental abundance plots as shown for [Mg/Fe] in Fig.\,\ref{fig:mgfe_sample1} but for
[Ni/Fe]. As can be seen, the resemblance is striking, further
indicating that our analysis and conclusions are robust.

\section{The age of Sample\,III based on asteroseismic data}
\label{app:age_samp3}

Table\,\ref{tab:agesIII} provides the date for stars from
\citet{2022MNRAS.514.2527B} that fall in our Sample\,III and fullfill
the quaility criteria we apply (Table\,\ref{tab:data}). The ages are
from \citet{2022MNRAS.514.2527B} while the elemental abundances are
taken from APOGEE\,DR17. We list the KIC/EPIC ID as given in
\citet{2022MNRAS.514.2527B} and provde a cross-match to the
\textit{Gaia} DR3 IDs.

\begin{table}
  \caption{Stars from \citet{2022MNRAS.514.2527B} that fulfill our
    quality criteria, Table\,\ref{tab:data}, and fall in our
    Sample\,III region in the Mg-Mn-Al-Fe-plane based on elemental
    abundances from APOGEE\,DR17. Ages are from
    \citet{2022MNRAS.514.2527B}, while [Fe/H] and [Mg/Fe] are taken
    from APOGEE\,DR17. The KIC/EPIC labelare taken from
    \citet{2022MNRAS.514.2527B}.}
   \begin{tabular}{lllllllllll}
    \hline
     KIC/EPIC & \textit{Gaia}\,ID & [Fe/H] & [Mg/Fe] & Age &  Age error \\
    \hline
     KIC 2571323  & 2051107025724208256  &   --0.77 &  0.34 & 9.81 & 2.87 /--0.96 \\
     KIC 2165615  & 2051825277390535552  &   --0.73 &  0.36 & 2.45 & 0.67/--0.38 \\
     KIC 2301577  & 2052544465376348416  &   --0.45 &  0.31 & 3.67 & 0.74/--0.33 \\
     KIC 5371173  & 2076546151381304448  &   --0.51 &  0.33 & 9.43 & 0.45/--1.12 \\
     KIC 7908109  & 2078761117553102208  &   --0.75 &  0.35 &  9.80  &1.50/--0.23 \\
    KIC 11774651  & 2087261373224223616  &   --0.43 &  0.32& 13.18 & 0.71/--1.62 \\
     KIC 5698156  & 2101432149662610816  &   --1.27 &  0.31 & 12.80 & 1.52/--2.15 \\
     KIC 5446927  & 2101503690937150464  &   --0.74 &  0.13 & 4.90 & 7.37 -3.25 \\
     KIC 6267115  & 2104059540072830336  &   --0.35 &  0.31 & 8.12 & 1.80 -1.86 \\
     KIC 7502070  & 2104862900816530432  &   --0.59 &  0.33 & 9.25&  2.68 -1.43 \\
     KIC 7946809  & 2105698602672618752  &   --0.54 &  0.33 & 7.04 & 2.88 -2.55 \\
     KIC 8544630   &2106715341689368320  &   --0.58 &  0.32 & 8.32 & 1.43 -0.55 \\
    KIC 10207078  & 2129106380594661248  &   --0.31 &  0.33&  9.40 & 0.66 -1.10\\
    KIC 12109442  & 2130163625443203328  &   --0.54 &  0.38& 12.05 & 1.76 -2.94\\
    KIC 10398120  & 2130915214660138624  &   --0.99 &  0.34&  8.59 & 1.35 -1.52\\
    KIC 12506245  & 2133443541646852864  &   --0.71 &  0.31& 12.75 & 0.78 -2.86\\
  EPIC 220387868  & 2551830805756981248 &   --1.07 &  0.36 & 8.20&  1.70 -1.88\\
  EPIC 220269276  & 2559320399792228096  &   --0.31 &  0.29& 4.33 & 2.87 -1.53\\
  EPIC 205997746  & 2596851370212990720 &   --1.06 &  0.32 &11.29 & 2.10 -1.90\\
  EPIC 205972576 & 2598768815412715776  &   --0.35 &  0.26 &11.62 & 2.30 -2.66\\
  EPIC 251512185 &  3684177626014911360  &   --0.67 &  0.37& 9.18 & 3.92 -4.38\\
 EPIC 204785972  &  4127168730546419072  &   --0.46 &  0.30& 7.34 & 4.39 -4.25\\
  EPIC 204298932 & 6050297413148822656   &   --0.68 &  0.35& 9.24 & 3.91 -3.11\\
  EPIC 205083494 &  6245695266654085888  &   --1.01 &  0.36&  8.17 & 1.80 -1.80\\
  EPIC 212297999 & 6293687295639821824   &   --0.57 &  0.33& 7.69 & 4.28 -3.30\\
  EPIC 213463719 & 6758726460165845248   &   --0.07 &  0.22& 13.19 & 1.40 -3.65\\
  EPIC 213523425 & 6759483817516180352   &   --0.37 &  0.30& 13.18 & 0.71 -2.35\\
  EPIC 213632986 &  6759511374026333568  &   --0.57 &  0.38& 6.98 & 5.01 -3.53\\
  EPIC 213853964 & 6759619023088186496   &   --0.70 &  0.29& 7.81 & 3.18 -1.98\\
  EPIC 213764390  &6759773577490110464   &   --0.36 &  0.25& 5.32 & 1.66 -1.41\\
\hline
  \end{tabular}
  \label{tab:agesIII}
\end{table}

\bibliography{FeltzingFeuillet_bibliography}{}

\begin{thebibliography}{}
\expandafter\ifx\csname natexlab\endcsname\relax\def\natexlab#1{#1}\fi
\providecommand{\url}[1]{\href{#1}{#1}}
\providecommand{\dodoi}[1]{doi:~\href{http://doi.org/#1}{\nolinkurl{#1}}}
\providecommand{\doeprint}[1]{\href{http://ascl.net/#1}{\nolinkurl{http://ascl.net/#1}}}
\providecommand{\doarXiv}[1]{\href{https://arxiv.org/abs/#1}{\nolinkurl{https://arxiv.org/abs/#1}}}

\bibitem[{{Abdurro'uf} {et~al.}(2022){Abdurro'uf}, {Accetta}, {Aerts}, {Silva
  Aguirre}, {Ahumada}, {Ajgaonkar}, {Filiz Ak}, {Alam}, {Allende Prieto},
  {Almeida}, {Anders}, {Anderson}, {Andrews}, {Anguiano}, {Aquino-Ortiz},
  {Aragon-Salamanca}, {Argudo-Fernandez}, {Ata}, {Aubert}, {Avila-Reese},
  {Badenes}, {Barba}, {Barger}, {Barrera-Ballesteros}, {Beaton}, {Beers},
  {Belfiore}, {Bender}, {Bernardi}, {Bershady}, {Beutler}, {Moni Bidin},
  {Bird}, {Bizyaev}, {Blanc}, {Blanton}, {Boardman}, {Bolton}, {Boquien},
  {Borissova}, {Bovy}, {Brandt}, {Brown}, {Brownstein}, {Brusa}, {Buchner},
  {Bundy}, {Burchett}, {Bureau}, {Burgasser}, {Cabang}, {Campbell},
  {Cappellari}, {Carlberg}, {Carneiro Wanderley}, {Carrera}, {Cash}, {Chen},
  {Chen}, {Cherinka}, {Chiappini}, {Choi}, {Chojnowski}, {Chung}, {Clerc},
  {Cohen}, {Comerford}, {Comparat}, {da Costa}, {Covey}, {Crane},
  {Cruz-Gonzalez}, {Culhane}, {Cunha}, {Dai}, {Damke}, {Darling}, {Davidson},
  {Davies}, {Dawson}, {De Lee}, {Diamond-Stanic}, {Cano-Diaz}, {Dominguez
  Sanchez}, {Donor}, {Duckworth}, {Dwelly}, {Eisenstein}, {Elsworth},
  {Emsellem}, {Eracleous}, {Escoffier}, {Fan}, {Farr}, {Feng},
  {Fernandez-Trincado}, {Feuillet}, {Filipp}, {Fillingham}, {Frinchaboy},
  {Fromenteau}, {Galbany}, {Garcia}, {Garcia-Hernandez}, {Ge}, {Geisler},
  {Gelfand}, {Geron}, {Gibson}, {Goddy}, {Godoy-Rivera}, {Grabowski}, {Green},
  {Greener}, {Grier}, {Griffith}, {Guo}, {Guy}, {Hadjara}, {Harding},
  {Hasselquist}, {Hayes}, {Hearty}, {Hernandez}, {Hill}, {Hogg}, {Holtzman},
  {Horta}, {Hsieh}, {Hsu}, {Hsu}, {Huber}, {Huertas-Company}, {Hutchinson},
  {Hwang}, {Ibarra-Medel}, {Ider Chitham}, {Ilha}, {Imig}, {Jaekle},
  {Jayasinghe}, {Ji}, {Johnson}, {Jones}, {Jonsson}, {Katkov}, {Khalatyan},
  {Kinemuchi}, {Kisku}, {Knapen}, {Kneib}, {Kollmeier}, {Kong}, {Kounkel},
  {Kreckel}, {Krishnarao}, {Lacerna}, {Lane}, {Langgin}, {Lavender}, {Law},
  {Lazarz}, {Leung}, {Leung}, {Lewis}, {Li}, {Li}, {Lian}, {Liang}, {Lin},
  {Lin}, {Lin}, {Lintott}, {Long}, {Longa-Pena}, {Lopez-Coba}, {Lu},
  {Lundgren}, {Luo}, {Mackereth}, {de la Macorra}, {Mahadevan}, {Majewski},
  {Manchado}, {Mandeville}, {Maraston}, {Margalef-Bentabol}, {Masseron},
  {Masters}, {Mathur}, {McDermid}, {Mckay}, {Merloni}, {Merrifield},
  {Meszaros}, {Miglio}, {Di Mille}, {Minniti}, {Minsley}, {Monachesi}, {Moon},
  {Mosser}, {Mulchaey}, {Muna}, {Munoz}, {Myers}, {Myers}, {Nadathur}, {Nair},
  {Nandra}, {Neumann}, {Newman}, {Nidever}, {Nikakhtar}, {Nitschelm},
  {O'Connell}, {Garma-Oehmichen}, {Luan Souza de Oliveira}, {Olney}, {Oravetz},
  {Ortigoza-Urdaneta}, {Osorio}, {Otter}, {Pace}, {Padilla}, {Pan}, {Pan},
  {Parikh}, {Parker}, {Peirani}, {Pena Ramirez}, {Penny}, {Percival},
  {Perez-Fournon}, {Pinsonneault}, {Poidevin}, {Poovelil}, {Price-Whelan},
  {Queiroz}, {Raddick}, {Ray}, {Barboza Rembold}, {Riddle}, {Riffel}, {Riffel},
  {Rix}, {Robin}, {Rodriguez-Puebla}, {Roman-Lopes}, {Roman-Zuniga}, {Rose},
  {Ross}, {Rossi}, {Rubin}, {Salvato}, {Sanchez}, {Sanchez-Gallego},
  {Sanderson}, {Santana Rojas}, {Sarceno}, {Sarmiento}, {Sayres}, {Sazonova},
  {Schaefer}, {Schiavon}, {Schlegel}, {Schneider}, {Schultheis}, {Schwope},
  {Serenelli}, {Serna}, {Shao}, {Shapiro}, {Sharma}, {Shen}, {Shetrone}, {Shu},
  {Simon}, {Skrutskie}, {Smethurst}, {Smith}, {Sobeck}, {Spoo}, {Sprague},
  {Stark}, {Stassun}, {Steinmetz}, {Stello}, {Stone-Martinez},
  {Storchi-Bergmann}, {Stringfellow}, {Stutz}, {Su}, {Taghizadeh-Popp},
  {Talbot}, {Tayar}, {Telles}, {Teske}, {Thakar}, {Theissen}, {Thomas},
  {Tkachenko}, {Tojeiro}, {Hernandez Toledo}, {Troup}, {Trump}, {Trussler},
  {Turner}, {Tuttle}, {Unda-Sanzana}, {Vazquez-Mata}, {Valentini},
  {Valenzuela}, {Vargas-Gonzalez}, {Vargas-Magana}, {Alfaro}, {Villanova},
  {Vincenzo}, {Wake}, {Warfield}, {Washington}, {Weaver}, {Weijmans},
  {Weinberg}, {Weiss}, {Westfall}, {Wild}, {Wilde}, {Wilson}, {Wilson},
  {Wilson}, {Wolf}, {Wood-Vasey}, {Yan}, {Zamora}, {Zasowski}, {Zhang}, {Zhao},
  {Zheng}, {Zheng}, \& {Zhu}}]{2021arXiv211202026A}
{Abdurro'uf}, {Accetta}, K., {Aerts}, C., {et~al.} 2022, \apjs, 259, 35,
  \dodoi{10.3847/1538-4365/ac4414}

\bibitem[{{Ahumada} {et~al.}(2020){Ahumada}, {Prieto}, {Almeida}, {Anders},
  {Anderson}, {Andrews}, {Anguiano}, {Arcodia}, {Armengaud}, {Aubert}, {Avila},
  {Avila-Reese}, {Badenes}, {Balland}, {Barger}, {Barrera-Ballesteros}, {Basu},
  {Bautista}, {Beaton}, {Beers}, {Benavides}, {Bender}, {Bernardi}, {Bershady},
  {Beutler}, {Bidin}, {Bird}, {Bizyaev}, {Blanc}, {Blanton}, {Boquien},
  {Borissova}, {Bovy}, {Brandt}, {Brinkmann}, {Brownstein}, {Bundy}, {Bureau},
  {Burgasser}, {Burtin}, {Cano-D{\'\i}az}, {Capasso}, {Cappellari}, {Carrera},
  {Chabanier}, {Chaplin}, {Chapman}, {Cherinka}, {Chiappini}, {Doohyun Choi},
  {Chojnowski}, {Chung}, {Clerc}, {Coffey}, {Comerford}, {Comparat}, {da
  Costa}, {Cousinou}, {Covey}, {Crane}, {Cunha}, {Ilha}, {Dai}, {Damsted},
  {Darling}, {Davidson}, {Davies}, {Dawson}, {De}, {de la Macorra}, {De Lee},
  {Queiroz}, {Deconto Machado}, {de la Torre}, {Dell'Agli}, {du Mas des
  Bourboux}, {Diamond-Stanic}, {Dillon}, {Donor}, {Drory}, {Duckworth},
  {Dwelly}, {Ebelke}, {Eftekharzadeh}, {Davis Eigenbrot}, {Elsworth},
  {Eracleous}, {Erfanianfar}, {Escoffier}, {Fan}, {Farr},
  {Fern{\'a}ndez-Trincado}, {Feuillet}, {Finoguenov}, {Fofie},
  {Fraser-McKelvie}, {Frinchaboy}, {Fromenteau}, {Fu}, {Galbany}, {Garcia},
  {Garc{\'\i}a-Hern{\'a}ndez}, {Oehmichen}, {Ge}, {Maia}, {Geisler}, {Gelfand},
  {Goddy}, {Gonzalez-Perez}, {Grabowski}, {Green}, {Grier}, {Guo}, {Guy},
  {Harding}, {Hasselquist}, {Hawken}, {Hayes}, {Hearty}, {Hekker}, {Hogg},
  {Holtzman}, {Horta}, {Hou}, {Hsieh}, {Huber}, {Hunt}, {Chitham}, {Imig},
  {Jaber}, {Angel}, {Johnson}, {Jones}, {J{\"o}nsson}, {Jullo}, {Kim},
  {Kinemuchi}, {Kirkpatrick}, {Kite}, {Klaene}, {Kneib}, {Kollmeier}, {Kong},
  {Kounkel}, {Krishnarao}, {Lacerna}, {Lan}, {Lane}, {Law}, {Le Goff}, {Leung},
  {Lewis}, {Li}, {Lian}, {Lin}, {Long}, {Longa-Pe{\~n}a}, {Lundgren}, {Lyke},
  {Ted Mackereth}, {MacLeod}, {Majewski}, {Manchado}, {Maraston}, {Martini},
  {Masseron}, {Masters}, {Mathur}, {McDermid}, {Merloni}, {Merrifield},
  {M{\'e}sz{\'a}ros}, {Miglio}, {Minniti}, {Minsley}, {Miyaji}, {Mohammad},
  {Mosser}, {Mueller}, {Muna}, {Mu{\~n}oz-Guti{\'e}rrez}, {Myers}, {Nadathur},
  {Nair}, {Nandra}, {do Nascimento}, {Nevin}, {Newman}, {Nidever}, {Nitschelm},
  {Noterdaeme}, {O'Connell}, {Olmstead}, {Oravetz}, {Oravetz}, {Osorio},
  {Pace}, {Padilla}, {Palanque-Delabrouille}, {Palicio}, {Pan}, {Pan},
  {Parker}, {Paviot}, {Peirani}, {Ram{\'r}ez}, {Penny}, {Percival},
  {Perez-Fournon}, {P{\'e}rez-R{\`a}fols}, {Petitjean}, {Pieri},
  {Pinsonneault}, {Poovelil}, {Povick}, {Prakash}, {Price-Whelan}, {Raddick},
  {Raichoor}, {Ray}, {Rembold}, {Rezaie}, {Riffel}, {Riffel}, {Rix}, {Robin},
  {Roman-Lopes}, {Rom{\'a}n-Z{\'u}{\~n}iga}, {Rose}, {Ross}, {Rossi},
  {Rowlands}, {Rubin}, {Salvato}, {S{\'a}nchez}, {S{\'a}nchez-Menguiano},
  {S{\'a}nchez-Gallego}, {Sayres}, {Schaefer}, {Schiavon}, {Schimoia},
  {Schlafly}, {Schlegel}, {Schneider}, {Schultheis}, {Schwope}, {Seo},
  {Serenelli}, {Shafieloo}, {Shamsi}, {Shao}, {Shen}, {Shetrone}, {Shirley},
  {Aguirre}, {Simon}, {Skrutskie}, {Slosar}, {Smethurst}, {Sobeck}, {Sodi},
  {Souto}, {Stark}, {Stassun}, {Steinmetz}, {Stello}, {Stermer},
  {Storchi-Bergmann}, {Streblyanska}, {Stringfellow}, {Stutz}, {Su{\'a}rez},
  {Sun}, {Taghizadeh-Popp}, {Talbot}, {Tayar}, {Thakar}, {Theriault}, {Thomas},
  {Thomas}, {Tinker}, {Tojeiro}, {Toledo}, {Tremonti}, {Troup}, {Tuttle},
  {Unda-Sanzana}, {Valentini}, {Vargas-Gonz{\'a}lez}, {Vargas-Maga{\~n}a},
  {V{\'a}zquez-Mata}, {Vivek}, {Wake}, {Wang}, {Weaver}, {Weijmans}, {Wild},
  {Wilson}, {Wilson}, {Wolthuis}, {Wood-Vasey}, {Yan}, {Yang}, {Y{\`e}che},
  {Zamora}, {Zarrouk}, {Zasowski}, {Zhang}, {Zhao}, {Zhao}, {Zheng}, {Zheng},
  {Zhu}, \& {Zou}}]{2020ApJS..249....3A}
{Ahumada}, R., {Prieto}, C.~A., {Almeida}, A., {et~al.} 2020, \apjs, 249, 3,
  \dodoi{10.3847/1538-4365/ab929e}

\bibitem[{{Alexeeva} {et~al.}(2018){Alexeeva}, {Ryabchikova}, {Mashonkina}, \&
  {Hu}}]{2018ApJ...866..153A}
{Alexeeva}, S., {Ryabchikova}, T., {Mashonkina}, L., \& {Hu}, S. 2018, \apj,
  866, 153, \dodoi{10.3847/1538-4357/aae1a8}

\bibitem[{{Amarante} {et~al.}(2020){Amarante}, {Beraldo e Silva}, {Debattista},
  \& {Smith}}]{2020ApJ...891L..30A}
{Amarante}, J. A.~S., {Beraldo e Silva}, L., {Debattista}, V.~P., \& {Smith},
  M.~C. 2020, \apjl, 891, L30, \dodoi{10.3847/2041-8213/ab78a4}

\bibitem[{{Amarsi} {et~al.}(2022){Amarsi}, {Liljegren}, \&
  {Nissen}}]{2022A&A...668A..68A}
{Amarsi}, A.~M., {Liljegren}, S., \& {Nissen}, P.~E. 2022, \aap, 668, A68,
  \dodoi{10.1051/0004-6361/202244542}

\bibitem[{{Amarsi} {et~al.}(2016){Amarsi}, {Lind}, {Asplund}, {Barklem}, \&
  {Collet}}]{2016MNRAS.463.1518A}
{Amarsi}, A.~M., {Lind}, K., {Asplund}, M., {Barklem}, P.~S., \& {Collet}, R.
  2016, \mnras, 463, 1518, \dodoi{10.1093/mnras/stw2077}

\bibitem[{{Andrews} {et~al.}(2017){Andrews}, {Weinberg}, {Sch{\"o}nrich}, \&
  {Johnson}}]{2017ApJ...835..224A}
{Andrews}, B.~H., {Weinberg}, D.~H., {Sch{\"o}nrich}, R., \& {Johnson}, J.~A.
  2017, \apj, 835, 224, \dodoi{10.3847/1538-4357/835/2/224}

\bibitem[{{Arnett}(1996)}]{1996snih.book.....A}
{Arnett}, D. 1996, {Supernovae and Nucleosynthesis: An Investigation of the
  History of Matter from the Big Bang to the Present}

\bibitem[{{Astropy Collaboration} {et~al.}(2013){Astropy Collaboration},
  {Robitaille}, {Tollerud}, {Greenfield}, {Droettboom}, {Bray}, {Aldcroft},
  {Davis}, {Ginsburg}, {Price-Whelan}, {Kerzendorf}, {Conley}, {Crighton},
  {Barbary}, {Muna}, {Ferguson}, {Grollier}, {Parikh}, {Nair}, {Unther},
  {Deil}, {Woillez}, {Conseil}, {Kramer}, {Turner}, {Singer}, {Fox}, {Weaver},
  {Zabalza}, {Edwards}, {Azalee Bostroem}, {Burke}, {Casey}, {Crawford},
  {Dencheva}, {Ely}, {Jenness}, {Labrie}, {Lim}, {Pierfederici}, {Pontzen},
  {Ptak}, {Refsdal}, {Servillat}, \& {Streicher}}]{2013A&A...558A..33A}
{Astropy Collaboration}, {Robitaille}, T.~P., {Tollerud}, E.~J., {et~al.} 2013,
  \aap, 558, A33, \dodoi{10.1051/0004-6361/201322068}

\bibitem[{{Astropy Collaboration} {et~al.}(2018){Astropy Collaboration},
  {Price-Whelan}, {Sip{\H{o}}cz}, {G{\"u}nther}, {Lim}, {Crawford}, {Conseil},
  {Shupe}, {Craig}, {Dencheva}, {Ginsburg}, {VanderPlas}, {Bradley},
  {P{\'e}rez-Su{\'a}rez}, {de Val-Borro}, {Aldcroft}, {Cruz}, {Robitaille},
  {Tollerud}, {Ardelean}, {Babej}, {Bach}, {Bachetti}, {Bakanov}, {Bamford},
  {Barentsen}, {Barmby}, {Baumbach}, {Berry}, {Biscani}, {Boquien}, {Bostroem},
  {Bouma}, {Brammer}, {Bray}, {Breytenbach}, {Buddelmeijer}, {Burke},
  {Calderone}, {Cano Rodr{\'\i}guez}, {Cara}, {Cardoso}, {Cheedella}, {Copin},
  {Corrales}, {Crichton}, {D'Avella}, {Deil}, {Depagne}, {Dietrich}, {Donath},
  {Droettboom}, {Earl}, {Erben}, {Fabbro}, {Ferreira}, {Finethy}, {Fox},
  {Garrison}, {Gibbons}, {Goldstein}, {Gommers}, {Greco}, {Greenfield},
  {Groener}, {Grollier}, {Hagen}, {Hirst}, {Homeier}, {Horton}, {Hosseinzadeh},
  {Hu}, {Hunkeler}, {Ivezi{\'c}}, {Jain}, {Jenness}, {Kanarek}, {Kendrew},
  {Kern}, {Kerzendorf}, {Khvalko}, {King}, {Kirkby}, {Kulkarni}, {Kumar},
  {Lee}, {Lenz}, {Littlefair}, {Ma}, {Macleod}, {Mastropietro}, {McCully},
  {Montagnac}, {Morris}, {Mueller}, {Mumford}, {Muna}, {Murphy}, {Nelson},
  {Nguyen}, {Ninan}, {N{\"o}the}, {Ogaz}, {Oh}, {Parejko}, {Parley}, {Pascual},
  {Patil}, {Patil}, {Plunkett}, {Prochaska}, {Rastogi}, {Reddy Janga},
  {Sabater}, {Sakurikar}, {Seifert}, {Sherbert}, {Sherwood-Taylor}, {Shih},
  {Sick}, {Silbiger}, {Singanamalla}, {Singer}, {Sladen}, {Sooley},
  {Sornarajah}, {Streicher}, {Teuben}, {Thomas}, {Tremblay}, {Turner},
  {Terr{\'o}n}, {van Kerkwijk}, {de la Vega}, {Watkins}, {Weaver}, {Whitmore},
  {Woillez}, {Zabalza}, \& {Astropy Contributors}}]{2018AJ....156..123A}
{Astropy Collaboration}, {Price-Whelan}, A.~M., {Sip{\H{o}}cz}, B.~M., {et~al.}
  2018, \aj, 156, 123, \dodoi{10.3847/1538-3881/aabc4f}

\bibitem[{{Bailer-Jones} {et~al.}(2021){Bailer-Jones}, {Rybizki}, {Fouesneau},
  {Demleitner}, \& {Andrae}}]{2021AJ....161..147B}
{Bailer-Jones}, C.~A.~L., {Rybizki}, J., {Fouesneau}, M., {Demleitner}, M., \&
  {Andrae}, R. 2021, \aj, 161, 147, \dodoi{10.3847/1538-3881/abd806}

\bibitem[{{Battistini} \& {Bensby}(2016)}]{2016A&A...586A..49B}
{Battistini}, C., \& {Bensby}, T. 2016, \aap, 586, A49,
  \dodoi{10.1051/0004-6361/201527385}

\bibitem[{{Belokurov}(2013)}]{2013NewAR..57..100B}
{Belokurov}, V. 2013, \nar, 57, 100, \dodoi{10.1016/j.newar.2013.07.001}

\bibitem[{{Belokurov} {et~al.}(2018){Belokurov}, {Erkal}, {Evans}, {Koposov},
  \& {Deason}}]{2018MNRAS.478..611B}
{Belokurov}, V., {Erkal}, D., {Evans}, N.~W., {Koposov}, S.~E., \& {Deason},
  A.~J. 2018, \mnras, 478, 611, \dodoi{10.1093/mnras/sty982}

\bibitem[{{Belokurov} {et~al.}(2020){Belokurov}, {Sanders}, {Fattahi}, {Smith},
  {Deason}, {Evans}, \& {Grand}}]{2020MNRAS.494.3880B}
{Belokurov}, V., {Sanders}, J.~L., {Fattahi}, A., {et~al.} 2020, \mnras, 494,
  3880, \dodoi{10.1093/mnras/staa876}

\bibitem[{{Belokurov} {et~al.}(2006){Belokurov}, {Zucker}, {Evans}, {Gilmore},
  {Vidrih}, {Bramich}, {Newberg}, {Wyse}, {Irwin}, {Fellhauer}, {Hewett},
  {Walton}, {Wilkinson}, {Cole}, {Yanny}, {Rockosi}, {Beers}, {Bell},
  {Brinkmann}, {Ivezi{\'c}}, \& {Lupton}}]{2006ApJ...642L.137B}
{Belokurov}, V., {Zucker}, D.~B., {Evans}, N.~W., {et~al.} 2006, \apjl, 642,
  L137, \dodoi{10.1086/504797}

\bibitem[{{Bensby} {et~al.}(2014){Bensby}, {Feltzing}, \&
  {Oey}}]{2014A&A...562A..71B}
{Bensby}, T., {Feltzing}, S., \& {Oey}, M.~S. 2014, \aap, 562, A71,
  \dodoi{10.1051/0004-6361/201322631}

\bibitem[{{Bergemann} {et~al.}(2017){Bergemann}, {Collet}, {Amarsi}, {Kovalev},
  {Ruchti}, \& {Magic}}]{2017ApJ...847...15B}
{Bergemann}, M., {Collet}, R., {Amarsi}, A.~M., {et~al.} 2017, \apj, 847, 15,
  \dodoi{10.3847/1538-4357/aa88cb}

\bibitem[{{Bergemann} {et~al.}(2012){Bergemann}, {Lind}, {Collet}, {Magic}, \&
  {Asplund}}]{2012MNRAS.427...27B}
{Bergemann}, M., {Lind}, K., {Collet}, R., {Magic}, Z., \& {Asplund}, M. 2012,
  \mnras, 427, 27, \dodoi{10.1111/j.1365-2966.2012.21687.x}

\bibitem[{{Bergemann} {et~al.}(2019){Bergemann}, {Gallagher}, {Eitner},
  {Bautista}, {Collet}, {Yakovleva}, {Mayriedl}, {Plez}, {Carlsson},
  {Leenaarts}, {Belyaev}, \& {Hansen}}]{2019A&A...631A..80B}
{Bergemann}, M., {Gallagher}, A.~J., {Eitner}, P., {et~al.} 2019, \aap, 631,
  A80, \dodoi{10.1051/0004-6361/201935811}

\bibitem[{{Binney}(2012)}]{2012MNRAS.426.1324B}
{Binney}, J. 2012, \mnras, 426, 1324, \dodoi{10.1111/j.1365-2966.2012.21757.x}

\bibitem[{{Bland-Hawthorn} \& {Gerhard}(2016)}]{2016ARA&A..54..529B}
{Bland-Hawthorn}, J., \& {Gerhard}, O. 2016, \araa, 54, 529,
  \dodoi{10.1146/annurev-astro-081915-023441}

\bibitem[{{Bonifacio} {et~al.}(2009){Bonifacio}, {Spite}, {Cayrel}, {Hill},
  {Spite}, {Fran{\c{c}}ois}, {Plez}, {Ludwig}, {Caffau}, {Molaro}, {Depagne},
  {Andersen}, {Barbuy}, {Beers}, {Nordstr{\"o}m}, \&
  {Primas}}]{2009A&A...501..519B}
{Bonifacio}, P., {Spite}, M., {Cayrel}, R., {et~al.} 2009, \aap, 501, 519,
  \dodoi{10.1051/0004-6361/200810610}

\bibitem[{{Borre} {et~al.}(2022){Borre}, {Aguirre B{\o}rsen-Koch}, {Helmi},
  {Koppelman}, {Nielsen}, {R{\o}rsted}, {Stello}, {Stokholm}, {Winther},
  {Davies}, {Hon}, {Kruijssen}, {Laporte}, {Reyes}, \&
  {Yu}}]{2022MNRAS.514.2527B}
{Borre}, C.~C., {Aguirre B{\o}rsen-Koch}, V., {Helmi}, A., {et~al.} 2022,
  \mnras, 514, 2527, \dodoi{10.1093/mnras/stac1498}

\bibitem[{{Bovy}(2015)}]{2015ApJS..216...29B}
{Bovy}, J. 2015, \apjs, 216, 29, \dodoi{10.1088/0067-0049/216/2/29}

\bibitem[{{Bovy} \& {Rix}(2013)}]{2013ApJ...779..115B}
{Bovy}, J., \& {Rix}, H.-W. 2013, \apj, 779, 115,
  \dodoi{10.1088/0004-637X/779/2/115}

\bibitem[{{Buder} {et~al.}(2018){Buder}, {Asplund}, {Duong}, {Kos}, {Lind},
  {Ness}, {Sharma}, {Bland-Hawthorn}, {Casey}, {de Silva}, {D'Orazi},
  {Freeman}, {Lewis}, {Lin}, {Martell}, {Schlesinger}, {Simpson}, {Zucker},
  {Zwitter}, {Amarsi}, {Anguiano}, {Carollo}, {Casagrande}, {{\v{C}}otar},
  {Cottrell}, {da Costa}, {Gao}, {Hayden}, {Horner}, {Ireland}, {Kafle},
  {Munari}, {Nataf}, {Nordlander}, {Stello}, {Ting}, {Traven}, {Watson},
  {Wittenmyer}, {Wyse}, {Yong}, {Zinn}, {{\v{Z}}erjal}, \& {Galah
  Collaboration}}]{2018MNRAS.478.4513B}
{Buder}, S., {Asplund}, M., {Duong}, L., {et~al.} 2018, \mnras, 478, 4513,
  \dodoi{10.1093/mnras/sty1281}

\bibitem[{{Buder} {et~al.}(2021){Buder}, {Sharma}, {Kos}, {Amarsi},
  {Nordlander}, {Lind}, {Martell}, {Asplund}, {Bland-Hawthorn}, {Casey}, {de
  Silva}, {D'Orazi}, {Freeman}, {Hayden}, {Lewis}, {Lin}, {Schlesinger},
  {Simpson}, {Stello}, {Zucker}, {Zwitter}, {Beeson}, {Buck}, {Casagrande},
  {Clark}, {{\v{C}}otar}, {da Costa}, {de Grijs}, {Feuillet}, {Horner},
  {Kafle}, {Khanna}, {Kobayashi}, {Liu}, {Montet}, {Nandakumar}, {Nataf},
  {Ness}, {Spina}, {Tepper-Garc{\'\i}a}, {Ting}, {Traven},
  {Vogrin{\v{c}}i{\v{c}}}, {Wittenmyer}, {Wyse}, {{\v{Z}}erjal}, \& {GALAH
  Collaboration}}]{2021MNRAS.506..150B}
{Buder}, S., {Sharma}, S., {Kos}, J., {et~al.} 2021, \mnras, 506, 150,
  \dodoi{10.1093/mnras/stab1242}

\bibitem[{{Chiappini} {et~al.}(2015){Chiappini}, {Anders}, {Rodrigues},
  {Miglio}, {Montalb{\'a}n}, {Mosser}, {Girardi}, {Valentini}, {Noels},
  {Morel}, {Minchev}, {Steinmetz}, {Santiago}, {Schultheis}, {Martig}, {da
  Costa}, {Maia}, {Allende Prieto}, {de Assis Peralta}, {Hekker},
  {Theme{\ss}l}, {Kallinger}, {Garc{\'\i}a}, {Mathur}, {Baudin}, {Beers},
  {Cunha}, {Harding}, {Holtzman}, {Majewski}, {M{\'e}sz{\'a}ros}, {Nidever},
  {Pan}, {Schiavon}, {Shetrone}, {Schneider}, \&
  {Stassun}}]{2015A&A...576L..12C}
{Chiappini}, C., {Anders}, F., {Rodrigues}, T.~S., {et~al.} 2015, \aap, 576,
  L12, \dodoi{10.1051/0004-6361/201525865}

\bibitem[{{Das} {et~al.}(2020){Das}, {Hawkins}, \&
  {Jofr{\'e}}}]{2020MNRAS.493.5195D}
{Das}, P., {Hawkins}, K., \& {Jofr{\'e}}, P. 2020, \mnras, 493, 5195,
  \dodoi{10.1093/mnras/stz3537}

\bibitem[{{Di Matteo} {et~al.}(2019){Di Matteo}, {Haywood}, {Lehnert}, {Katz},
  {Khoperskov}, {Snaith}, {G{\'o}mez}, \& {Robichon}}]{2019A&A...632A...4D}
{Di Matteo}, P., {Haywood}, M., {Lehnert}, M.~D., {et~al.} 2019, \aap, 632, A4,
  \dodoi{10.1051/0004-6361/201834929}

\bibitem[{{Fernandes} {et~al.}(2023){Fernandes}, {Mason}, {Horta}, {Schiavon},
  {Hayes}, {Hasselquist}, {Feuillet}, {Beaton}, {J{\"o}nsson}, {Kisku},
  {Lacerna}, {Lian}, {Minniti}, \& {Villanova}}]{2023MNRAS.519.3611F}
{Fernandes}, L., {Mason}, A.~C., {Horta}, D., {et~al.} 2023, \mnras, 519, 3611,
  \dodoi{10.1093/mnras/stac3543}

\bibitem[{{Feuillet} {et~al.}(2020){Feuillet}, {Feltzing}, {Sahlholdt}, \&
  {Casagrande}}]{2020MNRAS.497..109F}
{Feuillet}, D.~K., {Feltzing}, S., {Sahlholdt}, C.~L., \& {Casagrande}, L.
  2020, \mnras, 497, 109,
  \dodoi{10.1093/mnras/staa188810.48550/arXiv.2003.11039}

\bibitem[{{Feuillet} {et~al.}(2021){Feuillet}, {Sahlholdt}, {Feltzing}, \&
  {Casagrande}}]{2021MNRAS.508.1489F}
{Feuillet}, D.~K., {Sahlholdt}, C.~L., {Feltzing}, S., \& {Casagrande}, L.
  2021, \mnras, 508, 1489, \dodoi{10.1093/mnras/stab2614}

\bibitem[{{Forbes}(2020)}]{2020MNRAS.493..847F}
{Forbes}, D.~A. 2020, \mnras, 493, 847, \dodoi{10.1093/mnras/staa245}

\bibitem[{{Freeman} \& {Bland-Hawthorn}(2002)}]{2002ARA&A..40..487F}
{Freeman}, K., \& {Bland-Hawthorn}, J. 2002, \araa, 40, 487,
  \dodoi{10.1146/annurev.astro.40.060401.093840}

\bibitem[{{Gaia Collaboration} {et~al.}(2016{\natexlab{a}}){Gaia
  Collaboration}, {Brown}, {Vallenari}, {Prusti}, {de Bruijne}, {Mignard},
  {Drimmel}, {Babusiaux}, {Bailer-Jones}, {Bastian}, {Biermann}, {Evans},
  {Eyer}, {Jansen}, {Jordi}, {Katz}, {Klioner}, {Lammers}, {Lindegren}, {Luri},
  {O'Mullane}, {Panem}, {Pourbaix}, {Randich}, {Sartoretti}, {Siddiqui},
  {Soubiran}, {Valette}, {van Leeuwen}, {Walton}, {Aerts}, {Arenou}, {Cropper},
  {H{\o}g}, {Lattanzi}, {Grebel}, {Holland}, {Huc}, {Passot}, {Perryman},
  {Bramante}, {Cacciari}, {Casta{\~n}eda}, {Chaoul}, {Cheek}, {De Angeli},
  {Fabricius}, {Guerra}, {Hern{\'a}ndez}, {Jean-Antoine-Piccolo}, {Masana},
  {Messineo}, {Mowlavi}, {Nienartowicz}, {Ord{\'o}{\~n}ez-Blanco}, {Panuzzo},
  {Portell}, {Richards}, {Riello}, {Seabroke}, {Tanga}, {Th{\'e}venin},
  {Torra}, {Els}, {Gracia-Abril}, {Comoretto}, {Garcia-Reinaldos}, {Lock},
  {Mercier}, {Altmann}, {Andrae}, {Astraatmadja}, {Bellas-Velidis}, {Benson},
  {Berthier}, {Blomme}, {Busso}, {Carry}, {Cellino}, {Clementini}, {Cowell},
  {Creevey}, {Cuypers}, {Davidson}, {De Ridder}, {de Torres}, {Delchambre},
  {Dell'Oro}, {Ducourant}, {Fr{\'e}mat}, {Garc{\'\i}a-Torres}, {Gosset},
  {Halbwachs}, {Hambly}, {Harrison}, {Hauser}, {Hestroffer}, {Hodgkin},
  {Huckle}, {Hutton}, {Jasniewicz}, {Jordan}, {Kontizas}, {Korn}, {Lanzafame},
  {Manteiga}, {Moitinho}, {Muinonen}, {Osinde}, {Pancino}, {Pauwels}, {Petit},
  {Recio-Blanco}, {Robin}, {Sarro}, {Siopis}, {Smith}, {Smith}, {Sozzetti},
  {Thuillot}, {van Reeven}, {Viala}, {Abbas}, {Abreu Aramburu}, {Accart},
  {Aguado}, {Allan}, {Allasia}, {Altavilla}, {{\'A}lvarez}, {Alves},
  {Anderson}, {Andrei}, {Anglada Varela}, {Antiche}, {Antoja}, {Ant{\'o}n},
  {Arcay}, {Bach}, {Baker}, {Balaguer-N{\'u}{\~n}ez}, {Barache}, {Barata},
  {Barbier}, {Barblan}, {Barrado y Navascu{\'e}s}, {Barros}, {Barstow},
  {Becciani}, {Bellazzini}, {Bello Garc{\'\i}a}, {Belokurov}, {Bendjoya},
  {Berihuete}, {Bianchi}, {Bienaym{\'e}}, {Billebaud}, {Blagorodnova},
  {Blanco-Cuaresma}, {Boch}, {Bombrun}, {Borrachero}, {Bouquillon}, {Bourda},
  {Bouy}, {Bragaglia}, {Breddels}, {Brouillet}, {Br{\"u}semeister},
  {Bucciarelli}, {Burgess}, {Burgon}, {Burlacu}, {Busonero}, {Buzzi}, {Caffau},
  {Cambras}, {Campbell}, {Cancelliere}, {Cantat-Gaudin}, {Carlucci},
  {Carrasco}, {Castellani}, {Charlot}, {Charnas}, {Chiavassa}, {Clotet},
  {Cocozza}, {Collins}, {Costigan}, {Crifo}, {Cross}, {Crosta}, {Crowley},
  {Dafonte}, {Damerdji}, {Dapergolas}, {David}, {David}, {De Cat}, {de Felice},
  {de Laverny}, {De Luise}, {De March}, {de Martino}, {de Souza}, {Debosscher},
  {del Pozo}, {Delbo}, {Delgado}, {Delgado}, {Di Matteo}, {Diakite},
  {Distefano}, {Dolding}, {Dos Anjos}, {Drazinos}, {Duran}, {Dzigan},
  {Edvardsson}, {Enke}, {Evans}, {Eynard Bontemps}, {Fabre}, {Fabrizio},
  {Faigler}, {Falc{\~a}o}, {Farr{\`a}s Casas}, {Federici}, {Fedorets},
  {Fern{\'a}ndez-Hern{\'a}ndez}, {Fernique}, {Fienga}, {Figueras}, {Filippi},
  {Findeisen}, {Fonti}, {Fouesneau}, {Fraile}, {Fraser}, {Fuchs}, {Gai},
  {Galleti}, {Galluccio}, {Garabato}, {Garc{\'\i}a-Sedano}, {Garofalo},
  {Garralda}, {Gavras}, {Gerssen}, {Geyer}, {Gilmore}, {Girona}, {Giuffrida},
  {Gomes}, {Gonz{\'a}lez-Marcos}, {Gonz{\'a}lez-N{\'u}{\~n}ez},
  {Gonz{\'a}lez-Vidal}, {Granvik}, {Guerrier}, {Guillout}, {Guiraud},
  {G{\'u}rpide}, {Guti{\'e}rrez-S{\'a}nchez}, {Guy}, {Haigron},
  {Hatzidimitriou}, {Haywood}, {Heiter}, {Helmi}, {Hobbs}, {Hofmann}, {Holl},
  {Holland}, {Hunt}, {Hypki}, {Icardi}, {Irwin}, {Jevardat de Fombelle},
  {Jofr{\'e}}, {Jonker}, {Jorissen}, {Julbe}, {Karampelas}, {Kochoska},
  {Kohley}, {Kolenberg}, {Kontizas}, {Koposov}, {Kordopatis}, {Koubsky},
  {Krone-Martins}, {Kudryashova}, {Kull}, {Bachchan}, {Lacoste-Seris}, {Lanza},
  {Lavigne}, {Le Poncin-Lafitte}, {Lebreton}, {Lebzelter}, {Leccia}, {Leclerc},
  {Lecoeur-Taibi}, {Lemaitre}, {Lenhardt}, {Leroux}, {Liao}, {Licata},
  {Lindstr{\o}m}, {Lister}, {Livanou}, {Lobel}, {L{\"o}ffler}, {L{\'o}pez},
  {Lorenz}, {MacDonald}, {Magalh{\~a}es Fernandes}, {Managau}, {Mann},
  {Mantelet}, {Marchal}, {Marchant}, {Marconi}, {Marinoni}, {Marrese},
  {Marschalk{\'o}}, {Marshall}, {Mart{\'\i}n-Fleitas}, {Martino}, {Mary},
  {Matijevi{\v{c}}}, {Mazeh}, {McMillan}, {Messina}, {Michalik}, {Millar},
  {Miranda}, {Molina}, {Molinaro}, {Molinaro}, {Moln{\'a}r}, {Moniez},
  {Montegriffo}, {Mor}, {Mora}, {Morbidelli}, {Morel}, {Morgenthaler},
  {Morris}, {Mulone}, {Muraveva}, {Musella}, {Narbonne}, {Nelemans},
  {Nicastro}, {Noval}, {Ord{\'e}novic}, {Ordieres-Mer{\'e}}, {Osborne},
  {Pagani}, {Pagano}, {Pailler}, {Palacin}, {Palaversa}, {Parsons}, {Pecoraro},
  {Pedrosa}, {Pentik{\"a}inen}, {Pichon}, {Piersimoni}, {Pineau}, {Plachy},
  {Plum}, {Poujoulet}, {Pr{\v{s}}a}, {Pulone}, {Ragaini}, {Rago}, {Rambaux},
  {Ramos-Lerate}, {Ranalli}, {Rauw}, {Read}, {Regibo}, {Reyl{\'e}}, {Ribeiro},
  {Rimoldini}, {Ripepi}, {Riva}, {Rixon}, {Roelens}, {Romero-G{\'o}mez},
  {Rowell}, {Royer}, {Ruiz-Dern}, {Sadowski}, {Sagrist{\`a} Sell{\'e}s},
  {Sahlmann}, {Salgado}, {Salguero}, {Sarasso}, {Savietto}, {Schultheis},
  {Sciacca}, {Segol}, {Segovia}, {Segransan}, {Shih}, {Smareglia}, {Smart},
  {Solano}, {Solitro}, {Sordo}, {Soria Nieto}, {Souchay}, {Spagna}, {Spoto},
  {Stampa}, {Steele}, {Steidelm{\"u}ller}, {Stephenson}, {Stoev}, {Suess},
  {S{\"u}veges}, {Surdej}, {Szabados}, {Szegedi-Elek}, {Tapiador}, {Taris},
  {Tauran}, {Taylor}, {Teixeira}, {Terrett}, {Tingley}, {Trager}, {Turon},
  {Ulla}, {Utrilla}, {Valentini}, {van Elteren}, {Van Hemelryck}, {van
  Leeuwen}, {Varadi}, {Vecchiato}, {Veljanoski}, {Via}, {Vicente}, {Vogt},
  {Voss}, {Votruba}, {Voutsinas}, {Walmsley}, {Weiler}, {Weingrill}, {Wevers},
  {Wyrzykowski}, {Yoldas}, {{\v{Z}}erjal}, {Zucker}, {Zurbach}, {Zwitter},
  {Alecu}, {Allen}, {Allende Prieto}, {Amorim}, {Anglada-Escud{\'e}},
  {Arsenijevic}, {Azaz}, {Balm}, {Beck}, {Bernstein}, {Bigot}, {Bijaoui},
  {Blasco}, {Bonfigli}, {Bono}, {Boudreault}, {Bressan}, {Brown}, {Brunet},
  {Bunclark}, {Buonanno}, {Butkevich}, {Carret}, {Carrion}, {Chemin},
  {Ch{\'e}reau}, {Corcione}, {Darmigny}, {de Boer}, {de Teodoro}, {de Zeeuw},
  {Delle Luche}, {Domingues}, {Dubath}, {Fodor}, {Fr{\'e}zouls}, {Fries},
  {Fustes}, {Fyfe}, {Gallardo}, {Gallegos}, {Gardiol}, {Gebran}, {Gomboc},
  {G{\'o}mez}, {Grux}, {Gueguen}, {Heyrovsky}, {Hoar}, {Iannicola}, {Isasi
  Parache}, {Janotto}, {Joliet}, {Jonckheere}, {Keil}, {Kim}, {Klagyivik},
  {Klar}, {Knude}, {Kochukhov}, {Kolka}, {Kos}, {Kutka}, {Lainey}, {LeBouquin},
  {Liu}, {Loreggia}, {Makarov}, {Marseille}, {Martayan}, {Martinez-Rubi},
  {Massart}, {Meynadier}, {Mignot}, {Munari}, {Nguyen}, {Nordlander}, {Ocvirk},
  {O'Flaherty}, {Olias Sanz}, {Ortiz}, {Osorio}, {Oszkiewicz}, {Ouzounis},
  {Palmer}, {Park}, {Pasquato}, {Peltzer}, {Peralta}, {P{\'e}turaud},
  {Pieniluoma}, {Pigozzi}, {Poels}, {Prat}, {Prod'homme}, {Raison}, {Rebordao},
  {Risquez}, {Rocca-Volmerange}, {Rosen}, {Ruiz-Fuertes}, {Russo}, {Sembay},
  {Serraller Vizcaino}, {Short}, {Siebert}, {Silva}, {Sinachopoulos}, {Slezak},
  {Soffel}, {Sosnowska}, {Strai{\v{z}}ys}, {ter Linden}, {Terrell}, {Theil},
  {Tiede}, {Troisi}, {Tsalmantza}, {Tur}, {Vaccari}, {Vachier}, {Valles}, {Van
  Hamme}, {Veltz}, {Virtanen}, {Wallut}, {Wichmann}, {Wilkinson}, {Ziaeepour},
  \& {Zschocke}}]{2016A&A...595A...2G}
{Gaia Collaboration}, {Brown}, A.~G.~A., {Vallenari}, A., {et~al.}
  2016{\natexlab{a}}, \aap, 595, A2, \dodoi{10.1051/0004-6361/201629512}

\bibitem[{{Gaia Collaboration} {et~al.}(2016{\natexlab{b}}){Gaia
  Collaboration}, {Prusti}, {de Bruijne}, {Brown}, {Vallenari}, {Babusiaux},
  {Bailer-Jones}, {Bastian}, {Biermann}, {Evans}, {Eyer}, {Jansen}, {Jordi},
  {Klioner}, {Lammers}, {Lindegren}, {Luri}, {Mignard}, {Milligan}, {Panem},
  {Poinsignon}, {Pourbaix}, {Randich}, {Sarri}, {Sartoretti}, {Siddiqui},
  {Soubiran}, {Valette}, {van Leeuwen}, {Walton}, {Aerts}, {Arenou}, {Cropper},
  {Drimmel}, {H{\o}g}, {Katz}, {Lattanzi}, {O'Mullane}, {Grebel}, {Holland},
  {Huc}, {Passot}, {Bramante}, {Cacciari}, {Casta{\~n}eda}, {Chaoul}, {Cheek},
  {De Angeli}, {Fabricius}, {Guerra}, {Hern{\'a}ndez}, {Jean-Antoine-Piccolo},
  {Masana}, {Messineo}, {Mowlavi}, {Nienartowicz}, {Ord{\'o}{\~n}ez-Blanco},
  {Panuzzo}, {Portell}, {Richards}, {Riello}, {Seabroke}, {Tanga},
  {Th{\'e}venin}, {Torra}, {Els}, {Gracia-Abril}, {Comoretto},
  {Garcia-Reinaldos}, {Lock}, {Mercier}, {Altmann}, {Andrae}, {Astraatmadja},
  {Bellas-Velidis}, {Benson}, {Berthier}, {Blomme}, {Busso}, {Carry},
  {Cellino}, {Clementini}, {Cowell}, {Creevey}, {Cuypers}, {Davidson}, {De
  Ridder}, {de Torres}, {Delchambre}, {Dell'Oro}, {Ducourant}, {Fr{\'e}mat},
  {Garc{\'\i}a-Torres}, {Gosset}, {Halbwachs}, {Hambly}, {Harrison}, {Hauser},
  {Hestroffer}, {Hodgkin}, {Huckle}, {Hutton}, {Jasniewicz}, {Jordan},
  {Kontizas}, {Korn}, {Lanzafame}, {Manteiga}, {Moitinho}, {Muinonen},
  {Osinde}, {Pancino}, {Pauwels}, {Petit}, {Recio-Blanco}, {Robin}, {Sarro},
  {Siopis}, {Smith}, {Smith}, {Sozzetti}, {Thuillot}, {van Reeven}, {Viala},
  {Abbas}, {Abreu Aramburu}, {Accart}, {Aguado}, {Allan}, {Allasia},
  {Altavilla}, {{\'A}lvarez}, {Alves}, {Anderson}, {Andrei}, {Anglada Varela},
  {Antiche}, {Antoja}, {Ant{\'o}n}, {Arcay}, {Atzei}, {Ayache}, {Bach},
  {Baker}, {Balaguer-N{\'u}{\~n}ez}, {Barache}, {Barata}, {Barbier}, {Barblan},
  {Baroni}, {Barrado y Navascu{\'e}s}, {Barros}, {Barstow}, {Becciani},
  {Bellazzini}, {Bellei}, {Bello Garc{\'\i}a}, {Belokurov}, {Bendjoya},
  {Berihuete}, {Bianchi}, {Bienaym{\'e}}, {Billebaud}, {Blagorodnova},
  {Blanco-Cuaresma}, {Boch}, {Bombrun}, {Borrachero}, {Bouquillon}, {Bourda},
  {Bouy}, {Bragaglia}, {Breddels}, {Brouillet}, {Br{\"u}semeister},
  {Bucciarelli}, {Budnik}, {Burgess}, {Burgon}, {Burlacu}, {Busonero}, {Buzzi},
  {Caffau}, {Cambras}, {Campbell}, {Cancelliere}, {Cantat-Gaudin}, {Carlucci},
  {Carrasco}, {Castellani}, {Charlot}, {Charnas}, {Charvet}, {Chassat},
  {Chiavassa}, {Clotet}, {Cocozza}, {Collins}, {Collins}, {Costigan}, {Crifo},
  {Cross}, {Crosta}, {Crowley}, {Dafonte}, {Damerdji}, {Dapergolas}, {David},
  {David}, {De Cat}, {de Felice}, {de Laverny}, {De Luise}, {De March}, {de
  Martino}, {de Souza}, {Debosscher}, {del Pozo}, {Delbo}, {Delgado},
  {Delgado}, {di Marco}, {Di Matteo}, {Diakite}, {Distefano}, {Dolding}, {Dos
  Anjos}, {Drazinos}, {Dur{\'a}n}, {Dzigan}, {Ecale}, {Edvardsson}, {Enke},
  {Erdmann}, {Escolar}, {Espina}, {Evans}, {Eynard Bontemps}, {Fabre},
  {Fabrizio}, {Faigler}, {Falc{\~a}o}, {Farr{\`a}s Casas}, {Faye}, {Federici},
  {Fedorets}, {Fern{\'a}ndez-Hern{\'a}ndez}, {Fernique}, {Fienga}, {Figueras},
  {Filippi}, {Findeisen}, {Fonti}, {Fouesneau}, {Fraile}, {Fraser}, {Fuchs},
  {Furnell}, {Gai}, {Galleti}, {Galluccio}, {Garabato}, {Garc{\'\i}a-Sedano},
  {Gar{\'e}}, {Garofalo}, {Garralda}, {Gavras}, {Gerssen}, {Geyer}, {Gilmore},
  {Girona}, {Giuffrida}, {Gomes}, {Gonz{\'a}lez-Marcos},
  {Gonz{\'a}lez-N{\'u}{\~n}ez}, {Gonz{\'a}lez-Vidal}, {Granvik}, {Guerrier},
  {Guillout}, {Guiraud}, {G{\'u}rpide}, {Guti{\'e}rrez-S{\'a}nchez}, {Guy},
  {Haigron}, {Hatzidimitriou}, {Haywood}, {Heiter}, {Helmi}, {Hobbs},
  {Hofmann}, {Holl}, {Holland}, {Hunt}, {Hypki}, {Icardi}, {Irwin}, {Jevardat
  de Fombelle}, {Jofr{\'e}}, {Jonker}, {Jorissen}, {Julbe}, {Karampelas},
  {Kochoska}, {Kohley}, {Kolenberg}, {Kontizas}, {Koposov}, {Kordopatis},
  {Koubsky}, {Kowalczyk}, {Krone-Martins}, {Kudryashova}, {Kull}, {Bachchan},
  {Lacoste-Seris}, {Lanza}, {Lavigne}, {Le Poncin-Lafitte}, {Lebreton},
  {Lebzelter}, {Leccia}, {Leclerc}, {Lecoeur-Taibi}, {Lemaitre}, {Lenhardt},
  {Leroux}, {Liao}, {Licata}, {Lindstr{\o}m}, {Lister}, {Livanou}, {Lobel},
  {L{\"o}ffler}, {L{\'o}pez}, {Lopez-Lozano}, {Lorenz}, {Loureiro},
  {MacDonald}, {Magalh{\~a}es Fernandes}, {Managau}, {Mann}, {Mantelet},
  {Marchal}, {Marchant}, {Marconi}, {Marie}, {Marinoni}, {Marrese},
  {Marschalk{\'o}}, {Marshall}, {Mart{\'\i}n-Fleitas}, {Martino}, {Mary},
  {Matijevi{\v{c}}}, {Mazeh}, {McMillan}, {Messina}, {Mestre}, {Michalik},
  {Millar}, {Miranda}, {Molina}, {Molinaro}, {Molinaro}, {Moln{\'a}r},
  {Moniez}, {Montegriffo}, {Monteiro}, {Mor}, {Mora}, {Morbidelli}, {Morel},
  {Morgenthaler}, {Morley}, {Morris}, {Mulone}, {Muraveva}, {Musella},
  {Narbonne}, {Nelemans}, {Nicastro}, {Noval}, {Ord{\'e}novic},
  {Ordieres-Mer{\'e}}, {Osborne}, {Pagani}, {Pagano}, {Pailler}, {Palacin},
  {Palaversa}, {Parsons}, {Paulsen}, {Pecoraro}, {Pedrosa}, {Pentik{\"a}inen},
  {Pereira}, {Pichon}, {Piersimoni}, {Pineau}, {Plachy}, {Plum}, {Poujoulet},
  {Pr{\v{s}}a}, {Pulone}, {Ragaini}, {Rago}, {Rambaux}, {Ramos-Lerate},
  {Ranalli}, {Rauw}, {Read}, {Regibo}, {Renk}, {Reyl{\'e}}, {Ribeiro},
  {Rimoldini}, {Ripepi}, {Riva}, {Rixon}, {Roelens}, {Romero-G{\'o}mez},
  {Rowell}, {Royer}, {Rudolph}, {Ruiz-Dern}, {Sadowski}, {Sagrist{\`a}
  Sell{\'e}s}, {Sahlmann}, {Salgado}, {Salguero}, {Sarasso}, {Savietto},
  {Schnorhk}, {Schultheis}, {Sciacca}, {Segol}, {Segovia}, {Segransan},
  {Serpell}, {Shih}, {Smareglia}, {Smart}, {Smith}, {Solano}, {Solitro},
  {Sordo}, {Soria Nieto}, {Souchay}, {Spagna}, {Spoto}, {Stampa}, {Steele},
  {Steidelm{\"u}ller}, {Stephenson}, {Stoev}, {Suess}, {S{\"u}veges}, {Surdej},
  {Szabados}, {Szegedi-Elek}, {Tapiador}, {Taris}, {Tauran}, {Taylor},
  {Teixeira}, {Terrett}, {Tingley}, {Trager}, {Turon}, {Ulla}, {Utrilla},
  {Valentini}, {van Elteren}, {Van Hemelryck}, {van Leeuwen}, {Varadi},
  {Vecchiato}, {Veljanoski}, {Via}, {Vicente}, {Vogt}, {Voss}, {Votruba},
  {Voutsinas}, {Walmsley}, {Weiler}, {Weingrill}, {Werner}, {Wevers},
  {Whitehead}, {Wyrzykowski}, {Yoldas}, {{\v{Z}}erjal}, {Zucker}, {Zurbach},
  {Zwitter}, {Alecu}, {Allen}, {Allende Prieto}, {Amorim},
  {Anglada-Escud{\'e}}, {Arsenijevic}, {Azaz}, {Balm}, {Beck}, {Bernstein},
  {Bigot}, {Bijaoui}, {Blasco}, {Bonfigli}, {Bono}, {Boudreault}, {Bressan},
  {Brown}, {Brunet}, {Bunclark}, {Buonanno}, {Butkevich}, {Carret}, {Carrion},
  {Chemin}, {Ch{\'e}reau}, {Corcione}, {Darmigny}, {de Boer}, {de Teodoro}, {de
  Zeeuw}, {Delle Luche}, {Domingues}, {Dubath}, {Fodor}, {Fr{\'e}zouls},
  {Fries}, {Fustes}, {Fyfe}, {Gallardo}, {Gallegos}, {Gardiol}, {Gebran},
  {Gomboc}, {G{\'o}mez}, {Grux}, {Gueguen}, {Heyrovsky}, {Hoar}, {Iannicola},
  {Isasi Parache}, {Janotto}, {Joliet}, {Jonckheere}, {Keil}, {Kim},
  {Klagyivik}, {Klar}, {Knude}, {Kochukhov}, {Kolka}, {Kos}, {Kutka}, {Lainey},
  {LeBouquin}, {Liu}, {Loreggia}, {Makarov}, {Marseille}, {Martayan},
  {Martinez-Rubi}, {Massart}, {Meynadier}, {Mignot}, {Munari}, {Nguyen},
  {Nordlander}, {Ocvirk}, {O'Flaherty}, {Olias Sanz}, {Ortiz}, {Osorio},
  {Oszkiewicz}, {Ouzounis}, {Palmer}, {Park}, {Pasquato}, {Peltzer}, {Peralta},
  {P{\'e}turaud}, {Pieniluoma}, {Pigozzi}, {Poels}, {Prat}, {Prod'homme},
  {Raison}, {Rebordao}, {Risquez}, {Rocca-Volmerange}, {Rosen}, {Ruiz-Fuertes},
  {Russo}, {Sembay}, {Serraller Vizcaino}, {Short}, {Siebert}, {Silva},
  {Sinachopoulos}, {Slezak}, {Soffel}, {Sosnowska}, {Strai{\v{z}}ys}, {ter
  Linden}, {Terrell}, {Theil}, {Tiede}, {Troisi}, {Tsalmantza}, {Tur},
  {Vaccari}, {Vachier}, {Valles}, {Van Hamme}, {Veltz}, {Virtanen}, {Wallut},
  {Wichmann}, {Wilkinson}, {Ziaeepour}, \& {Zschocke}}]{2016A&A...595A...1G}
{Gaia Collaboration}, {Prusti}, T., {de Bruijne}, J.~H.~J., {et~al.}
  2016{\natexlab{b}}, \aap, 595, A1, \dodoi{10.1051/0004-6361/201629272}

\bibitem[{{Gaia Collaboration} {et~al.}(2018{\natexlab{a}}){Gaia
  Collaboration}, {Brown}, {Vallenari}, {Prusti}, {de Bruijne}, {Babusiaux},
  {Bailer-Jones}, {Biermann}, {Evans}, {Eyer}, {Jansen}, {Jordi}, {Klioner},
  {Lammers}, {Lindegren}, {Luri}, {Mignard}, {Panem}, {Pourbaix}, {Randich},
  {Sartoretti}, {Siddiqui}, {Soubiran}, {van Leeuwen}, {Walton}, {Arenou},
  {Bastian}, {Cropper}, {Drimmel}, {Katz}, {Lattanzi}, {Bakker}, {Cacciari},
  {Casta{\~n}eda}, {Chaoul}, {Cheek}, {De Angeli}, {Fabricius}, {Guerra},
  {Holl}, {Masana}, {Messineo}, {Mowlavi}, {Nienartowicz}, {Panuzzo},
  {Portell}, {Riello}, {Seabroke}, {Tanga}, {Th{\'e}venin}, {Gracia-Abril},
  {Comoretto}, {Garcia-Reinaldos}, {Teyssier}, {Altmann}, {Andrae}, {Audard},
  {Bellas-Velidis}, {Benson}, {Berthier}, {Blomme}, {Burgess}, {Busso},
  {Carry}, {Cellino}, {Clementini}, {Clotet}, {Creevey}, {Davidson}, {De
  Ridder}, {Delchambre}, {Dell'Oro}, {Ducourant},
  {Fern{\'a}ndez-Hern{\'a}ndez}, {Fouesneau}, {Fr{\'e}mat}, {Galluccio},
  {Garc{\'\i}a-Torres}, {Gonz{\'a}lez-N{\'u}{\~n}ez}, {Gonz{\'a}lez-Vidal},
  {Gosset}, {Guy}, {Halbwachs}, {Hambly}, {Harrison}, {Hern{\'a}ndez},
  {Hestroffer}, {Hodgkin}, {Hutton}, {Jasniewicz}, {Jean-Antoine-Piccolo},
  {Jordan}, {Korn}, {Krone-Martins}, {Lanzafame}, {Lebzelter}, {L{\"o}ffler},
  {Manteiga}, {Marrese}, {Mart{\'\i}n-Fleitas}, {Moitinho}, {Mora}, {Muinonen},
  {Osinde}, {Pancino}, {Pauwels}, {Petit}, {Recio-Blanco}, {Richards},
  {Rimoldini}, {Robin}, {Sarro}, {Siopis}, {Smith}, {Sozzetti}, {S{\"u}veges},
  {Torra}, {van Reeven}, {Abbas}, {Abreu Aramburu}, {Accart}, {Aerts},
  {Altavilla}, {{\'A}lvarez}, {Alvarez}, {Alves}, {Anderson}, {Andrei},
  {Anglada Varela}, {Antiche}, {Antoja}, {Arcay}, {Astraatmadja}, {Bach},
  {Baker}, {Balaguer-N{\'u}{\~n}ez}, {Balm}, {Barache}, {Barata}, {Barbato},
  {Barblan}, {Barklem}, {Barrado}, {Barros}, {Barstow}, {Bartholom{\'e}
  Mu{\~n}oz}, {Bassilana}, {Becciani}, {Bellazzini}, {Berihuete}, {Bertone},
  {Bianchi}, {Bienaym{\'e}}, {Blanco-Cuaresma}, {Boch}, {Boeche}, {Bombrun},
  {Borrachero}, {Bossini}, {Bouquillon}, {Bourda}, {Bragaglia}, {Bramante},
  {Breddels}, {Bressan}, {Brouillet}, {Br{\"u}semeister}, {Brugaletta},
  {Bucciarelli}, {Burlacu}, {Busonero}, {Butkevich}, {Buzzi}, {Caffau},
  {Cancelliere}, {Cannizzaro}, {Cantat-Gaudin}, {Carballo}, {Carlucci},
  {Carrasco}, {Casamiquela}, {Castellani}, {Castro-Ginard}, {Charlot},
  {Chemin}, {Chiavassa}, {Cocozza}, {Costigan}, {Cowell}, {Crifo}, {Crosta},
  {Crowley}, {Cuypers}, {Dafonte}, {Damerdji}, {Dapergolas}, {David}, {David},
  {de Laverny}, {De Luise}, {De March}, {de Martino}, {de Souza}, {de Torres},
  {Debosscher}, {del Pozo}, {Delbo}, {Delgado}, {Delgado}, {Di Matteo},
  {Diakite}, {Diener}, {Distefano}, {Dolding}, {Drazinos}, {Dur{\'a}n},
  {Edvardsson}, {Enke}, {Eriksson}, {Esquej}, {Eynard Bontemps}, {Fabre},
  {Fabrizio}, {Faigler}, {Falc{\~a}o}, {Farr{\`a}s Casas}, {Federici},
  {Fedorets}, {Fernique}, {Figueras}, {Filippi}, {Findeisen}, {Fonti},
  {Fraile}, {Fraser}, {Fr{\'e}zouls}, {Gai}, {Galleti}, {Garabato},
  {Garc{\'\i}a-Sedano}, {Garofalo}, {Garralda}, {Gavel}, {Gavras}, {Gerssen},
  {Geyer}, {Giacobbe}, {Gilmore}, {Girona}, {Giuffrida}, {Glass}, {Gomes},
  {Granvik}, {Gueguen}, {Guerrier}, {Guiraud}, {Guti{\'e}rrez-S{\'a}nchez},
  {Haigron}, {Hatzidimitriou}, {Hauser}, {Haywood}, {Heiter}, {Helmi}, {Heu},
  {Hilger}, {Hobbs}, {Hofmann}, {Holland}, {Huckle}, {Hypki}, {Icardi},
  {Jan{\ss}en}, {Jevardat de Fombelle}, {Jonker}, {Juh{\'a}sz}, {Julbe},
  {Karampelas}, {Kewley}, {Klar}, {Kochoska}, {Kohley}, {Kolenberg},
  {Kontizas}, {Kontizas}, {Koposov}, {Kordopatis}, {Kostrzewa-Rutkowska},
  {Koubsky}, {Lambert}, {Lanza}, {Lasne}, {Lavigne}, {Le Fustec}, {Le
  Poncin-Lafitte}, {Lebreton}, {Leccia}, {Leclerc}, {Lecoeur-Taibi},
  {Lenhardt}, {Leroux}, {Liao}, {Licata}, {Lindstr{\o}m}, {Lister}, {Livanou},
  {Lobel}, {L{\'o}pez}, {Managau}, {Mann}, {Mantelet}, {Marchal}, {Marchant},
  {Marconi}, {Marinoni}, {Marschalk{\'o}}, {Marshall}, {Martino}, {Marton},
  {Mary}, {Massari}, {Matijevi{\v{c}}}, {Mazeh}, {McMillan}, {Messina},
  {Michalik}, {Millar}, {Molina}, {Molinaro}, {Moln{\'a}r}, {Montegriffo},
  {Mor}, {Morbidelli}, {Morel}, {Morris}, {Mulone}, {Muraveva}, {Musella},
  {Nelemans}, {Nicastro}, {Noval}, {O'Mullane}, {Ord{\'e}novic},
  {Ord{\'o}{\~n}ez-Blanco}, {Osborne}, {Pagani}, {Pagano}, {Pailler},
  {Palacin}, {Palaversa}, {Panahi}, {Pawlak}, {Piersimoni}, {Pineau}, {Plachy},
  {Plum}, {Poggio}, {Poujoulet}, {Pr{\v{s}}a}, {Pulone}, {Racero}, {Ragaini},
  {Rambaux}, {Ramos-Lerate}, {Regibo}, {Reyl{\'e}}, {Riclet}, {Ripepi}, {Riva},
  {Rivard}, {Rixon}, {Roegiers}, {Roelens}, {Romero-G{\'o}mez}, {Rowell},
  {Royer}, {Ruiz-Dern}, {Sadowski}, {Sagrist{\`a} Sell{\'e}s}, {Sahlmann},
  {Salgado}, {Salguero}, {Sanna}, {Santana-Ros}, {Sarasso}, {Savietto},
  {Schultheis}, {Sciacca}, {Segol}, {Segovia}, {S{\'e}gransan}, {Shih},
  {Siltala}, {Silva}, {Smart}, {Smith}, {Solano}, {Solitro}, {Sordo}, {Soria
  Nieto}, {Souchay}, {Spagna}, {Spoto}, {Stampa}, {Steele},
  {Steidelm{\"u}ller}, {Stephenson}, {Stoev}, {Suess}, {Surdej}, {Szabados},
  {Szegedi-Elek}, {Tapiador}, {Taris}, {Tauran}, {Taylor}, {Teixeira},
  {Terrett}, {Teyssandier}, {Thuillot}, {Titarenko}, {Torra Clotet}, {Turon},
  {Ulla}, {Utrilla}, {Uzzi}, {Vaillant}, {Valentini}, {Valette}, {van Elteren},
  {Van Hemelryck}, {van Leeuwen}, {Vaschetto}, {Vecchiato}, {Veljanoski},
  {Viala}, {Vicente}, {Vogt}, {von Essen}, {Voss}, {Votruba}, {Voutsinas},
  {Walmsley}, {Weiler}, {Wertz}, {Wevers}, {Wyrzykowski}, {Yoldas},
  {{\v{Z}}erjal}, {Ziaeepour}, {Zorec}, {Zschocke}, {Zucker}, {Zurbach}, \&
  {Zwitter}}]{2018A&A...616A...1G}
{Gaia Collaboration}, {Brown}, A.~G.~A., {Vallenari}, A., {et~al.}
  2018{\natexlab{a}}, \aap, 616, A1, \dodoi{10.1051/0004-6361/201833051}

\bibitem[{{Gaia Collaboration} {et~al.}(2018{\natexlab{b}}){Gaia
  Collaboration}, {Babusiaux}, {van Leeuwen}, {Barstow}, {Jordi}, {Vallenari},
  {Bossini}, {Bressan}, {Cantat-Gaudin}, {van Leeuwen}, {Brown}, {Prusti}, {de
  Bruijne}, {Bailer-Jones}, {Biermann}, {Evans}, {Eyer}, {Jansen}, {Klioner},
  {Lammers}, {Lindegren}, {Luri}, {Mignard}, {Panem}, {Pourbaix}, {Randich},
  {Sartoretti}, {Siddiqui}, {Soubiran}, {Walton}, {Arenou}, {Bastian},
  {Cropper}, {Drimmel}, {Katz}, {Lattanzi}, {Bakker}, {Cacciari},
  {Casta{\~n}eda}, {Chaoul}, {Cheek}, {De Angeli}, {Fabricius}, {Guerra},
  {Holl}, {Masana}, {Messineo}, {Mowlavi}, {Nienartowicz}, {Panuzzo},
  {Portell}, {Riello}, {Seabroke}, {Tanga}, {Th{\'e}venin}, {Gracia-Abril},
  {Comoretto}, {Garcia-Reinaldos}, {Teyssier}, {Altmann}, {Andrae}, {Audard},
  {Bellas-Velidis}, {Benson}, {Berthier}, {Blomme}, {Burgess}, {Busso},
  {Carry}, {Cellino}, {Clementini}, {Clotet}, {Creevey}, {Davidson}, {De
  Ridder}, {Delchambre}, {Dell'Oro}, {Ducourant},
  {Fern{\'a}ndez-Hern{\'a}ndez}, {Fouesneau}, {Fr{\'e}mat}, {Galluccio},
  {Garc{\'\i}a-Torres}, {Gonz{\'a}lez-N{\'u}{\~n}ez}, {Gonz{\'a}lez-Vidal},
  {Gosset}, {Guy}, {Halbwachs}, {Hambly}, {Harrison}, {Hern{\'a}ndez},
  {Hestroffer}, {Hodgkin}, {Hutton}, {Jasniewicz}, {Jean-Antoine-Piccolo},
  {Jordan}, {Korn}, {Krone-Martins}, {Lanzafame}, {Lebzelter}, {L{\"o}ffler},
  {Manteiga}, {Marrese}, {Mart{\'\i}n-Fleitas}, {Moitinho}, {Mora}, {Muinonen},
  {Osinde}, {Pancino}, {Pauwels}, {Petit}, {Recio-Blanco}, {Richards},
  {Rimoldini}, {Robin}, {Sarro}, {Siopis}, {Smith}, {Sozzetti}, {S{\"u}veges},
  {Torra}, {van Reeven}, {Abbas}, {Abreu Aramburu}, {Accart}, {Aerts},
  {Altavilla}, {{\'A}lvarez}, {Alvarez}, {Alves}, {Anderson}, {Andrei},
  {Anglada Varela}, {Antiche}, {Antoja}, {Arcay}, {Astraatmadja}, {Bach},
  {Baker}, {Balaguer-N{\'u}{\~n}ez}, {Balm}, {Barache}, {Barata}, {Barbato},
  {Barblan}, {Barklem}, {Barrado}, {Barros}, {Bartholom{\'e} Mu{\~n}oz},
  {Bassilana}, {Becciani}, {Bellazzini}, {Berihuete}, {Bertone}, {Bianchi},
  {Bienaym{\'e}}, {Blanco-Cuaresma}, {Boch}, {Boeche}, {Bombrun}, {Borrachero},
  {Bouquillon}, {Bourda}, {Bragaglia}, {Bramante}, {Breddels}, {Brouillet},
  {Br{\"u}semeister}, {Brugaletta}, {Bucciarelli}, {Burlacu}, {Busonero},
  {Butkevich}, {Buzzi}, {Caffau}, {Cancelliere}, {Cannizzaro}, {Carballo},
  {Carlucci}, {Carrasco}, {Casamiquela}, {Castellani}, {Castro-Ginard},
  {Charlot}, {Chemin}, {Chiavassa}, {Cocozza}, {Costigan}, {Cowell}, {Crifo},
  {Crosta}, {Crowley}, {Cuypers}, {Dafonte}, {Damerdji}, {Dapergolas}, {David},
  {David}, {de Laverny}, {De Luise}, {De March}, {de Martino}, {de Souza}, {de
  Torres}, {Debosscher}, {del Pozo}, {Delbo}, {Delgado}, {Delgado}, {Diakite},
  {Diener}, {Distefano}, {Dolding}, {Drazinos}, {Dur{\'a}n}, {Edvardsson},
  {Enke}, {Eriksson}, {Esquej}, {Eynard Bontemps}, {Fabre}, {Fabrizio},
  {Faigler}, {Falc{\~a}o}, {Farr{\`a}s Casas}, {Federici}, {Fedorets},
  {Fernique}, {Figueras}, {Filippi}, {Findeisen}, {Fonti}, {Fraile}, {Fraser},
  {Fr{\'e}zouls}, {Gai}, {Galleti}, {Garabato}, {Garc{\'\i}a-Sedano},
  {Garofalo}, {Garralda}, {Gavel}, {Gavras}, {Gerssen}, {Geyer}, {Giacobbe},
  {Gilmore}, {Girona}, {Giuffrida}, {Glass}, {Gomes}, {Granvik}, {Gueguen},
  {Guerrier}, {Guiraud}, {Guti{\'e}}, {Haigron}, {Hatzidimitriou}, {Hauser},
  {Haywood}, {Heiter}, {Helmi}, {Heu}, {Hilger}, {Hobbs}, {Hofmann}, {Holland},
  {Huckle}, {Hypki}, {Icardi}, {Jan{\ss}en}, {Jevardat de Fombelle}, {Jonker},
  {Juh{\'a}sz}, {Julbe}, {Karampelas}, {Kewley}, {Klar}, {Kochoska}, {Kohley},
  {Kolenberg}, {Kontizas}, {Kontizas}, {Koposov}, {Kordopatis},
  {Kostrzewa-Rutkowska}, {Koubsky}, {Lambert}, {Lanza}, {Lasne}, {Lavigne}, {Le
  Fustec}, {Le Poncin-Lafitte}, {Lebreton}, {Leccia}, {Leclerc},
  {Lecoeur-Taibi}, {Lenhardt}, {Leroux}, {Liao}, {Licata}, {Lindstr{\o}m},
  {Lister}, {Livanou}, {Lobel}, {L{\'o}pez}, {Managau}, {Mann}, {Mantelet},
  {Marchal}, {Marchant}, {Marconi}, {Marinoni}, {Marschalk{\'o}}, {Marshall},
  {Martino}, {Marton}, {Mary}, {Massari}, {Matijevi{\v{c}}}, {Mazeh},
  {McMillan}, {Messina}, {Michalik}, {Millar}, {Molina}, {Molinaro},
  {Moln{\'a}r}, {Montegriffo}, {Mor}, {Morbidelli}, {Morel}, {Morris},
  {Mulone}, {Muraveva}, {Musella}, {Nelemans}, {Nicastro}, {Noval},
  {O'Mullane}, {Ord{\'e}novic}, {Ord{\'o}{\~n}ez-Blanco}, {Osborne}, {Pagani},
  {Pagano}, {Pailler}, {Palacin}, {Palaversa}, {Panahi}, {Pawlak},
  {Piersimoni}, {Pineau}, {Plachy}, {Plum}, {Poggio}, {Poujoulet},
  {Pr{\v{s}}a}, {Pulone}, {Racero}, {Ragaini}, {Rambaux}, {Ramos-Lerate},
  {Regibo}, {Reyl{\'e}}, {Riclet}, {Ripepi}, {Riva}, {Rivard}, {Rixon},
  {Roegiers}, {Roelens}, {Romero-G{\'o}mez}, {Rowell}, {Royer}, {Ruiz-Dern},
  {Sadowski}, {Sagrist{\`a} Sell{\'e}s}, {Sahlmann}, {Salgado}, {Salguero},
  {Sanna}, {Santana-Ros}, {Sarasso}, {Savietto}, {Schultheis}, {Sciacca},
  {Segol}, {Segovia}, {S{\'e}gransan}, {Shih}, {Siltala}, {Silva}, {Smart},
  {Smith}, {Solano}, {Solitro}, {Sordo}, {Soria Nieto}, {Souchay}, {Spagna},
  {Spoto}, {Stampa}, {Steele}, {Steidelm{\"u}ller}, {Stephenson}, {Stoev},
  {Suess}, {Surdej}, {Szabados}, {Szegedi-Elek}, {Tapiador}, {Taris}, {Tauran},
  {Taylor}, {Teixeira}, {Terrett}, {Teyssandier}, {Thuillot}, {Titarenko},
  {Torra Clotet}, {Turon}, {Ulla}, {Utrilla}, {Uzzi}, {Vaillant}, {Valentini},
  {Valette}, {van Elteren}, {Van Hemelryck}, {Vaschetto}, {Vecchiato},
  {Veljanoski}, {Viala}, {Vicente}, {Vogt}, {von Essen}, {Voss}, {Votruba},
  {Voutsinas}, {Walmsley}, {Weiler}, {Wertz}, {Wevers}, {Wyrzykowski},
  {Yoldas}, {{\v{Z}}erjal}, {Ziaeepour}, {Zorec}, {Zschocke}, {Zucker},
  {Zurbach}, \& {Zwitter}}]{2018A&A...616A..10G}
{Gaia Collaboration}, {Babusiaux}, C., {van Leeuwen}, F., {et~al.}
  2018{\natexlab{b}}, \aap, 616, A10, \dodoi{10.1051/0004-6361/201832843}

\bibitem[{{Gaia Collaboration} {et~al.}(2021){Gaia Collaboration}, {Brown},
  {Vallenari}, {Prusti}, {de Bruijne}, {Babusiaux}, {Biermann}, {Creevey},
  {Evans}, {Eyer}, {Hutton}, {Jansen}, {Jordi}, {Klioner}, {Lammers},
  {Lindegren}, {Luri}, {Mignard}, {Panem}, {Pourbaix}, {Randich}, {Sartoretti},
  {Soubiran}, {Walton}, {Arenou}, {Bailer-Jones}, {Bastian}, {Cropper},
  {Drimmel}, {Katz}, {Lattanzi}, {van Leeuwen}, {Bakker}, {Cacciari},
  {Casta{\~n}eda}, {De Angeli}, {Ducourant}, {Fabricius}, {Fouesneau},
  {Fr{\'e}mat}, {Guerra}, {Guerrier}, {Guiraud}, {Jean-Antoine Piccolo},
  {Masana}, {Messineo}, {Mowlavi}, {Nicolas}, {Nienartowicz}, {Pailler},
  {Panuzzo}, {Riclet}, {Roux}, {Seabroke}, {Sordo}, {Tanga}, {Th{\'e}venin},
  {Gracia-Abril}, {Portell}, {Teyssier}, {Altmann}, {Andrae}, {Bellas-Velidis},
  {Benson}, {Berthier}, {Blomme}, {Brugaletta}, {Burgess}, {Busso}, {Carry},
  {Cellino}, {Cheek}, {Clementini}, {Damerdji}, {Davidson}, {Delchambre},
  {Dell'Oro}, {Fern{\'a}ndez-Hern{\'a}ndez}, {Galluccio}, {Garc{\'\i}a-Lario},
  {Garcia-Reinaldos}, {Gonz{\'a}lez-N{\'u}{\~n}ez}, {Gosset}, {Haigron},
  {Halbwachs}, {Hambly}, {Harrison}, {Hatzidimitriou}, {Heiter},
  {Hern{\'a}ndez}, {Hestroffer}, {Hodgkin}, {Holl}, {Jan{\ss}en}, {Jevardat de
  Fombelle}, {Jordan}, {Krone-Martins}, {Lanzafame}, {L{\"o}ffler}, {Lorca},
  {Manteiga}, {Marchal}, {Marrese}, {Moitinho}, {Mora}, {Muinonen}, {Osborne},
  {Pancino}, {Pauwels}, {Petit}, {Recio-Blanco}, {Richards}, {Riello},
  {Rimoldini}, {Robin}, {Roegiers}, {Rybizki}, {Sarro}, {Siopis}, {Smith},
  {Sozzetti}, {Ulla}, {Utrilla}, {van Leeuwen}, {van Reeven}, {Abbas}, {Abreu
  Aramburu}, {Accart}, {Aerts}, {Aguado}, {Ajaj}, {Altavilla}, {{\'A}lvarez},
  {{\'A}lvarez Cid-Fuentes}, {Alves}, {Anderson}, {Anglada Varela}, {Antoja},
  {Audard}, {Baines}, {Baker}, {Balaguer-N{\'u}{\~n}ez}, {Balbinot}, {Balog},
  {Barache}, {Barbato}, {Barros}, {Barstow}, {Bartolom{\'e}}, {Bassilana},
  {Bauchet}, {Baudesson-Stella}, {Becciani}, {Bellazzini}, {Bernet}, {Bertone},
  {Bianchi}, {Blanco-Cuaresma}, {Boch}, {Bombrun}, {Bossini}, {Bouquillon},
  {Bragaglia}, {Bramante}, {Breedt}, {Bressan}, {Brouillet}, {Bucciarelli},
  {Burlacu}, {Busonero}, {Butkevich}, {Buzzi}, {Caffau}, {Cancelliere},
  {C{\'a}novas}, {Cantat-Gaudin}, {Carballo}, {Carlucci}, {Carnerero},
  {Carrasco}, {Casamiquela}, {Castellani}, {Castro-Ginard}, {Castro Sampol},
  {Chaoul}, {Charlot}, {Chemin}, {Chiavassa}, {Cioni}, {Comoretto}, {Cooper},
  {Cornez}, {Cowell}, {Crifo}, {Crosta}, {Crowley}, {Dafonte}, {Dapergolas},
  {David}, {David}, {de Laverny}, {De Luise}, {De March}, {De Ridder}, {de
  Souza}, {de Teodoro}, {de Torres}, {del Peloso}, {del Pozo}, {Delbo},
  {Delgado}, {Delgado}, {Delisle}, {Di Matteo}, {Diakite}, {Diener},
  {Distefano}, {Dolding}, {Eappachen}, {Edvardsson}, {Enke}, {Esquej}, {Fabre},
  {Fabrizio}, {Faigler}, {Fedorets}, {Fernique}, {Fienga}, {Figueras},
  {Fouron}, {Fragkoudi}, {Fraile}, {Franke}, {Gai}, {Garabato},
  {Garcia-Gutierrez}, {Garc{\'\i}a-Torres}, {Garofalo}, {Gavras}, {Gerlach},
  {Geyer}, {Giacobbe}, {Gilmore}, {Girona}, {Giuffrida}, {Gomel}, {Gomez},
  {Gonzalez-Santamaria}, {Gonz{\'a}lez-Vidal}, {Granvik},
  {Guti{\'e}rrez-S{\'a}nchez}, {Guy}, {Hauser}, {Haywood}, {Helmi}, {Hidalgo},
  {Hilger}, {H{\l}adczuk}, {Hobbs}, {Holland}, {Huckle}, {Jasniewicz},
  {Jonker}, {Juaristi Campillo}, {Julbe}, {Karbevska}, {Kervella}, {Khanna},
  {Kochoska}, {Kontizas}, {Kordopatis}, {Korn}, {Kostrzewa-Rutkowska},
  {Kruszy{\'n}ska}, {Lambert}, {Lanza}, {Lasne}, {Le Campion}, {Le Fustec},
  {Lebreton}, {Lebzelter}, {Leccia}, {Leclerc}, {Lecoeur-Taibi}, {Liao},
  {Licata}, {Lindstr{\o}m}, {Lister}, {Livanou}, {Lobel}, {Madrero Pardo},
  {Managau}, {Mann}, {Marchant}, {Marconi}, {Marcos Santos}, {Marinoni},
  {Marocco}, {Marshall}, {Martin Polo}, {Mart{\'\i}n-Fleitas}, {Masip},
  {Massari}, {Mastrobuono-Battisti}, {Mazeh}, {McMillan}, {Messina},
  {Michalik}, {Millar}, {Mints}, {Molina}, {Molinaro}, {Moln{\'a}r},
  {Montegriffo}, {Mor}, {Morbidelli}, {Morel}, {Morris}, {Mulone}, {Munoz},
  {Muraveva}, {Murphy}, {Musella}, {Noval}, {Ord{\'e}novic}, {Orr{\`u}},
  {Osinde}, {Pagani}, {Pagano}, {Palaversa}, {Palicio}, {Panahi}, {Pawlak},
  {Pe{\~n}alosa Esteller}, {Penttil{\"a}}, {Piersimoni}, {Pineau}, {Plachy},
  {Plum}, {Poggio}, {Poretti}, {Poujoulet}, {Pr{\v{s}}a}, {Pulone}, {Racero},
  {Ragaini}, {Rainer}, {Raiteri}, {Rambaux}, {Ramos}, {Ramos-Lerate}, {Re
  Fiorentin}, {Regibo}, {Reyl{\'e}}, {Ripepi}, {Riva}, {Rixon}, {Robichon},
  {Robin}, {Roelens}, {Rohrbasser}, {Romero-G{\'o}mez}, {Rowell}, {Royer},
  {Rybicki}, {Sadowski}, {Sagrist{\`a} Sell{\'e}s}, {Sahlmann}, {Salgado},
  {Salguero}, {Samaras}, {Sanchez Gimenez}, {Sanna}, {Santove{\~n}a},
  {Sarasso}, {Schultheis}, {Sciacca}, {Segol}, {Segovia}, {S{\'e}gransan},
  {Semeux}, {Shahaf}, {Siddiqui}, {Siebert}, {Siltala}, {Slezak}, {Smart},
  {Solano}, {Solitro}, {Souami}, {Souchay}, {Spagna}, {Spoto}, {Steele},
  {Steidelm{\"u}ller}, {Stephenson}, {S{\"u}veges}, {Szabados}, {Szegedi-Elek},
  {Taris}, {Tauran}, {Taylor}, {Teixeira}, {Thuillot}, {Tonello}, {Torra},
  {Torra}, {Turon}, {Unger}, {Vaillant}, {van Dillen}, {Vanel}, {Vecchiato},
  {Viala}, {Vicente}, {Voutsinas}, {Weiler}, {Wevers}, {Wyrzykowski}, {Yoldas},
  {Yvard}, {Zhao}, {Zorec}, {Zucker}, {Zurbach}, \&
  {Zwitter}}]{2021A&A...649A...1G}
{Gaia Collaboration}, {Brown}, A.~G.~A., {Vallenari}, A., {et~al.} 2021, \aap,
  649, A1, \dodoi{10.1051/0004-6361/202039657}

\bibitem[{{Gallart} {et~al.}(2019){Gallart}, {Bernard}, {Brook}, {Ruiz-Lara},
  {Cassisi}, {Hill}, \& {Monelli}}]{2019NatAs...3..932G}
{Gallart}, C., {Bernard}, E.~J., {Brook}, C.~B., {et~al.} 2019, Nature
  Astronomy, 3, 932, \dodoi{10.1038/s41550-019-0829-5}

\bibitem[{{Hammer} {et~al.}(2023){Hammer}, {Li}, {Mamon}, {Pawlowski},
  {Bonifacio}, {Jiao}, {Wang}, {Wang}, \& {Yang}}]{2022arXiv221207441H}
{Hammer}, F., {Li}, H., {Mamon}, G.~A., {et~al.} 2023, \mnras, 519, 5059,
  \dodoi{10.1093/mnras/stac3758}

\bibitem[{{Hasselquist} {et~al.}(2021){Hasselquist}, {Hayes}, {Lian},
  {Weinberg}, {Zasowski}, {Horta}, {Beaton}, {Feuillet}, {Garro}, {Gallart},
  {Smith}, {Holtzman}, {Minniti}, {Lacerna}, {Shetrone}, {J{\"o}nsson},
  {Cioni}, {Fillingham}, {Cunha}, {O{\'C}onnell}, {Fern{\'a}ndez-Trincado},
  {Mu{\~n}oz}, {Schiavon}, {Almeida}, {Anguiano}, {Beers}, {Bizyaev},
  {Brownstein}, {Cohen}, {Frinchaboy}, {Garc{\'\i}a-Hern{\'a}ndez}, {Geisler},
  {Lane}, {Majewski}, {Nidever}, {Nitschelm}, {Povick}, {Price-Whelan},
  {Roman-Lopes}, {Rosado}, {Sobeck}, {Stringfellow}, {Valenzuela}, {Villanova},
  \& {Vincenzo}}]{2021arXiv210905130H}
{Hasselquist}, S., {Hayes}, C.~R., {Lian}, J., {et~al.} 2021, \apj, 923, 172,
  \dodoi{10.3847/1538-4357/ac25f9}

\bibitem[{{Hawkins} {et~al.}(2015){Hawkins}, {Jofr{\'e}}, {Masseron}, \&
  {Gilmore}}]{2015MNRAS.453..758H}
{Hawkins}, K., {Jofr{\'e}}, P., {Masseron}, T., \& {Gilmore}, G. 2015, \mnras,
  453, 758, \dodoi{10.1093/mnras/stv1586}

\bibitem[{{Hayes} {et~al.}(2018){Hayes}, {Majewski}, {Shetrone},
  {Fern{\'a}ndez-Alvar}, {Allende Prieto}, {Schuster}, {Carigi}, {Cunha},
  {Smith}, {Sobeck}, {Almeida}, {Beers}, {Carrera}, {Fern{\'a}ndez-Trincado},
  {Garc{\'\i}a-Hern{\'a}ndez}, {Geisler}, {Lane}, {Lucatello}, {Matthews},
  {Minniti}, {Nitschelm}, {Tang}, {Tissera}, \& {Zamora}}]{2018ApJ...852...49H}
{Hayes}, C.~R., {Majewski}, S.~R., {Shetrone}, M., {et~al.} 2018, \apj, 852,
  49, \dodoi{10.3847/1538-4357/aa9cec}

\bibitem[{{Hayes} {et~al.}(2020){Hayes}, {Majewski}, {Hasselquist}, {Anguiano},
  {Shetrone}, {Law}, {Schiavon}, {Cunha}, {Smith}, {Beaton}, {Price-Whelan},
  {Allende Prieto}, {Battaglia}, {Bizyaev}, {Brownstein}, {Cohen},
  {Frinchaboy}, {Garc{\'\i}a-Hern{\'a}ndez}, {Lacerna}, {Lane},
  {M{\'e}sz{\'a}ros}, {Bidin}, {M{\~{u}}noz}, {Nidever}, {Oravetz}, {Oravetz},
  {Pan}, {Roman-Lopes}, {Sobeck}, \& {Stringfellow}}]{2020ApJ...889...63H}
{Hayes}, C.~R., {Majewski}, S.~R., {Hasselquist}, S., {et~al.} 2020, \apj, 889,
  63, \dodoi{10.3847/1538-4357/ab62ad}

\bibitem[{{Haywood} {et~al.}(2018){Haywood}, {Di Matteo}, {Lehnert}, {Snaith},
  {Khoperskov}, \& {G{\'o}mez}}]{2018ApJ...863..113H}
{Haywood}, M., {Di Matteo}, P., {Lehnert}, M.~D., {et~al.} 2018, \apj, 863,
  113, \dodoi{10.3847/1538-4357/aad235}

\bibitem[{{Heiter} \& {Eriksson}(2006)}]{2006A&A...452.1039H}
{Heiter}, U., \& {Eriksson}, K. 2006, \aap, 452, 1039,
  \dodoi{10.1051/0004-6361:20064925}

\bibitem[{{Helmi}(2020)}]{2020ARA&A..58..205H}
{Helmi}, A. 2020, \araa, 58, 205, \dodoi{10.1146/annurev-astro-032620-021917}

\bibitem[{{Helmi} {et~al.}(2018){Helmi}, {Babusiaux}, {Koppelman}, {Massari},
  {Veljanoski}, \& {Brown}}]{2018Natur.563...85H}
{Helmi}, A., {Babusiaux}, C., {Koppelman}, H.~H., {et~al.} 2018, \nat, 563, 85,
  \dodoi{10.1038/s41586-018-0625-x}

\bibitem[{{Helmi} {et~al.}(1999){Helmi}, {White}, {de Zeeuw}, \&
  {Zhao}}]{1999Natur.402...53H}
{Helmi}, A., {White}, S. D.~M., {de Zeeuw}, P.~T., \& {Zhao}, H. 1999, \nat,
  402, 53, \dodoi{10.1038/46980}

\bibitem[{{Horta} {et~al.}(2021){Horta}, {Schiavon}, {Mackereth}, {Pfeffer},
  {Mason}, {Kisku}, {Fragkoudi}, {Allende Prieto}, {Cunha}, {Hasselquist},
  {Holtzman}, {Majewski}, {Nataf}, {O'Connell}, {Schultheis}, \&
  {Smith}}]{2021MNRAS.500.1385H}
{Horta}, D., {Schiavon}, R.~P., {Mackereth}, J.~T., {et~al.} 2021, \mnras, 500,
  1385, \dodoi{10.1093/mnras/staa2987}

\bibitem[{{Horta} {et~al.}(2022){Horta}, {Schiavon}, {Mackereth}, {Weinberg},
  {Hasselquist}, {Feuillet}, {O'Connell}, {Anguiano}, {Allende-Prieto},
  {Beaton}, {Bizyaev}, {Cunha}, {Geisler}, {Garc{\'\i}a-Hern{\'a}ndez},
  {Holtzman}, {J{\"o}nsson}, {Lane}, {Majewski}, {M{\'e}sz{\'a}ros}, {Minniti},
  {Nitschelm}, {Shetrone}, {Smith}, \& {Zasowski}}]{2022MNRAS.tmp.3011H}
---. 2022, \mnras, \dodoi{10.1093/mnras/stac3179}

\bibitem[{{Jofr{\'e}} {et~al.}(2016){Jofr{\'e}}, {Jorissen}, {Van Eck},
  {Izzard}, {Masseron}, {Hawkins}, {Gilmore}, {Paladini}, {Escorza},
  {Blanco-Cuaresma}, \& {Manick}}]{2016A&A...595A..60J}
{Jofr{\'e}}, P., {Jorissen}, A., {Van Eck}, S., {et~al.} 2016, \aap, 595, A60,
  \dodoi{10.1051/0004-6361/201629356}

\bibitem[{{J{\"o}nsson} {et~al.}(2020){J{\"o}nsson}, {Holtzman}, {Allende
  Prieto}, {Cunha}, {Garc{\'\i}a-Hern{\'a}ndez}, {Hasselquist}, {Masseron},
  {Osorio}, {Shetrone}, {Smith}, {Stringfellow}, {Bizyaev}, {Edvardsson},
  {Majewski}, {M{\'e}sz{\'a}ros}, {Souto}, {Zamora}, {Beaton}, {Bovy}, {Donor},
  {Pinsonneault}, {Poovelil}, \& {Sobeck}}]{2020AJ....160..120J}
{J{\"o}nsson}, H., {Holtzman}, J.~A., {Allende Prieto}, C., {et~al.} 2020, \aj,
  160, 120, \dodoi{10.3847/1538-3881/aba592}

\bibitem[{{Khoperskov} {et~al.}(2022{\natexlab{a}}){Khoperskov}, {Minchev},
  {Libeskind}, {Haywood}, {Di Matteo}, {Belokurov}, {Steinmetz}, {Gomez},
  {Grand}, {Knebe}, {Sorce}, {Sparre}, {Tempel}, \&
  {Vogelsberger}}]{2022arXiv220604521K}
{Khoperskov}, S., {Minchev}, I., {Libeskind}, N., {et~al.} 2022{\natexlab{a}},
  arXiv e-prints, arXiv:2206.04521, \dodoi{10.48550/arXiv.2206.04521}

\bibitem[{{Khoperskov} {et~al.}(2022{\natexlab{b}}){Khoperskov}, {Minchev},
  {Libeskind}, {Haywood}, {Di Matteo}, {Belokurov}, {Steinmetz}, {Gomez},
  {Grand}, {Knebe}, {Sorce}, {Sparre}, {Tempel}, \&
  {Vogelsberger}}]{2022arXiv220604522K}
---. 2022{\natexlab{b}}, arXiv e-prints, arXiv:2206.04522,
  \dodoi{10.48550/arXiv.2206.04522}

\bibitem[{{Khoperskov} {et~al.}(2022{\natexlab{c}}){Khoperskov}, {Minchev},
  {Libeskind}, {Belokurov}, {Steinmetz}, {Gomez}, {Grand}, {Knebe}, {Sorce},
  {Sparre}, {Tempel}, \& {Vogelsberger}}]{2022arXiv220605491K}
---. 2022{\natexlab{c}}, arXiv e-prints, arXiv:2206.05491.
\newblock \doarXiv{2206.05491}

\bibitem[{{Kobayashi} {et~al.}(2020){Kobayashi}, {Karakas}, \&
  {Lugaro}}]{2020ApJ...900..179K}
{Kobayashi}, C., {Karakas}, A.~I., \& {Lugaro}, M. 2020, \apj, 900, 179,
  \dodoi{10.3847/1538-4357/abae65}

\bibitem[{{Koch} {et~al.}(2013){Koch}, {Feltzing}, {Ad{\'e}n}, \&
  {Matteucci}}]{2013A&A...554A...5K}
{Koch}, A., {Feltzing}, S., {Ad{\'e}n}, D., \& {Matteucci}, F. 2013, \aap, 554,
  A5, \dodoi{10.1051/0004-6361/201220742}

\bibitem[{{Koppelman} {et~al.}(2019){Koppelman}, {Helmi}, {Massari},
  {Price-Whelan}, \& {Starkenburg}}]{2019A&A...631L...9K}
{Koppelman}, H.~H., {Helmi}, A., {Massari}, D., {Price-Whelan}, A.~M., \&
  {Starkenburg}, T.~K. 2019, \aap, 631, L9, \dodoi{10.1051/0004-6361/201936738}

\bibitem[{{Kruijssen} {et~al.}(2019){Kruijssen}, {Pfeffer}, {Reina-Campos},
  {Crain}, \& {Bastian}}]{2019MNRAS.486.3180K}
{Kruijssen}, J.~M.~D., {Pfeffer}, J.~L., {Reina-Campos}, M., {Crain}, R.~A., \&
  {Bastian}, N. 2019, \mnras, 486, 3180, \dodoi{10.1093/mnras/sty1609}

\bibitem[{{Lambert}(1989)}]{1989AIPC..183..168L}
{Lambert}, D.~L. 1989, in American Institute of Physics Conference Series, Vol.
  183, Cosmic Abundances of Matter, ed. C.~J. {Waddington}, 168--199,
  \dodoi{10.1063/1.38011}

\bibitem[{{Lane} {et~al.}(2022){Lane}, {Bovy}, \&
  {Mackereth}}]{2022MNRAS.510.5119L}
{Lane}, J. M.~M., {Bovy}, J., \& {Mackereth}, J.~T. 2022, \mnras, 510, 5119,
  \dodoi{10.1093/mnras/stab3755}

\bibitem[{{Libeskind} {et~al.}(2020){Libeskind}, {Carlesi}, {Grand},
  {Khalatyan}, {Knebe}, {Pakmor}, {Pilipenko}, {Pawlowski}, {Sparre}, {Tempel},
  {Wang}, {Courtois}, {Gottl{\"o}ber}, {Hoffman}, {Minchev}, {Pfrommer},
  {Sorce}, {Springel}, {Steinmetz}, {Tully}, {Vogelsberger}, \&
  {Yepes}}]{2020MNRAS.498.2968L}
{Libeskind}, N.~I., {Carlesi}, E., {Grand}, R. J.~J., {et~al.} 2020, \mnras,
  498, 2968, \dodoi{10.1093/mnras/staa2541}

\bibitem[{{Lind} {et~al.}(2012){Lind}, {Bergemann}, \&
  {Asplund}}]{2012MNRAS.427...50L}
{Lind}, K., {Bergemann}, M., \& {Asplund}, M. 2012, \mnras, 427, 50,
  \dodoi{10.1111/j.1365-2966.2012.21686.x}

\bibitem[{{Majewski} {et~al.}(2017){Majewski}, {Schiavon}, {Frinchaboy},
  {Allende Prieto}, {Barkhouser}, {Bizyaev}, {Blank}, {Brunner}, {Burton},
  {Carrera}, {Chojnowski}, {Cunha}, {Epstein}, {Fitzgerald}, {Garc{\'\i}a
  P{\'e}rez}, {Hearty}, {Henderson}, {Holtzman}, {Johnson}, {Lam}, {Lawler},
  {Maseman}, {M{\'e}sz{\'a}ros}, {Nelson}, {Nguyen}, {Nidever}, {Pinsonneault},
  {Shetrone}, {Smee}, {Smith}, {Stolberg}, {Skrutskie}, {Walker}, {Wilson},
  {Zasowski}, {Anders}, {Basu}, {Beland}, {Blanton}, {Bovy}, {Brownstein},
  {Carlberg}, {Chaplin}, {Chiappini}, {Eisenstein}, {Elsworth}, {Feuillet},
  {Fleming}, {Galbraith-Frew}, {Garc{\'\i}a}, {Garc{\'\i}a-Hern{\'a}ndez},
  {Gillespie}, {Girardi}, {Gunn}, {Hasselquist}, {Hayden}, {Hekker}, {Ivans},
  {Kinemuchi}, {Klaene}, {Mahadevan}, {Mathur}, {Mosser}, {Muna}, {Munn},
  {Nichol}, {O'Connell}, {Parejko}, {Robin}, {Rocha-Pinto}, {Schultheis},
  {Serenelli}, {Shane}, {Silva Aguirre}, {Sobeck}, {Thompson}, {Troup},
  {Weinberg}, \& {Zamora}}]{2017AJ....154...94M}
{Majewski}, S.~R., {Schiavon}, R.~P., {Frinchaboy}, P.~M., {et~al.} 2017, \aj,
  154, 94, \dodoi{10.3847/1538-3881/aa784d}

\bibitem[{{Malhan} {et~al.}(2022){Malhan}, {Ibata}, {Sharma}, {Famaey},
  {Bellazzini}, {Carlberg}, {D'Souza}, {Yuan}, {Martin}, \&
  {Thomas}}]{2022ApJ...926..107M}
{Malhan}, K., {Ibata}, R.~A., {Sharma}, S., {et~al.} 2022, \apj, 926, 107,
  \dodoi{10.3847/1538-4357/ac4d2a}

\bibitem[{{Massari} {et~al.}(2019){Massari}, {Koppelman}, \&
  {Helmi}}]{2019A&A...630L...4M}
{Massari}, D., {Koppelman}, H.~H., \& {Helmi}, A. 2019, \aap, 630, L4,
  \dodoi{10.1051/0004-6361/201936135}

\bibitem[{{Masseron} {et~al.}(2021){Masseron}, {Osorio},
  {Garc{\'\i}a-Hern{\'a}ndez}, {Allende Prieto}, {Zamora}, \&
  {M{\'e}sz{\'a}ros}}]{2021A&A...647A..24M}
{Masseron}, T., {Osorio}, Y., {Garc{\'\i}a-Hern{\'a}ndez}, D.~A., {et~al.}
  2021, \aap, 647, A24, \dodoi{10.1051/0004-6361/202039484}

\bibitem[{{Matteucci}(2012)}]{2012ceg..book.....M}
{Matteucci}, F. 2012, {Chemical Evolution of Galaxies},
  \dodoi{10.1007/978-3-642-22491-1}

\bibitem[{{McMillan}(2017)}]{2017MNRAS.465...76M}
{McMillan}, P.~J. 2017, \mnras, 465, 76, \dodoi{10.1093/mnras/stw2759}

\bibitem[{{McWilliam}(1997)}]{1997ARA&A..35..503M}
{McWilliam}, A. 1997, \araa, 35, 503, \dodoi{10.1146/annurev.astro.35.1.503}

\bibitem[{{Miglio} {et~al.}(2017){Miglio}, {Chiappini}, {Mosser}, {Davies},
  {Freeman}, {Girardi}, {Jofr{\'e}}, {Kawata}, {Rendle}, {Valentini},
  {Casagrande}, {Chaplin}, {Gilmore}, {Hawkins}, {Holl}, {Appourchaux},
  {Belkacem}, {Bossini}, {Brogaard}, {Goupil}, {Montalb{\'a}n}, {Noels},
  {Anders}, {Rodrigues}, {Piotto}, {Pollacco}, {Rauer}, {Prieto}, {Avelino},
  {Babusiaux}, {Barban}, {Barbuy}, {Basu}, {Baudin}, {Benomar}, {Bienaym{\'e}},
  {Binney}, {Bland-Hawthorn}, {Bressan}, {Cacciari}, {Campante}, {Cassisi},
  {Christensen-Dalsgaard}, {Combes}, {Creevey}, {Cunha}, {Jong}, {Laverny},
  {Degl'Innocenti}, {Deheuvels}, {Depagne}, {Ridder}, {Matteo}, {Mauro},
  {Dupret}, {Eggenberger}, {Elsworth}, {Famaey}, {Feltzing}, {Garc{\'\i}a},
  {Gerhard}, {Gibson}, {Gizon}, {Haywood}, {Handberg}, {Heiter}, {Hekker},
  {Huber}, {Ibata}, {Katz}, {Kawaler}, {Kjeldsen}, {Kurtz}, {Lagarde},
  {Lebreton}, {Lund}, {Majewski}, {Marigo}, {Martig}, {Mathur}, {Minchev},
  {Morel}, {Ortolani}, {Pinsonneault}, {Plez}, {Moroni}, {Pricopi},
  {Recio-Blanco}, {Reyl{\'e}}, {Robin}, {Roxburgh}, {Salaris}, {Santiago},
  {Schiavon}, {Serenelli}, {Sharma}, {Aguirre}, {Soubiran}, {Steinmetz},
  {Stello}, {Strassmeier}, {Ventura}, {Ventura}, {Walton}, \&
  {Worley}}]{2017AN....338..644M}
{Miglio}, A., {Chiappini}, C., {Mosser}, B., {et~al.} 2017, Astronomische
  Nachrichten, 338, 644, \dodoi{10.1002/asna.201713385}

\bibitem[{{Miglio} {et~al.}(2021){Miglio}, {Chiappini}, {Mackereth}, {Davies},
  {Brogaard}, {Casagrande}, {Chaplin}, {Girardi}, {Kawata}, {Khan}, {Izzard},
  {Montalb{\'a}n}, {Mosser}, {Vincenzo}, {Bossini}, {Noels}, {Rodrigues},
  {Valentini}, \& {Mandel}}]{2021A&A...645A..85M}
{Miglio}, A., {Chiappini}, C., {Mackereth}, J.~T., {et~al.} 2021, \aap, 645,
  A85, \dodoi{10.1051/0004-6361/202038307}

\bibitem[{{Mints} \& {Hekker}(2019)}]{2019A&A...621A..17M}
{Mints}, A., \& {Hekker}, S. 2019, \aap, 621, A17,
  \dodoi{10.1051/0004-6361/201834256}

\bibitem[{{Montalb{\'a}n} {et~al.}(2021){Montalb{\'a}n}, {Mackereth}, {Miglio},
  {Vincenzo}, {Chiappini}, {Buldgen}, {Mosser}, {Noels}, {Scuflaire}, {Vrard},
  {Willett}, {Davies}, {Hall}, {Nielsen}, {Khan}, {Rendle}, {van Rossem},
  {Ferguson}, \& {Chaplin}}]{2021NatAs...5..640M}
{Montalb{\'a}n}, J., {Mackereth}, J.~T., {Miglio}, A., {et~al.} 2021, Nature
  Astronomy, 5, 640, \dodoi{10.1038/s41550-021-01347-7}

\bibitem[{{Myeong} {et~al.}(2019){Myeong}, {Vasiliev}, {Iorio}, {Evans}, \&
  {Belokurov}}]{2019MNRAS.488.1235M}
{Myeong}, G.~C., {Vasiliev}, E., {Iorio}, G., {Evans}, N.~W., \& {Belokurov},
  V. 2019, \mnras, 488, 1235, \dodoi{10.1093/mnras/stz1770}

\bibitem[{{Naidu} {et~al.}(2020){Naidu}, {Conroy}, {Bonaca}, {Johnson}, {Ting},
  {Caldwell}, {Zaritsky}, \& {Cargile}}]{2020ApJ...901...48N}
{Naidu}, R.~P., {Conroy}, C., {Bonaca}, A., {et~al.} 2020, \apj, 901, 48,
  \dodoi{10.3847/1538-4357/abaef4}

\bibitem[{{Naidu} {et~al.}(2022){Naidu}, {Ji}, {Conroy}, {Bonaca}, {Ting},
  {Zaritsky}, {van Son}, {Broekgaarden}, {Tacchella}, {Chandra}, {Caldwell},
  {Cargile}, \& {Speagle}}]{2022ApJ...926L..36N}
{Naidu}, R.~P., {Ji}, A.~P., {Conroy}, C., {et~al.} 2022, \apjl, 926, L36,
  \dodoi{10.3847/2041-8213/ac5589}

\bibitem[{{Nissen} \& {Schuster}(1997)}]{1997A&A...326..751N}
{Nissen}, P.~E., \& {Schuster}, W.~J. 1997, \aap, 326, 751

\bibitem[{{Nissen} \& {Schuster}(2010)}]{2010A&A...511L..10N}
---. 2010, \aap, 511, L10, \dodoi{10.1051/0004-6361/200913877}

\bibitem[{{Nomoto} {et~al.}(1997){Nomoto}, {Iwamoto}, {Nakasato}, {Thielemann},
  {Brachwitz}, {Tsujimoto}, {Kubo}, \& {Kishimoto}}]{1997NuPhA.621..467N}
{Nomoto}, K., {Iwamoto}, K., {Nakasato}, N., {et~al.} 1997, \nphysa, 621, 467,
  \dodoi{10.1016/S0375-9474(97)00291-1}

\bibitem[{{Nordlander} \& {Lind}(2017)}]{2017A&A...607A..75N}
{Nordlander}, T., \& {Lind}, K. 2017, \aap, 607, A75,
  \dodoi{10.1051/0004-6361/201730427}

\bibitem[{{Osorio} {et~al.}(2020){Osorio}, {Allende Prieto}, {Hubeny},
  {M{\'e}sz{\'a}ros}, \& {Shetrone}}]{2020A&A...637A..80O}
{Osorio}, Y., {Allende Prieto}, C., {Hubeny}, I., {M{\'e}sz{\'a}ros}, S., \&
  {Shetrone}, M. 2020, \aap, 637, A80, \dodoi{10.1051/0004-6361/201937054}

\bibitem[{{Pagnini} {et~al.}(2023){Pagnini}, {Di Matteo}, {Khoperskov},
  {Mastrobuono-Battisti}, {Haywood}, {Renaud}, \&
  {Combes}}]{2022arXiv221004245P}
{Pagnini}, G., {Di Matteo}, P., {Khoperskov}, S., {et~al.} 2023, \aap, 673,
  A86, \dodoi{10.1051/0004-6361/202245128}

\bibitem[{{Roederer} {et~al.}(2016){Roederer}, {Mateo}, {Bailey}, {Song},
  {Bell}, {Crane}, {Loebman}, {Nidever}, {Olszewski}, {Shectman}, {Thompson},
  {Valluri}, \& {Walker}}]{2016AJ....151...82R}
{Roederer}, I.~U., {Mateo}, M., {Bailey}, John~I., I., {et~al.} 2016, \aj, 151,
  82, \dodoi{10.3847/0004-6256/151/3/82}

\bibitem[{{Sahlholdt} {et~al.}(2019{\natexlab{a}}){Sahlholdt}, {Casagrande}, \&
  {Feltzing}}]{2019ApJ...881L..10S}
{Sahlholdt}, C.~L., {Casagrande}, L., \& {Feltzing}, S. 2019{\natexlab{a}},
  \apjl, 881, L10, \dodoi{10.3847/2041-8213/ab321e}

\bibitem[{{Sahlholdt} {et~al.}(2019{\natexlab{b}}){Sahlholdt}, {Feltzing},
  {Lindegren}, \& {Church}}]{2019MNRAS.482..895S}
{Sahlholdt}, C.~L., {Feltzing}, S., {Lindegren}, L., \& {Church}, R.~P.
  2019{\natexlab{b}}, \mnras, 482, 895, \dodoi{10.1093/mnras/sty2732}

\bibitem[{{Shetrone} {et~al.}(2015){Shetrone}, {Bizyaev}, {Lawler}, {Allende
  Prieto}, {Johnson}, {Smith}, {Cunha}, {Holtzman}, {Garc{\'\i}a P{\'e}rez},
  {M{\'e}sz{\'a}ros}, {Sobeck}, {Zamora}, {Garc{\'\i}a-Hern{\'a}ndez}, {Souto},
  {Chojnowski}, {Koesterke}, {Majewski}, \& {Zasowski}}]{2015ApJS..221...24S}
{Shetrone}, M., {Bizyaev}, D., {Lawler}, J.~E., {et~al.} 2015, \apjs, 221, 24,
  \dodoi{10.1088/0067-0049/221/2/24}

\bibitem[{{Smith} {et~al.}(2021){Smith}, {Bizyaev}, {Cunha}, {Shetrone},
  {Souto}, {Allende Prieto}, {Masseron}, {M{\'e}sz{\'a}ros}, {J{\"o}nsson},
  {Hasselquist}, {Osorio}, {Garc{\'\i}a-Hern{\'a}ndez}, {Plez}, {Beaton},
  {Holtzman}, {Majewski}, {Stringfellow}, \& {Sobeck}}]{2021AJ....161..254S}
{Smith}, V.~V., {Bizyaev}, D., {Cunha}, K., {et~al.} 2021, \aj, 161, 254,
  \dodoi{10.3847/1538-3881/abefdc}

\bibitem[{{Soderblom}(2010)}]{2010ARA&A..48..581S}
{Soderblom}, D.~R. 2010, \araa, 48, 581,
  \dodoi{10.1146/annurev-astro-081309-130806}

\bibitem[{{Stonkut{\.{e}}} {et~al.}(2016){Stonkut{\.{e}}}, {Koposov}, {Howes},
  {Feltzing}, {Worley}, {Gilmore}, {Ruchti}, {Kordopatis}, {Randich},
  {Zwitter}, {Bensby}, {Bragaglia}, {Smiljanic}, {Costado},
  {Tautvai{\v{s}}ien{\.{e}}}, {Casey}, {Korn}, {Lanzafame}, {Pancino},
  {Franciosini}, {Hourihane}, {Jofr{\'e}}, {Lardo}, {Lewis}, {Magrini},
  {Monaco}, {Morbidelli}, {Sacco}, \& {Sbordone}}]{2016MNRAS.460.1131S}
{Stonkut{\.{e}}}, E., {Koposov}, S.~E., {Howes}, L.~M., {et~al.} 2016, \mnras,
  460, 1131, \dodoi{10.1093/mnras/stw1011}

\bibitem[{{Tolstoy} {et~al.}(2009){Tolstoy}, {Hill}, \&
  {Tosi}}]{2009ARA&A..47..371T}
{Tolstoy}, E., {Hill}, V., \& {Tosi}, M. 2009, \araa, 47, 371,
  \dodoi{10.1146/annurev-astro-082708-101650}

\bibitem[{{Vasiliev}(2019)}]{2019MNRAS.484.2832V}
{Vasiliev}, E. 2019, \mnras, 484, 2832, \dodoi{10.1093/mnras/stz171}

\bibitem[{{Weinberg} {et~al.}(2019){Weinberg}, {Holtzman}, {Hasselquist},
  {Bird}, {Johnson}, {Shetrone}, {Sobeck}, {Allende Prieto}, {Bizyaev},
  {Carrera}, {Cohen}, {Cunha}, {Ebelke}, {Fernandez-Trincado},
  {Garc{\'\i}a-Hern{\'a}ndez}, {Hayes}, {J{\"o}nsson}, {Lane}, {Majewski},
  {Malanushenko}, {M{\'e}sz{\'a}ros}, {Nidever}, {Nitschelm}, {Pan}, {Rix},
  {Rybizki}, {Schiavon}, {Schneider}, {Wilson}, \&
  {Zamora}}]{2019ApJ...874..102W}
{Weinberg}, D.~H., {Holtzman}, J.~A., {Hasselquist}, S., {et~al.} 2019, \apj,
  874, 102, \dodoi{10.3847/1538-4357/ab07c7}

\bibitem[{{Zhang} {et~al.}(2017){Zhang}, {Shi}, {Pan}, {Allende Prieto}, \&
  {Liu}}]{2017ApJ...835...90Z}
{Zhang}, J., {Shi}, J., {Pan}, K., {Allende Prieto}, C., \& {Liu}, C. 2017,
  \apj, 835, 90, \dodoi{10.3847/1538-4357/835/1/90}

\end{thebibliography}
\bibliographystyle{aasjournal}

\end{document}